\documentclass[%
 aip,jcp,
 amsmath,amssymb,
 preprint,
]{revtex4-1}

\usepackage{graphicx}
\usepackage{dcolumn}
\usepackage{bm}

\usepackage[utf8]{inputenc}
\usepackage[T1]{fontenc}
\usepackage{mathptmx}
\usepackage{listings}
\usepackage{rotating} 

\usepackage{color}
\usepackage{dcolumn} 
\usepackage{bm} 
\usepackage{graphicx}
\usepackage{multirow} 
\usepackage{pifont} 
\usepackage{epsfig}
\usepackage{amsmath} 
\usepackage{subfigure}
\usepackage{float}
\usepackage{booktabs}
\usepackage{tabularx}
\usepackage{natbib}
\usepackage{gensymb}
\usepackage{hyperref}
\usepackage{rotating}
\usepackage[normalem]{ulem}
\newcommand{\bhd}{\hat{b}^{\dagger}}
\newcommand{\bh}{\hat{b}}
\newcommand{\omc}{\hbar \omega_{{\rm cav}}}

\begin{document}

\author{Jonathan J. Foley IV}
\affiliation{
             Department of Chemistry, 
             University of North Carolina Charlotte, 
             Charlotte, North Carolina, 28223}
\email{jfoley19@charlotte.edu}
\author{Jonathan F. McTague\textsuperscript{\textbar \textbar}}
\affiliation{Department of Chemistry
             William Paterson University
             Wayne, New Jersey, 07470} 
\author{A. Eugene DePrince III}
\affiliation{
             Department of Chemistry and Biochemistry,
             Florida State University,
             Tallahassee, FL 32306-4390}
\email{adeprince@fsu.edu}

\title{{\em Ab initio} methods for polariton chemistry}

\begin{abstract}
Polariton chemistry exploits the strong interaction between quantized excitations in molecules and quantized photon states in optical cavities to affect chemical reactivity.  Molecular polaritons have been experimentally realized by the coupling of electronic, vibrational, and rovibrational transitions to photon modes, which has spurred tremendous theoretical effort to model and explain how polariton formation can influence chemistry.  This tutorial
review focuses on {\color{black}computational approaches for the electronic strong coupling problem through the combination of }
familiar techniques from \textit{ab initio} electronic structure theory {\color{black}and} cavity quantum electrodynamics, toward the goal of supplying predictive theories for polariton chemistry. Our aim is to emphasize the relevant theoretical details with enough clarity for newcomers to the field to follow, and to present simple and practical code examples to catalyze further development work.

\end{abstract}

\maketitle

\section{Introduction}

Strong interactions between nanoconfined photons and molecular systems\cite{Nori19_19,Narang18_1479,Barnes14_013901} can lead to the creation of hybrid light--matter states known as polaritons that may display remarkably different chemical and physical properties than their parent components.\cite{Whittaker98_6697,Mugnier04_036404,Ebbesen12_1592,Smith14_5561,Ebbesen15_1123,Baumberg16_127,Ebbesen16_2403,Nitzan17_443003,Ebbesen17_9034,Bellessa19_173902,Forrest20_371,Weichman23_JACS} The technological and chemical applications of these strongly-coupled light--matter states are wide ranging. Recent examples of cavity control of chemical reactivity and catalysis,\cite{George19_10635,Feist19_8698, Simpkins_ARPC_2023} polariton lasing,\cite{Forrest20_371} manipulation of non-linear optical effects in organic molecules~\cite{CG_NatureCommun_2022}, optical energy propagation,\cite{Basu20_5043, Musser_AS_2022}  plasmon-based photostabilization,\cite{Shegai18_eaas9552} plasmon-based multimode vibrational strong couplingz~,\cite{Sheldon_JCP_2021} Bose-Einstein condensation of molecular exciton-polaritons,\cite{SKC_AOM_2021} and protection against decoherence processes\cite{Franco_JPCL_2022} offer only a glimpse into the transformative potential of polaritonic approaches to chemistry and materials science. In order for the field to fully live up to its promise, the experimental realization of strong and ultra-strong light--matter coupling must be accompanied by high-quality theoretical descriptions of the emergence and properties of molecular polaritons.

There have been several excellent review and perspective articles focusing on theoretical advances
related to polaritonic chemistry.  Theoretical challenges in polaritonic chemistry bridge most domains of chemical physics, including polaritonic structure, dynamics, statistical thermodynamics, and rate theories as pointed out by a recent comprehensive review 
by Huo and co-workers~\cite{Huo22_Chemrxiv} and an incisive perspective by Feist and co-workers~\cite{Feist22_ACSPhoton}.  Ruggenthaler \textit{et al.} have contributed a rigorous review of several promising directions in {\it ab initio} cavity quantum electrodynamics (QED) methods with a particular emphasis on real-space approaches to bridge density functional theory and its real-time extensions with cavity QED; the resulting QEDFT approach\cite{Rubio22_Arxiv} has played an important role in simulating polaritonic structure.  In this tutorial review, we also focus on the problem of simulating polaritonic structure through the lens of \textit{ab initio} cavity QED, but we emphasize emerging methods implemented with Gaussian basis sets. Throughout, we refer to \textit{ab initio} cavity QED methods (whether in Guassian or real-space grid bases) as those where the starting point is a single time-independent Schr\"odinger equation for charged particles comprising a molecular system coupled to quantized photonic degrees of freedom.  These methods can be seen to be complementary to parameterized cavity QED (pCQED) methods where one essentially considers solving two Schr\"odinger equations in series: a first for the molecular system, and the second for the coupled molecular-photonic system that is parameterized by the solutions to the 
first.\cite{Huo22_Chemrxiv,Huo23_Chemrxiv, Huo23_JCTC}  As a tutorial review, our aim is to provide a level of technical detail sufficient for newcomers to the field to implement some of the more introductory methods and start applying them as-is, or to leverage these implementations to seed new or more elaborate methodological developments.  In addition to the discussion of the related theory in text, we provide example code in a tutorial style that utilizes the Psi4Numpy framework for a QED-Hartree-Fock self-consistent field method and a QED-Configuration Interaction Singles method. We will present some illustrative calculations utilizing these methods, and also discuss results from the literature for methods beyond those for which we have provided tutorial implementations.

Historically, theoretical descriptions of strong light--matter interactions have been built upon simple model Hamiltonians that describe interactions between two- or few-level quantum emitters and a single photon modes. For electronic strong coupling in polariton chemistry, the Jaynes-Cummings model provides such an example.  Here two states of the quantum emitter are parameterized by the ground- and excited-state energies, and these states couple to the photon mode through a dipolar transition;
see, for example, Ref.~\onlinecite{Huo22_Chemrxiv} for a derivation and detailed discussion of this model.  Such models are powerful tools for simulating qualitative changes to properties of molecular systems strongly coupled to nanoconfined photons,\cite{Carusotto06_033811,Genes19_203602,Nori19_19} offering essential insight, for example, into optical changes that can be induced by manipulating the energy content of an external field\cite{Gray13_075411} or into changes in chemical reactivity\cite{Huo19_5519} or rates of electron transfer reactions.\cite{Huo20_6321} While such simulations improve our qualitative understanding of many problems, quantitative predictions of chemical reactivity or orbital-specific quantities ({\em e.g.}, ionization potentials) within optical cavities or other nanoconfined environments necessitate an {\em ab inito} approach to light--matter interactions or a pCQED treatement with a sufficiently large basis of molecular and photonic eigenstates\cite{Huo22_Chemrxiv}.

The most conceptually straightforward strategy to realize an {\em ab initio} polaritonic model is to generalize an existing methodology to treat more than one type of quantum-mechanical particle -- namely, for the description of both electrons and photons. Following this scheme, approaches based on quantum electrodynamics generalizations of density functional theory (QEDFT\cite{Bauer11_042107,Tokatly13_233001,Rubio14_012508,Rubio15_093001,Rubio18_992,Appel19_225,Narang20_094116,Rubio22_Arxiv} and QED-DFT\cite{DePrince22_9303,Rubio22_094101,DePrince23_5264}), configuration interaction (QED-CIS)~\cite{Foley_154103}, and coupled cluster (QED-CC)~\cite{Koch21_094113} have emerged.   An alternative and perhaps more direct description of polaritonic structure could be obtained from a theory designed from the outset with a different particle type, the polariton, in mind.\cite{Rubio18_arxiv,Rubio19_2694,Rubio20_5601} This approach could be the more natural one, but, in the framework outlined in Ref.~\citenum{Rubio20_5601}, the technical challenge of designing algorithms for treating multiple types of quantum-mechanical particles is supplanted by a new problem: enforcing the correct Fermi-Bose statistics on the polaritonic wave function. In either case, the vast majoriy of polaritonic quantum chemical models are built upon density functional theory (DFT). For many applications, DFT offers an excellent balance of accuracy and computational affordability. However, DFT suffers from a number of well-known deficiencies\cite{Yang08_792} that are no doubt inherited by polaritonic extensions of the model and potentially limit its applicability to arbitrary polaritonic problems. Hence, while this review article touches on QED generalizations of DFT, the main focus is wave function methods.

\section{The Pauli-Fierz Hamiltonian}
The starting point for our presentation of {\em ab initio} polaritonic structure theory is the Pauli-Fierz  (PF) Hamiltonian,\cite{Spohn04_book,Rubio18_0118} represented in the length gauge and within the dipole and Born-Oppenheimer approximations. An excellent pedagogical discussion and derivation of this Hamiltonian from the minimal coupling Hamiltonian in the Coulomb gauge can found in a recent papers and reviews by Huo and co-workers.\cite{Huo22_Chemrxiv,Huo20_9215} Here, we briefly outline some key details, assuming a single photon mode for simplicity, but the Hamiltonian we derive can be generalized for multiple modes.  It has been shown that the inclusion of multiple modes can profoundly impact ground-state and excited-state polariton surfaces, and physichemical process in model systems.\cite{Hoffman_JCP_2020}  Most {\it ab inito} cavity QED studies to date have considered only a single mode, so multi-mode effects represent an important area to explore in future work. 

\begin{figure}
    \centering
    \includegraphics[width=0.3\textwidth]{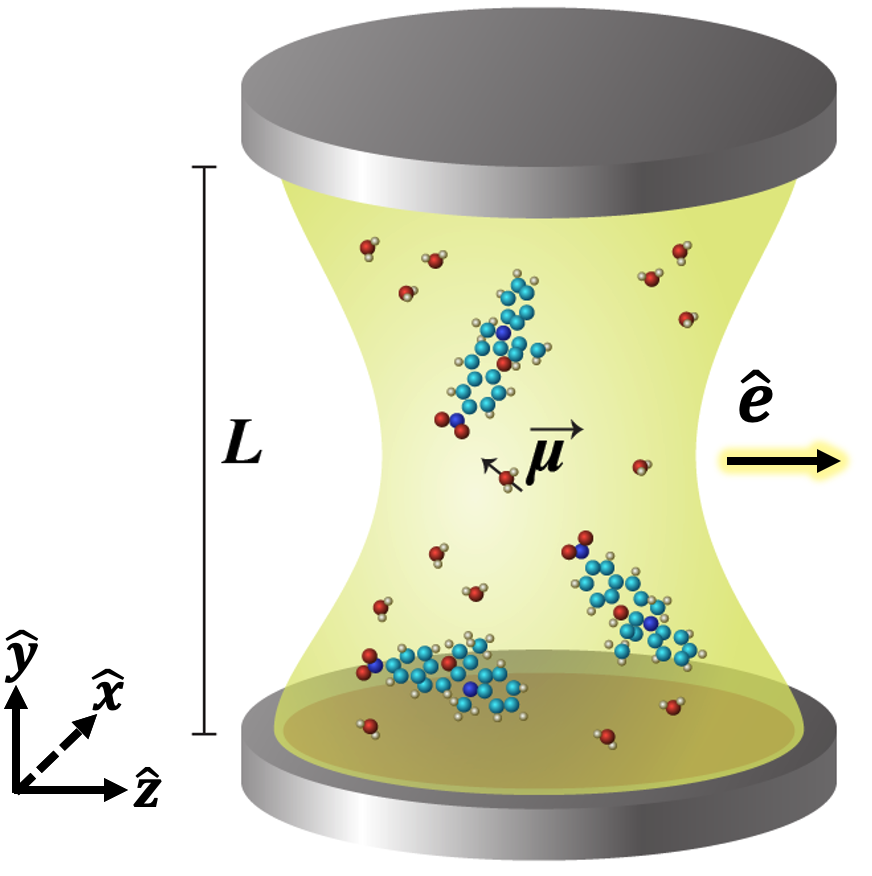}
    \caption{Schematic of a Fabry-Perot cavity containing molecular subsystems.  The
    transverse component of the vector potential and the associated electric field component points along the vector $\bf{\hat{e}}$, which is parallel to the $z$-axis in this scheme. The cavity mode volume is determined by $L^3$. }
    \label{fig:cavity-schematic}
\end{figure}

We begin with the minimal coupling Hamiltonian in the Coulomb gauge,
\begin{equation}
\label{EQN:PdotA}
\hat{H}_{{\rm p \cdot A}} = \sum_i^N \frac{1}{2m_i}\left(\hat{p}_i - z_i \hat{{\bf A}}_{\perp} \right)^2 + \hat{V}({\bf \hat{x}}) + \hbar \omega_{\rm cav} \bhd \bh,
\end{equation}
where the subscript on the Hamiltonian denotes that this operator is also referred to as the "${\rm p} \cdot {\rm A}$" Hamiltonian.\cite{Huo20_9215}  The sum runs over all charged particles (electrons and nuclei in molecular systems), $\hat{p}_i$ and $z_i$ are the momentum operator and charge for particle $i$, respectively, $\hat{{\bf A}}_{\perp}$ is the transverse component of the vector potential \textcolor{black}{which points along $\bf{\hat{e}}$ shown in Figure~\ref{fig:cavity-schematic}}, $ \hat{V}({\bf \hat{x}})$ is the Coulomb potential operator for all pairs of charged particles, and $\hbar \omega_{\rm cav} \bhd \bh$ captures the photon energy. The symbols $\hat{b}^\dagger$ and $\hat{b}$ are photonic creation and annihilation operators, respectively.  Important properties of the photonic creation and annihilation operators include their action on photon number states,
\begin{align}
    \bhd |n\rangle &= \sqrt{n+1}|n+1\rangle \\
    \bh |n\rangle &= \sqrt{n}|n-1\rangle \\
    \bhd \bh |n\rangle &= n|n\rangle,
\end{align}
and their commutation relations
\begin{align}
    [\bh, \bhd] &= 1 \\
    [\bhd, \bh] &= -1.
\end{align}

In Eq.~\ref{EQN:PdotA}, the coupling between light and matter is captured by first term, which includes the matter momenta and the product of the matter charges and the vector potential; note that, in the Coulomb gauge, the vector potential is purely transverse.  The "${\rm p} \cdot {\rm A}$" Hamiltonian in the Coulomb gauge is quite natural for formulations of {\it ab inito} QED represented in 
a real-space grid basis, and so approaches such as QEDFT are formulated in this gauge.\cite{Rubio22_Arxiv}  However,
because momentum eigenfunctions are delocalized functions, capturing the coupling matrix elements between in 
the "${\rm p} \cdot {\rm A}$" representation is challenging for formulations that utilize Gaussian basis sets, which 
are inherently localized in space.  Therefore, PF Hamiltonian in the length gauge that we seek may be obtained from $\hat{H}_{{\rm p \cdot A}}$ via a gauge transformation, known as the Power-Zienau-Wooley (PZW) transformation, followed by a unitary phase transformation. The PZW transformation operator is
\begin{equation}\label{EQN:UPZW}
    \hat{U}_{{\rm PZW}} = {\rm exp}\left( -\frac{i}{\hbar} \hat{{\bm \mu}} \cdot \hat{{\bf A}} \right),
\end{equation}
where $\hat{{\bf A}} ={\bf A}_{\rm 0} \left(\bh + \bhd \right) $ and ${\bf A}_{\rm 0} = \sqrt{\frac{\hbar}{2\omega_{{\rm cav}} \epsilon_0 V}}\hat{{\bf e}}$ is the vector potential of the cavity photon, which is still purely transverse but we are dropping the $\perp$ for simplicity.  \textcolor{black}{We can see that the cavity mode volume derives from the cavity length $L$ shown in Figure~\ref{fig:cavity-schematic} as $V = L^3$, and sets the magnitude of the vector potential.}

Let's consider the PZW transform of each term in Eq.~\ref{EQN:PdotA}.  As noted in Ref.~\onlinecite{Huo20_9215}, the 
PZW operator boosts the momentum operator by an amount $z \hat{{\bf A}} $.  To see why this is the case, consider the BCH expansion of
this transformation for the light-matter coupling term for a single particle
with charge $z$:
\begin{equation}
\hat{U}_{{\rm PZW}} \left(\hat{p} - z \hat{{\bf A}} \right) \hat{U}^{\dagger}_{{\rm PZW}} = e^{\hat{B}} \hat{C} e^{-\hat{B}} = \hat{C} + [\hat{B},\hat{C}] + \frac{1}{2} [\hat{B},[\hat{B},\hat{C}]] + ... 
\end{equation}
where $\hat{C} = \left(\hat{p} - z \hat{{\bf A}}\right)$ and $\hat{B} = -\frac{i}{\hbar}z\hat{{\bf A}} \hat{x}$,
and we have used the fact that the dipole operator $\hat{\mu} = z\hat{x}$.  Because $\hat{{\bf A}}$ commutes with itself, we have
$[\hat{C}, \hat{B}] = -\frac{i}{\hbar}z\hat{{\bf A}} [\hat{x},\hat{p}] = z\hat{{\bf A}}$, and all subsequent commutators equal to zero.
Thus, we can see that
\begin{equation}
 \hat{U}_{{\rm PZW}} \left(\hat{p} - z \hat{{\bf A}} \right) \hat{U}^{\dagger}_{{\rm PZW}} =    \left(\hat{p} - z \hat{{\bf A}} \right) + z\hat{{\bf A}} = \hat{p}.
\end{equation}
Consequently, the first term in the PZW trasnformation of Eq.~\ref{EQN:PdotA} becomes
\begin{align}
\hat{U}_{{\rm PZW}} \sum_i^N \frac{1}{2m_i}\left(\hat{p}_i - z_i \hat{{\bf A}} \right)^2 \hat{U}^{\dagger}_{{\rm PZW}} = \sum_i^N \frac{1}{2m_i}\hat{p}_i.
\end{align}
Both $\hat{{\bf A}}$ and $\hat{x}$ commute with $\hat{V}({\bf \hat{x}})$, so we
have
\begin{equation}
 \hat{U}_{{\rm PZW}} \hat{V}({\bf \hat{x}}) \hat{U}^{\dagger}_{{\rm PZW}} = \hat{V}({\bf \hat{x}}).
\end{equation}
Finally, we have 
\begin{equation}
\hat{U}_{{\rm PZW}} \; \hbar \omega_{{\rm cav}} \bhd \bh  \; \hat{U}^{\dagger}_{{\rm PZW}} = e^{\hat{B}} \hat{C} e^{-\hat{B}} = \hat{C} + [\hat{B}, \hat{C}] + \frac{1}{2} [\hat{B},[\hat{B},\hat{C}]] + ... 
\end{equation}
where we will call $\hat{C} = \omc \bhd \bh$ and $\hat{B} = {\bf g}(\bh + \bhd)$, where ${\bf g} = -\frac{i}{\hbar}\hat{{\bm \mu}} \cdot {\bf A}_0$.
The first commutator gives
\begin{equation}
\omc {\bf g} \: [(\bh + \bhd),\bhd \bh] = -\omc {\bf g} (\bhd - \bh),
\end{equation}
and the second commutator gives
\begin{equation}
-\frac{1}{2} \omc {\bf g}^2 [(\bh + \bhd), (\bhd - \bh)] = -\omc {\bf g}^2, 
\end{equation}
so that this term overall reads
\begin{equation}\label{EQN:PZW_HP}
\hat{U}_{{\rm PZW}} \; \hbar \omega_{{\rm cav}} \bhd \bh  \; \hat{U}^{\dagger}_{{\rm PZW}} = \omc \bhd \bh + i\omega_{{\rm cav}}\hat{{\bm \mu}} \cdot {\bf A}_0(\bhd - \bh) +\frac{\omega_{{\rm cav}}}{\hbar} ( \hat{{\bm \mu}} \cdot {\bf A}_0)^2.
\end{equation}
Combining all terms gives the Hamiltonian in the dipole gauge, also called the "${\rm d} \cdot {\rm E}$" Hamiltonian:\cite{Huo22_Chemrxiv}
\begin{equation}\label{EQN:ddotE}
\hat{H}_{{\rm d} \cdot {\rm E}} = \sum_i^N \frac{\hat{p}_i^2}{2m_i} + \hat{V}(\hat{\bf x}) + \omc \bhd \bh + i\omega_{{\rm cav}}\hat{{\bm \mu}} \cdot {\bf A}_0(\bhd - \bh) +\frac{\omega_{{\rm cav}}}{\hbar} ( \hat{{\bm \mu}} \cdot {\bf A}_0)^2.
\end{equation}
To derive the Pauli-Fierz Hamiltonian from Eq.~\ref{EQN:ddotE} we apply a unitary phase transformation defined by the operator
\begin{equation}\label{EQN:UPHI}
\hat{U}_{\phi} =  {\rm exp}\left( i \frac{\pi}{2} \bhd \bh \right),
\end{equation}
which transforms the photonic operators as follows:
\begin{align}
\hat{U}_{\phi}\bhd \hat{U}^{\dagger}_{\phi} = i\bhd \\ \nonumber
\hat{U}_{\phi}\bh \hat{U}^{\dagger}_{\phi} = -i\bh \\
\hat{U}_{\phi}\bhd \bh \hat{U}^{\dagger}_{\phi} = \bhd \bh \nonumber.
\end{align}
Thus, the Pauli-Fierz Hamiltonian can be defined as
\begin{equation}\label{EQN:H_PF_General}
\hat{H}_{PF} = \hat{U}_{\phi} \hat{H}_{{\rm d} \cdot {\rm E}} \hat{U}^{\dagger}_{\phi} = \sum_i^N \frac{\hat{p}_i^2}{2m_i} + \hat{V}(\hat{\bf x}) + \omc \bhd \bh - \omega_{{\rm cav}}\hat{{\bm \mu}} \cdot {\bf A}_0(\bhd + \bh) +\frac{\omega_{{\rm cav}}}{\hbar} ( \hat{{\bm \mu}} \cdot {\bf A}_0)^2.
\end{equation}
It is common to define the coupling vector ${\bm \lambda} = \sqrt{\frac{\hbar}{\epsilon_0 V} } \hat{{\bf e}}$, and so after
recalling the definition 
${\bf A}_{\rm 0} = \sqrt{\frac{\hbar}{2\omega_{{\rm cav}} \epsilon_0 V}} \hat{{\bf e}}$, 
we can write ${\bf A}_{\rm 0} = \sqrt{\frac{1}{2\omega_{\rm cav}}} {\bm \lambda }$.  At this point, the sum $\sum_i^N \frac{\hat{p}_i^2}{2m_i} = \hat{T}_{\rm e} + \hat{T}_{\rm N}$ runs over the electrons and nuclei, and potential operator
$\hat{V}(\hat{\bf x}) = \hat{V}_{\rm ee} + \hat{V}_{\rm eN} + \hat{V}_{\rm NN}$ includes electron-electron repulsion, electron-nuclear attraction, and nuclear-nuclear repulsion operators.  We will invoke the Born-Oppenheimer approximation, which fixes the nuclei and eliminates the nuclear kinetic energy operator, and makes the nuclear-nuclear repulsion a constant for a given 
molecular geometry.
With these definitions in mind, we write the Pauli-Fierz Hamiltonian~\cite{Spohn04_book,Rubio18_0118} in the length gauge and within the dipole and Born-Oppenheimer approximations and in atomic units as follows:
\begin{equation}
    \label{EQN:PFH_LENGTH_GAUGE}
    \hat{H} = \hat{H}_{\rm e} + \omega_{\rm cav} \hat{b}^\dagger \hat{b} - \sqrt{ \frac{\omega_{\rm cav}}{2} } ({\bm \lambda} \cdot \hat{{\bm \mu}} )(\hat{b}^\dagger +\hat{b}) + \frac{1}{2} ({\bm \lambda} \cdot \hat{{\bm \mu}} )^2.
\end{equation}
Here, $\hat{H}_{\rm e}$ represents the electronic Hamiltonian that arises in standard electronic structure theories when
the Born-Oppenheimer approximation is imposed on the charged particles captured
by the $\sum_i^N \frac{\hat{p}_i^2}{2m_i} + \hat{V}(\hat{\bf x})$ term 
in Eq.~\ref{EQN:H_PF_General}.
The second term $\hat{H}_{\rm cav} = \omega_{\rm cav} \hat{b}^{\dagger} \hat{b}$ represents the Hamiltonian for the cavity mode, which is a harmonic oscillator with fundamental frequency $\omega_{\rm cav}$.
The last two terms are the bilinear coupling, $\hat{H}_{\rm blc} =  \sqrt{ \frac{\omega_{\rm cav}}{2} } ({\bm \lambda} \cdot \hat{{\bm \mu}} )(\hat{b}^\dagger +\hat{b})$, and dipole self-energy terms $\hat{H}_{\rm DSE} = \frac{1}{2}\left( {\bm \lambda} \cdot \hat{{\bm \mu}} \right)^2$, respectively. We will assume a Cartesian coordinate system where ${\bm \lambda}$ and
$\hat{{\bm \mu}}$ will have $x$, $y$, and $z$ components.  \textcolor{black}{As is shown in Figure~\ref{fig:cavity-schematic}, the projection of the cartesian components of the molecular dipole operator onto $\bf{\hat{e}}$ will arise from the orientation of the molecule relative to the cavity polarization.  We note that for molecules in the gas phase, the orientations sampled will depend on the distribution of rotational states occupied by the molecules at the temperature of setup.}  The molecular dipole operator $\hat{{\bm \mu}}$ has an electronic and a nuclear contributions, {\em i.e.}, 
$\hat{{\bm \mu}} = \hat{{\bm \mu}}_{\rm e} + {\bm \mu}_{\rm n}$. In the Born-Oppenheimer approximation, the nuclear contribution is a constant for a given geometry. {\color{black}While this review focuses on purely electronic coupling, a more complete description of the polaritonic structure could also consider nuclear effects, in which case the breakdown of the Born-Oppenheimer approximation may have considerable consequences on predicted spectra and dynamics of these systems. As an example, simulated vibrationally-resolved electronic spectra can vary dramatically depending on whether they are computed in the adiabatic or diabatic representation and also on the particular diabatization scheme.\cite{Knowles22_204119}  }

In the following sections, we use standard labeling notation for molecular spin orbitals, {\em i.e.}, labels $i$, $j$, $k$, and $l$ refer to electronic molecular spin-orbitals that are occupied in a reference configuration, and labels $a$, $b$, $c$, $d$ refer to unoccupied electronic molecular spin-orbitals. General electronic molecular orbitals will be indexed by $p$, $q$, $r$, and $s$, and electronic atomic orbitals will be indexed by Greek labels. Unless otherwise noted, all electronic orbital labels refer to spin-orbitals. The symbols $\hat{a}^\dagger$ and $\hat{a}$ will represent fermionic creation and annihilation operators, respectively, while $\hat{b}^\dagger$ and $\hat{b}$ will represent the bosonic equivalents.

\section{Mean-Field Cavity QED}
As our first step in approximating the energy eigenstates of Eq.~\eqref{EQN:PFH_LENGTH_GAUGE}, we introduce the cavity quantum electrodynamics Hartree-Fock (QED-HF) method based on the reference wavefunction
\begin{equation}
\label{EQN:QEDHF}
    |0^{\rm e}0^{\rm p}\rangle = |0^{\rm e}\rangle \otimes |0^{\rm p}\rangle
\end{equation}
which is a direct product of a Slater determinant of electronic spin orbitals ($|0^{\rm e}\rangle$) and a zero-photon state ($|0^{\rm p}\rangle$). This zero-photon state is defined as a linear combination of photon-number states
\begin{equation}
|0^{\rm p}\rangle = \sum_n (\hat{b}^\dagger)^n |0\rangle c_n
\end{equation}
where $|0\rangle$ represents the photon vacuum.
The functions $|0^{\rm e}\rangle$ and $|0^{\rm p}\rangle$ can be determined via the following modified Roothaan-Hall procedure. In the first step, the electronic wavefunction can be determined as the Slater determinant that minimizes the expectation value of Eq.~\eqref{EQN:PFH_LENGTH_GAUGE}, given a fixed zero-photon state. Second, given $|0^{\rm e}\rangle$, we integrate out the electronic degrees of freedom of Eq.~\eqref{EQN:PFH_LENGTH_GAUGE} to obtain a photon Hamiltonian 
\begin{equation}
    \hat{H}_{\rm p} = \langle 0^{\rm e} | \hat{H} | 0^{\rm e}\rangle
\end{equation}
the lowest eigenfunction of which is $|0^{\rm p}\rangle$. In practice, $|0^{\rm p}\rangle$ can be determined by expanding $\hat{H}_{\rm p}$ in a basis of photon-number states and bringing it to diagonal form. This two-step procedure should be repeated until self-consistency. 

\begin{figure}
    \centering
    \includegraphics{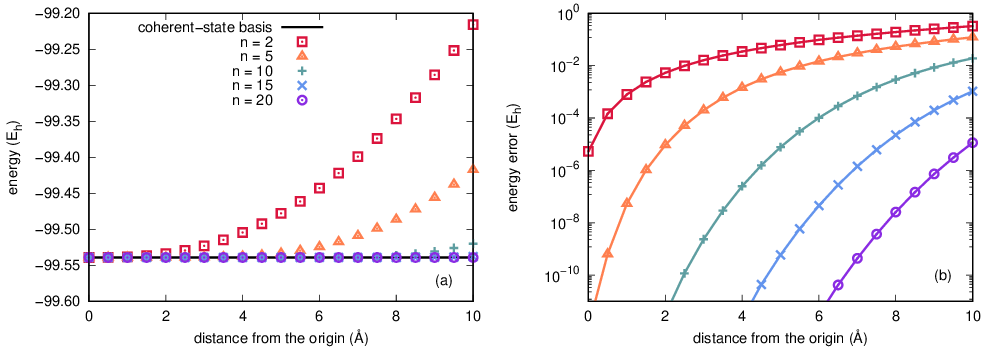}
    \caption{Origin dependence of the QED-Hartree-Fock energy when varying the number of photon-number states used to represent the photon Hamiltonian, $\hat{H}_{\rm p}$ for a hydrogen fluoride cation.} 
    \label{FIG:PHOTON_CONVERGENCE}
\end{figure}

One key detail in this procedure is that incorrect behavior can be recovered if the photon space is not fully converged. As an example, Fig.~\ref{FIG:PHOTON_CONVERGENCE}(a) illustrates the QED-HF energy for a cavity-bound hydrogen fluoride cation (described by the cc-pVQZ basis set) as the molecule is moved away from the origin. Here,  the cation is coupled to a single-mode cavity with a fundamental frequency of 2 eV,  the cavity mode is polarized along the molecular axis, the coupling strength, $\lambda$, is 0.05 atomic units, and the H--F distance is fixed at 0.917 \AA~throughout the translation. The QED-HF energy should be origin invariant, but, as is evident from the data, the correct invariance properties are only observed in the limit that the photon basis is complete. Figure \ref{FIG:PHOTON_CONVERGENCE}(b) illustrates the error in the QED-HF energy, with respect to calculations carried out in the so-called ``coherent-state basis,''\cite{Koch20_041043} which, as discussed below, yields results that are equivalent to those obtained with a complete photon basis. Here, we can see that even with 20 photon number states, the QED-HF energy is still not strictly origin invariant, and this issue is more pronounced the farther from the origin the molecule is placed.

\begin{figure}
    \centering
    \includegraphics{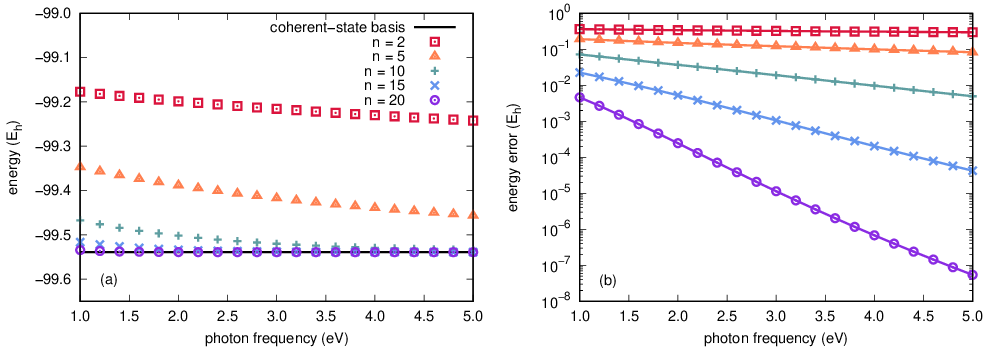}
    \caption{Frequency dependence of the QED-Hartree-Fock energy when varying the number of photon-number states used to represent the photon Hamiltonian, $\hat{H}_{\rm p}$ for hydrogen fluoride cation.}
    \label{FIG:PHOTON_FREQUENCY_DEPENDENCE}
\end{figure}

Aside from origin invariance, the QED-HF energy should be independent of the photon frequency;\cite{Koch20_041043} any polaritonic wave function that is factorizable as a product of an electronic wave function and a photonic wave function should have this property. Figure \ref{FIG:PHOTON_FREQUENCY_DEPENDENCE} illustrates the frequency dependence of the QED-HF energy for the same cavity-bound hydrogen fluoride cation when the molecule is placed 10 \AA~from the origin. Clearly, an incomplete photon basis leads to an incorrect frequency dependence in the QED-HF energy. The errors with respect to calculations carried out in the coherent-state basis depicted in Fig.~\ref{FIG:PHOTON_FREQUENCY_DEPENDENCE}(b) demonstrate that errors due to the incompleteness of the photon basis can be quite large, even when considering 20 photon number states. In this case, errors larger than $10^{-3}$ E$_{\rm h}$ are observed for cavity mode frequencies less than 1.5 eV; these errors become much smaller as the photon frequency increases.

As alluded to above, an equivalent representation of ground-state QED-HF involves representing the problem within the coherent-state basis,\cite{Koch20_041043} which is the basis that diagonalizes $\hat{H}_{\rm p}$. In this way, we avoid the need to solve the second step of the modified Roothaan-Hall procedure described above and automatically ensure convergence of the procedure with respect to the number of photon-number states. In the coherent-state basis, we need only solve the electronic problem with a transformed Hamiltonian, the form of which is derived in the next subsection.

\subsection{Coherent-State Transformation of the Hamiltonian}\label{SEC:CoherentStateBasis}

As noted in Ref.~\citenum{Koch20_041043}, $|0^{\rm p}\rangle$ can be exactly defined with a unitary coherent-state transformation operator of the form
\begin{equation}
\label{EQN:U_CS}
    \hat{U}_{\rm CS} = {\rm exp}\left( z(\hat{b}^{\dagger} - \hat{b}) \right) 
\end{equation}
were $z$ is a parameter defined such that $\hat{U}_{\rm CS} \hat{H}_{\rm p} \hat{U}^{\dagger}_{\rm CS}$ is a diagonal operator: 
\begin{equation}\label{EQN:z}
    z = \frac{-{\bm \lambda} \cdot \langle {\bm{\hat{\mu}}} \rangle }{\sqrt{2 \omega_{\rm cav}}}.
\end{equation}
The term $\langle {\bm{\hat{\mu}}} \rangle$ in Eq.~\ref{EQN:z} represents the expectation value of the molecular dipole moment (with respect to the Slater determinant, $|0^{\rm e}\rangle$), which is also a vector quantity.
We can relate the photon vacuum to the zero-photon state through the unitary transformation
defined in Eq.~\ref{EQN:U_CS},
\begin{equation}
\label{EQN:PHOTON_WFN_CS}
|0^{\rm p}\rangle = \hat{U}^{\dagger}_{\rm CS} |0\rangle
\end{equation}
where $|0\rangle$ represents the photon vacuum. Now, consider the expectation value of the PF Hamiltonian with respect to the QED-HF wavefunction:
\begin{equation}
    \langle 0^{\rm e}0^{\rm p} | \hat{H} | 0^{\rm e}0^{\rm p}\rangle = \langle 0^{\rm e} | \otimes \langle 0 | \hat{U}_{\rm CS} \hat{H} \hat{U}^\dagger_{\rm CS} |0 \rangle \otimes | 0^{\rm e}\rangle
\end{equation}
From the right-hand side of this expression, it is evident that the electronic wave function, $|0^{\rm e}\rangle$, could be determined by minimizing the expectation value of the transformed Hamiltonian, $\langle 0 | \hat{U}_{\rm CS} \hat{H} \hat{U}^\dagger_{\rm CS} | 0\rangle$, with respect to variations in the orbitals, without any explicit consideration of the photon degrees of freedom. Hence, by applying the coherent-state transformation to the full PF Hamiltonian, we avoid the second step of the modified Roothan-Hall procedure for QED-HF that is outlined above. 

To transform $\hat{H}_{PF}$ to the coherent-state basis, we note that 
\begin{align}
\hat{U}_{\rm CS}\bhd \hat{U}^{\dagger}_{\rm CS} &= \bhd - z[\bhd, (\bhd - \bh)] = \bhd - z \nonumber \\
\hat{U}_{\rm CS}\bh \hat{U}^{\dagger}_{\rm CS} &= \bh - z[\bh, (\bhd - \bh)] = \bh - z \\
\hat{U}_{\rm CS}\bhd \bh \hat{U}^{\dagger}_{\rm CS} &= 
\hat{U}_{\rm CS}\bhd \hat{U}^{\dagger}_{\rm CS} \hat{U}_{\rm CS}\bh \hat{U}^{\dagger}_{\rm CS} = (\bhd - z) (\bh - z). \nonumber
\end{align}
So, applying this transformation to Eq.~\ref{EQN:PFH_LENGTH_GAUGE} yields
\begin{equation}\label{EQN:PFH_CS}
\hat{H}_{\rm CS} = \hat{H}_e + \omega_{\rm cav} (\bhd - z) (\bh - z) - \sqrt{\frac{\omega_{\rm cav}}{2}} {\bm \lambda} \cdot \hat{\bm \mu} (\bhd + \bh - 2z) + \frac{1}{2} ({\bm \lambda} \cdot \hat{\bm \mu})^2,
\end{equation}
and substituting Eq.~\ref{EQN:z} gives the specific form of the Pauli-Fierz Hamiltonian in the coherent state basis:
\begin{equation}\label{EQN:PFH_CS}
\hat{H}_{\rm CS} = \hat{H}_e + \omega_{{\rm cav}} \bhd \bh  - \sqrt{\frac{\omega_{\rm cav}}{2}} [{\bm{\lambda}} \cdot ({\bm{\hat{\mu}}} - \langle {\bm{\hat{\mu}}} \rangle )] (\bhd + \bh) +  \frac{1}{2} [{\bm{\lambda}} \cdot ({\bm{\hat{\mu}}} - \langle {\bm{\hat{\mu}}} \rangle )]^2 .
\end{equation}

Although we see that in Figure~\ref{FIG:PHOTON_CONVERGENCE} the
total energy for charge systems remains origin invariant in the coherent state basis, the orbitals and the Fock matrix itself are not origin invariant for charged systems in this formulation.  This presents
challenges for introducing perturbative corrections for
electron-electron and electron-photon correlation.  This was recently 
observed by Riso {\it et al.} who developed a strong coupling quantum 
electrodynamics Hartree-Fock theory (SC-QED-HF) that leads to
a fully origin-invariant formluation~\cite{Koch22_NatComm} based on the 
following ansatz:
\begin{equation}
    |\Phi_{SCQEDHF}\rangle = {\rm exp}\left( -\frac{{\bm{\lambda}}}{\sqrt{2\omega_{\rm cav}}} \sum_{p \sigma} \eta_{p\sigma} \hat{a}^{\dagger}_{p\sigma} \hat{a}_{p\sigma} \left(\bh - \bhd \right) \right) | 0^{\rm e}\rangle |0\rangle
\end{equation}
where $\hat{a}^{\dagger}_{p\sigma}$ and $\hat{a}_{p\sigma}$ are fermionic creation and annihilation operators for spin orbital $p\sigma$ and $\eta_p$ are orbital-specific coherent state coefficients.

\subsection{Cavity QED Hartree-Fock (QED-HF) in the Coherent-State Basis}
\label{sec:cs-qed-hf}
Consider a QED-HF wave function of the form of Eq.~\ref{EQN:QEDHF}. We express the photon state using the coherent-state transformation (Eq.~\ref{EQN:PHOTON_WFN_CS}) and take the expectation value of the Pauli-Fierz Hamiltonian to give
\begin{eqnarray}
\label{EQN:QED_RHF_ENERGY}
    E_\text{QED-HF} &=& \sum_{\mu\nu} ( T_{\mu\nu} + V_{\mu\nu} + \frac{1}{2} J_{\mu\nu} - \frac{1}{2} K_{\mu\nu}) \gamma_{\rm \mu\nu}  + \langle \frac{1}{2} [{\bm{\lambda}} \cdot ({\bm{\hat{\mu}}}_{\rm e} - \langle {\bm{\hat{\mu}}}_{\rm e} )\rangle]^2\rangle
\end{eqnarray}
Here, $\mu$ and $\nu$ represent atomic basis functions, and
$T_{\mu\nu}$, $V_{\mu\nu}$, $J_{\mu\nu}$, and $K_{\mu\nu}$ are electron kinetic energy integrals, electron-nucleus potential energy integrals, elements of the Coulomb matrix, and elements of the exchange matrix, respectively. The elements of the Coulomb and exchange matrices are defined by
\begin{equation}
    J_{\mu\nu} = \sum_{\lambda \sigma} (\mu\nu|\lambda\sigma) \gamma_{\lambda \sigma} 
\end{equation}
and 
\begin{equation}
    K_{\mu\nu} = \sum_{\lambda \sigma} (\mu \lambda | \sigma \nu)  \gamma_{\lambda\sigma}
\end{equation}
where the symbol $(\mu\nu|\lambda\sigma)$ represents a two-electron repulsion integral in chemists' notation, and 
$\gamma_{\mu\nu} = \sum_i^{N_{\rm e}} c^*_{\mu i} c_{\nu i}$ is the one-particle reduced density matrix
(with $\{c_{\mu i}\}$ and $N_{\rm e}$ being molecular orbital coefficients and the number of electrons, respectively).
The last term in Eq.~\ref{EQN:QED_RHF_ENERGY} is the dipole self-energy; note that, in the coherent-state basis, this quantity depends on only electronic degrees of freedom. Note also that the bilinear coupling term in Eq.~\ref{EQN:PFH_CS} does not contribute to the QED-HF total energy when the Hamiltonian is represented in the coherent-state basis. This property is shared by all QED approaches where the wave function is represented as a product of electron and photon functions ({\em e.g.}, in the QED-DFT approach described in Ref.\citenum{DePrince22_9303} and in Sec.~\ref{SEC:QED-DFT}).

The implementation of the dipole self-energy term is not consistent across the literature, with the difference being the treatment of the square of the electric dipole operator. To appreciate these differences, we first expand the dipole self-energy operator as
\begin{eqnarray}
\label{EQN:DSE_OPERATOR}
    \frac{1}{2} [{\bm{\lambda}} \cdot ({\bm{\hat{\mu}}}_e - \langle {\bm{\hat{\mu}}}_e \rangle)]^2 &=& \frac{1}{2} ( {\bm{\lambda}}\cdot  {\bm{\hat{\mu}}}_{\rm e} ) ^2 -  ( {\bm{\lambda}}\cdot {\bm{\hat{\mu}}}_{\rm e} ) ( {\bm{\lambda}}\cdot \langle {\bm{\hat{\mu}}}_{\rm e}\rangle )+ \frac{1}{2} ( {\bm{\lambda}}\cdot \langle {\bm{\hat{\mu}}}_{\rm e}\rangle ) ^2.
\end{eqnarray}
Now, the square of the electric dipole operator (the first term on the right-hand side of Eq.~\ref{EQN:DSE_OPERATOR}) can be expanded in terms of one- and two-electron contributions as
\begin{equation}
\label{EQN:DIPOLE_SQUARED_FIRST_QUANTIZATION}
    ( {\bm{\lambda}}\cdot  {\bm{\hat{\mu}}}_{\rm e} ) ^2 = \sum_{i \neq j} [ {\bm{\lambda}}\cdot  {\bm{\hat{\mu}}}_{\rm e}(i) ][ {\bm{\lambda}}\cdot  {\bm{\hat{\mu}}}_{\rm e}(j)] + \sum_i [ {\bm{\lambda}}\cdot  {\bm{\hat{\mu}}}_{\rm e}(i) ]^2.
\end{equation}
where $i$ and $j$ represent different electrons.
The right-hand side of Eq.~\ref{EQN:DIPOLE_SQUARED_FIRST_QUANTIZATION} can be expressed in second-quantized notation as
\begin{eqnarray}
\label{EQN:DIPOLE_SQUARED}
    ( {\bm{\lambda}}\cdot  {\bm{\hat{\mu}}}_{\rm e} ) ^2 &=&  \sum_{\mu\nu\lambda\sigma} d_{\mu\nu} d_{\lambda\sigma} \hat{a}^\dagger_\mu \hat{a}^\dagger_\lambda \hat{a}_\sigma \hat{a}_\nu - \sum_{\mu\nu} q_{\mu\nu} \hat{a}^\dagger_\mu \hat{a}_\nu.
\end{eqnarray}
where $\hat{a}^\dagger$ and $\hat{a}$ represent fermionic creation and annihilation operators, respectively. The symbols $d_{\mu\nu}$ and $q_{\mu\nu}$ represent modified electric dipole and electric quadrupole integrals, which have the form
\begin{equation}
    d_{\mu\nu} = - \sum_{a \in \{x,y,z\}} \lambda_a \int \chi^*_\mu r_a \chi_{\nu} d\tau,
\end{equation}
and
\begin{equation}
    q_{\mu\nu} = - \sum_{ab \in \{x,y,z\}} \lambda_a \lambda_b \int \chi^*_\mu r_a r_b \chi_{\nu} d\tau.
\end{equation}
respectively, and are evaluated over atomic basis functions, $\chi_\mu$. Here, $\lambda_a$ is a cartesian component of ${\bm{\lambda}}$, and $r_a$ is a cartesian component of the position vector [{\em e.g.}, for ${\mathbf{r}} = (x, y, z)$, $r_x$ = $x$]. As is well known, the square of an operator expanded initially in first quantization and then represented in second quantization is not necessarily the same as the square of the second quantized form of the operator; these representations are only equivalent in the limit that the one-electron basis set is complete. Equation \ref{EQN:DIPOLE_SQUARED} makes no assumptions about the completeness of the one-particle basis set and is the form of the square of the dipole operator employed in Refs.~\citenum{DePrince21_094112,DePrince22_053710,DePrince22_054105,DePrince22_9303,Foley_154103}. On the other hand, many other studies take the second-quantized form of the square of the electric dipole operator to be the product of second-quantized electric dipole operators, which leads to
\begin{eqnarray}
\label{EQN:DIPOLE_SQUARED_WRONG}
    ( {\bm{\lambda}}\cdot  {\bm{\hat{\mu}}}_{\rm e} ) ^2 &=&  \sum_{\mu\nu\lambda\sigma} d_{\mu\nu} d_{\lambda\sigma} \hat{a}^\dagger_\mu \hat{a}_\nu \hat{a}^\dagger_\lambda \hat{a}_\sigma  \nonumber \\
         &=&  \sum_{\mu\nu\lambda\sigma} d_{\mu\nu} d_{\lambda\sigma} \hat{a}^\dagger_\mu \hat{a}^\dagger_\lambda \hat{a}_\sigma \hat{a}_\nu +\sum_{\mu\nu}  \hat{a}^\dagger_\mu \hat{a}_\nu \sum_\sigma d_{\mu\sigma} d_{\sigma\nu}.
\end{eqnarray}
In these studies, the assumption that the basis set is assumed to be complete is never stated, but this choice is evident in the form of the Fock matrix (see Eq.~30 of Ref.~\citenum{Koch20_041043}, for example). In this review, we choose the form of $( {\bm{\lambda}}\cdot  {\bm{\hat{\mu}}}_{\rm e} ) ^2$ given by Eq.~\ref{EQN:DIPOLE_SQUARED}. Given that choice, and the fact that
\begin{equation}
    ( {\bm{\lambda}}\cdot {\bm{\hat{\mu}}}_{\rm e} ) = \sum_{\mu\nu} d_{\mu\nu} \hat{a}^\dagger_\mu \hat{a}_\nu,
\end{equation}
we arrive at
\begin{eqnarray}
\label{EQN:DSE_FINAL}
\frac{1}{2} [{\bm{\lambda}} \cdot ({\bm{\hat{\mu}}}_e - \langle {\bm{\hat{\mu}}}_e \rangle)]^2 &=& \frac{1}{2} \sum_{\mu\nu\lambda\sigma} d_{\mu\nu} d_{\lambda\sigma} \hat{a}^\dagger_\mu \hat{a}^\dagger_\lambda \hat{a}_\sigma \hat{a}_\nu \nonumber \\ 
&+& \sum_{\mu\nu} O^{\rm DSE}_{\mu\nu} \hat{a}^\dagger_\mu \hat{a}_\nu + \frac{1}{2} ({\bm{\lambda}}\cdot\langle {\bm{\mu}}_{\rm e}\rangle)^2.
\end{eqnarray}
where 
\begin{equation}
\label{EQN:ODSE}
    O^{\rm DSE}_{\mu\nu} = -( {\bm{\lambda}}\cdot \langle {\bm{\hat{\mu}}}_{\rm e}\rangle ) d_{\mu\nu} - \frac{1}{2} q_{\mu\nu}. 
\end{equation}
Now, we can evaluate the expectation of Eq.~\ref{EQN:DSE_FINAL} with respect to a single determinant, which gives
\begin{eqnarray}
\label{EQN:DSE_FINAL_FINAL}
    \langle \frac{1}{2} [{\bm{\lambda}} \cdot ({\bm{\hat{\mu}}}_e - \langle {\bm{\hat{\mu}}}_e \rangle)]^2 \rangle &=& \sum_{\mu\nu} (\frac{1}{2} J^{\rm DSE}_{\mu\nu} - \frac{1}{2} K^{\rm DSE}_{\mu\nu} + O^{\rm DSE}_{\mu\nu} ) \gamma_{\mu\nu} \nonumber \\
    &+& \frac{1}{2} ({\bm{\lambda}}\cdot\langle {\bm{\mu}}_{\rm e}\rangle)^2
\end{eqnarray}
Here, $J^{\rm DSE}_{\mu\nu}$ and $K^{\rm DSE}_{\mu\nu}$ are elements of dipole self-energy matrices that are analogies of the usual Coulomb and exchange matrices:
\begin{equation}
\label{EQN:JDSE}
    J^{\rm DSE}_{\mu\nu} = d_{\mu\nu} \sum_{\lambda \sigma} d_{\lambda \sigma} \gamma_{\lambda\sigma} =   ({\bm \lambda} \cdot \langle {\bm{\hat{\mu}}}_{\rm e}  \rangle) d_{\mu\nu}
\end{equation}
\begin{equation}
\label{EQN:KDSE}
    K^{\rm DSE}_{\mu\nu} = \sum_{\lambda \sigma} d_{\mu\sigma} d_{\lambda \nu} \gamma_{\lambda\sigma}.
\end{equation}

With all of the components of the energy (Eq.~\ref{EQN:QED_RHF_ENERGY}) defined, we can make this energy stationary with respect to the molecular orbital expansion coefficients, $\{c_{\mu i}\}$, while enforcing orthogonality of the molecular orbitals, which leads to a set of Hartree-Fock equations that resembles those in the ordinary electronic problem, augmented by the dipole self-energy contributions. As such, QED-HF orbitals are eigenfunctions of a modified Fock matrix,
\begin{eqnarray}
    F_{\mu\nu} &=& T_{\mu\nu} + V_{\mu\nu} + J_{\mu\nu} - K_{\mu\nu} \nonumber \\
    &+& O^{\rm DSE}_{\rm \mu\nu} + J_{\mu\nu}^{\rm DSE} - K_{\mu\nu}^{\rm DSE} 
\end{eqnarray}
For organizational purposes, it will become convenient to partition the Fock matrix into contributions that define the canonical Fock operator,  $F^{\rm C}_{\mu\nu} = T_{\mu\nu} + V_{\mu\nu} + J_{\mu\nu} - K_{\mu\nu}$,
plus terms that derive from the dipole self energy,
$F^{{\rm DSE}}_{\mu\nu} = O^{\rm DSE}_{\rm \mu\nu} + J_{\mu\nu}^{\rm DSE} - K_{\mu\nu}^{\rm DSE}$.

Upon solving the QED-HF equations, one obtains a set of molecular orbitals corresponding to the (mean-field) ground state of a many-electron system coupled to an optical cavity. For sufficiently large coupling strengths, the cavity can induce significant changes in these orbitals, as compared to orbitals obtained from a standard HF procedure on the isolated many-electron system. Here, we examine such changes for a formaldehyde molecule that has been coupled to a single-mode optical cavity. Excited states of this system have been explored using QED generalizations of time-dependent density functional theory\cite{Narang18_113002,Shao21_064107} (see Sec.~\ref{SEC:QED-TDDFT} for a description of the relevant theory). Here, we adapt the results of Ref.~\citenum{Foley_154103} and focus on cavity-induced changes to the ground state ({\em i.e.}, to the molecular orbitals). We supplement this discussion with a tutorial implementation of QED-HF that the interested reader can find \href{https://github.com/FoleyLab/psi4polaritonic/blob/cpr/QED-HF_Tutorial.ipynb}{online}.\cite{Foley23_QED_HF_TUTORIAL}  The tutorial provides a benchmark calculation on the water molecule, and can be modified to study other systems.

As described in Ref.~\citenum{Foley_154103}, the geometry of isolated formaldehyde was optimized using restricted HF (RHF) theory and the cc-pVDZ basis set, and the principal symmetry axis of the molecule is aligned along the $z$-axis.  At this level, the RHF ground-state has a dipole moment oriented along the $z$-axis with $\langle \mu \rangle_z = -1.009$ a.u.
We consider solutions to the QED-HF equations for a coupling vector with fixed magnitude, 
({\em i.e.}, $|{\bm \lambda}| = 0.1$ a.u.), 
and three different cavity mode polarizations: 
${\bm \lambda}_y = 0.1~\hat{{\bf e}}_y$ a.u.,
${\bm \lambda}_z = 0.1~\hat{{\bf e}}_z$ a.u., and 
${\bm \lambda}_{yz} = \sqrt{\frac{1}{2}} ({\bm \lambda}_{y} + {\bm \lambda}_{z})$ a.u., with 
$\hat{{\bf e}}_y=(0,~1,~0)$ and
$\hat{{\bf e}}_z=(0,~0,~1)$. 
As compared to the HF energy, the QED-HF energy is higher in all cases, with the largest increase occurring for ${\bm \lambda}_z$ (see Table I). Going back to  the explicit expressions for the QED-HF dipole self energy derived above, we can see that this large change likely originates from the permanent dipole moment that is oriented along the $z$-axis, which contributes to the last term in Eq.~\ref{EQN:DSE_FINAL_FINAL}.
The cavity-induced changes to the energy for the other polarizations point to important effects arising from the other contributions to Eq.~\ref{EQN:DSE_FINAL_FINAL}. Specifically, in the case of ${\bm \lambda}_y$, we should see no permanent dipole moment contributions to the dipole self energy, which indicates that the cavity effects stem entirely from the quadrupolar contribution to $O^{\rm DSE}$ (Eq.~\ref{EQN:ODSE}) and the exchange-like contribution (Eq.~\ref{EQN:KDSE}).

To quantify cavity-induced changes to the energy, Ref.~\citenum{Foley_154103} considered how various contributions to the QED-HF energy change with and without coupling to the photon field. Specific formulae for these couplings are given in reference~\onlinecite{Foley_154103}.
\begin{table}[!ht]
\begin{tabular}{c|c|c|c|c|c|c}
Total & \multicolumn{2}{c|}{Canonical RHF} & \multicolumn{4}{c|}{Cavity Contributions} \\
\hline
 $\Delta E$ (eV) & \% $\Delta_{1E}$ & \% $\Delta_{2E}$ & \% $\Delta_{1de}$ & \% $\Delta_{1qe}$ & \% $\Delta_{2de}$ & \% $\Delta_{d_c}$ \\
 \hline
 \multicolumn{7}{c}{${\bm \lambda_y}$} \\
  \hline
  0.925 & -229 & 230 & 0 & 209 & -110 & 0 \\
 \hline
 \multicolumn{7}{c}{${\bm \lambda_z}$} \\
 \hline
 1.110 & -178 & 179 & -31 & 431 & -316 & 15 \\
  \hline
 \multicolumn{7}{c}{${\bm \lambda_{yz}}$} \\
 \hline
 1.034 & -200 & 201 & -16 & 329 & -222 & 8 \\
 \hline
\end{tabular}
\caption{Change in total QED-HF energy ($\Delta E$ in eV) and \% relative changes in different contributions
 to the total QED-HF energy for three different polarizations
 of a photonic mode with magnitude $|{\bm \lambda}| = 0.1$ a.u..
 The terms $\Delta_{1E}$ and $\Delta_{2E}$ denote changes in the RHF 1- and 2-electron
 energies, respectively, and the terms $\Delta_{1de}$, $\Delta_{2de}$, $\Delta_{1qe}$, $\Delta_{d_c}$ denote changes in the QED-HF 1-electron dipole, 2-electron dipole, 1-electron quadrupole, and dipole constant terms respectively.}
\end{table}
The quadrupolar contribution to $O^{\rm DSE}$ ($\Delta_{1qe}$) and the Coulomb-like and exchange-like contributions (Eqs.~\ref{EQN:JDSE} and \ref{EQN:KDSE}), the combination of which is denoted $\Delta_{2de}$ in Table I, 
typically account for the largest changes to the QED-HF energy
for the three polarizations considered in Table I.  However, the changes in the one- and two-electron contributions to the canonical RHF energy (denoted $\Delta_{1E}$ and $\Delta_{2E}$) suggest that cavity-induced changes to the orbitals themselves can have appreciable energetic consequences. We note that the various components of the energetic changes largely cancel with one other ({\em i.e.} $\Delta_{1E} \approx -\Delta_{2E}$ in all three
cases), leading to more modest changes in the total energy (see Table I).

\begin{figure}[!ht]\label{fig:QED-HF-orbitals}
\centering
\includegraphics[width=0.5\textwidth]{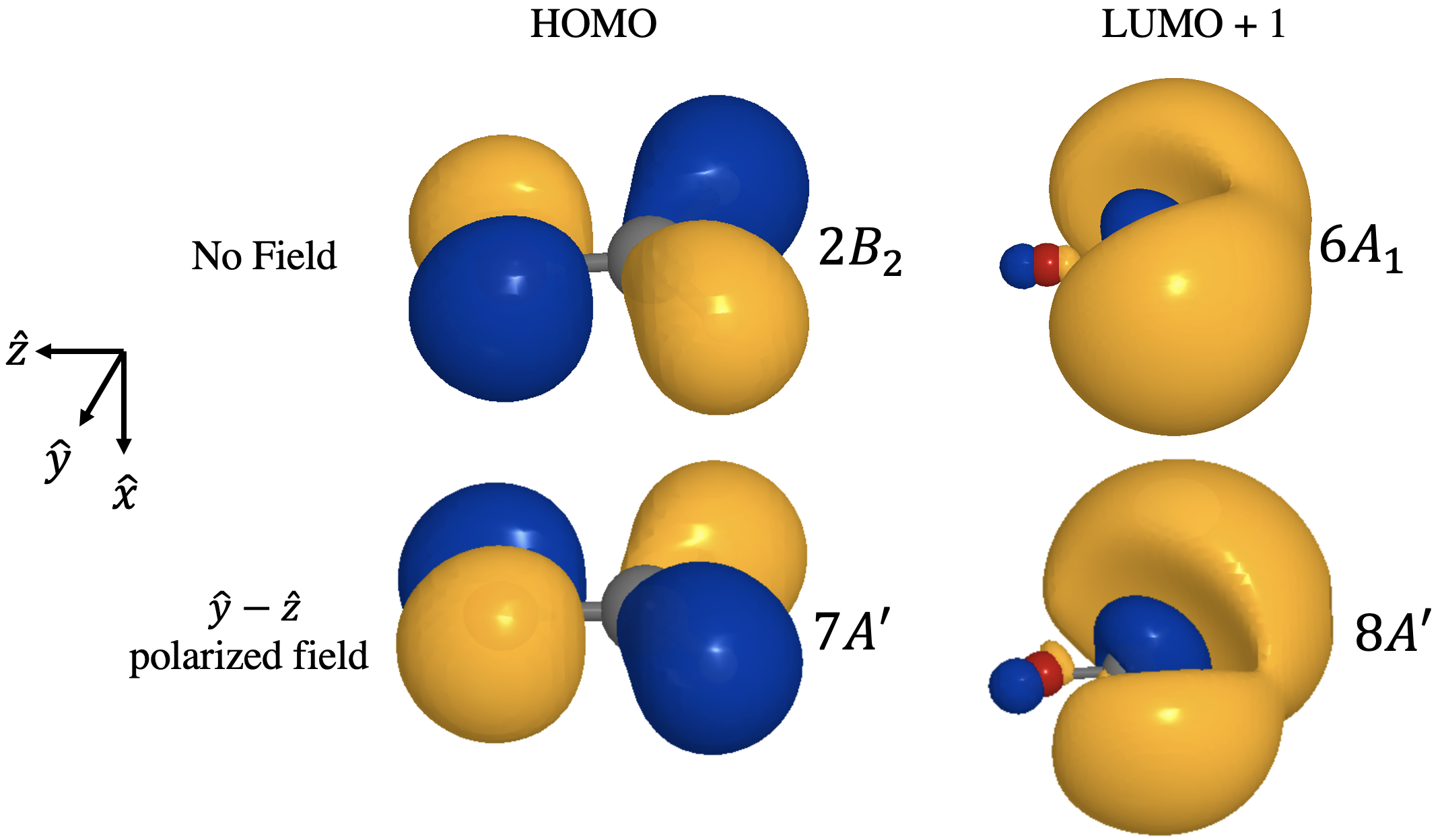}
\caption{Comparison of the HOMO and LUMO+1 orbitals of formaldehyde uncoupled to an photon mode (top) and strongly coupled to a photon mode polarized along the $y-z$ axis (bottom), where strong coupling results in a change in symmetry from $C_{2v}$ to $C_s$.  Adapted with permission from Ref.~\onlinecite{Foley_154103}}
\end{figure}

Aside from the energy, we can also visualize the impact that the cavity has on the real-space form of the molecular orbitals. As an example, Fig.~\ref{fig:QED-HF-orbitals} depicts HF orbitals for the highest occupied molecular orbital (HOMO, $2B_2$) and the second-lowest unoccupied molecular orbital (LUMO+1, $6A_1$) for an isolated formaldehyde molecule and the corresponding QED-HF orbitals for the
${\bm \lambda}_{yz}$ case ( $7A^{'}$ and $8A^{'}$ ). The QED-HF orbitals are noticably distorted compared to the HF ones, which results in a reduction of symmetry from
$C_{2v}$ to $C_s$ and impacts both ground-state energy and properties.
The direct inclusion of these cavity-induced effects on the orbital basis is  one appealing advantage of {\it ab initio} QED methods.  

\subsection{Cavity QED Density Functional Theory (QED-DFT)}

\label{SEC:QED-DFT}

The QED-HF theory outlined above can easily be adapted to develop a QED generalization of Kohn-Sham DFT, or QED-DFT.\cite{DePrince22_9303,Rubio22_094101,DePrince23_5264}  To do so, one can simply follows the basic premise of Kohn-Sham DFT:\cite{Sham65_A1133} there exists a fictitious system of non-interacting photons and electrons that has the same density as the fully-interacting system. The QED-DFT ground-state is then taken to have the form of Eq.~\ref{EQN:QEDHF}, except that $|0^{\rm e}\rangle$ now refers to a determinant of Kohn-Sham orbitals. As with QED-HF, the photon part of the wave function can be exactly represented using the coherent-state transformation operator, see Eq.~\ref{EQN:PHOTON_WFN_CS}. All electron-electron correlation and exchange effects and electron-photon correlation effects can then, in principle, be accounted for by appropriate functionals of the density (and gradient of the density, etc.), as in standard Kohn-Sham DFT. Historically, QED-DFT was predated by a different generalization of DFT for cavity QED applications, called QEDFT, \cite{Rubio15_15285,Rubio18_992,Rubio19_2757,Narang20_094116,Narang21_104109} 
which, rather than following the Kohn-Sham scheme, represents the electronic and photonic degrees of freedom directly in real space. 
QED-DFT studies typically employ standard exchange-correlation functionals used in electronic structure theory ({\em i.e.}, they ignore electron-photon correlation effects), while, for QEDFT, a few examples of electron-photon correlation functions have been put forward.\cite{Rubio15_093001, Rubio17_113036, Rubio18_992, Rubio21_e2110464118,Flick22_143201}

\section{Single-Particle Post-SCF Cavity QED Methods}

\subsection{Cavity QED-Configuration Interaction with Single Excitations (QED-CIS)}

A general correlated wave function for a many-electron system coupled to a single-mode cavity could take the form \begin{equation}
\label{EQN:FCI}
    |\Psi\rangle = \sum_{\mu}\sum_{A} c_{\mu}^{A} | \mu^{\rm e}\rangle \otimes | A^{\rm p}\rangle
\end{equation}
where $|\mu^{\rm e}\rangle$ represents a determinant of electronic orbitals, $|A^{\rm p}\rangle$ is a photon-number state corresponding to $A$ photons in the cavity mode, and $c_{\mu}^{A}$ is an expansion coefficient. If \{$|\mu^{\rm e}\rangle$\} includes all possible determinants and \{$|A^{\rm p}\rangle$\} includes all possible photon-number states, then this full configuration interaction (CI) wave function provides an exact description of the electronic/polaritonic structure, within a given one-electron basis set. However, as in the usual electronic case, a full CI description of a cavity-coupled many-electron system is, in general, an intractable prospect. The simplest solution to this problem is to truncate both the many-electron basis and the photon basis at some level.

McTague and Foley proposed\cite{Foley_154103} a truncated cavity QED-CI approach wherein the sum over Slater determinants, $\mu$, in Eq.~\ref{EQN:FCI} was restricted to include only the reference electronic configuration, $|0^{\rm e}\rangle$, and all single electronic excitations out of this configuration, and the sum over photon-number states was restricted to include only states representing zero or one photon in the cavity ($|0\rangle$ and $|1\rangle$, respectively). Those authors termed this approach cavity QED configuration interaction with single excitations, or CQED-CIS, but, following the naming convention used in some QED coupled-cluster approaches\cite{Koch20_041043} (see Sec.~\ref{sec:qed-cc}), we adopt the name QED-CIS-1. The QED-CIS-1 wave function for state $I$ takes the form
\begin{equation}\label{EQN:QED-CIS-1}
|\Psi_I\rangle = c_0^0 |0^{\rm e} \rangle \otimes |0\rangle + \sum_{i,a} c_{ia}^0 |\Phi_i^a\rangle \otimes |0\rangle + c_0^1 |0^{\rm e}\rangle \otimes |1\rangle + \sum_{i,a} c_{ia}^1 |\Phi_i^a\rangle \otimes |1\rangle. 
\end{equation}
Following Ref.~\citenum{Foley_154103}, 
$|\Phi_i^a\rangle = \frac{1}{\sqrt{2}}\left(|\Phi_{i_\alpha}^{a_\alpha} \rangle + |\Phi_{i_\beta}^{a_\beta}\rangle \right)$ represents a singlet spin-adapted basis function, where $|\Phi_{i_\sigma}^{a_\sigma}\rangle$  is a determinant generated by exciting an electron with spin $\sigma$ from a spatial orbital that is occupied in $|0^{\rm e}\rangle$, $\phi_i$, to an unoccupied spatial orbital, $\phi_a$. For multiple cavity modes, QED-CIS-1 is defined such that the photon basis includes all possible combinations zero or one photon in each of the modes.

The expansion coefficients in Eq.~\ref{EQN:QED-CIS-1} can be determined as the elements of the eigenvectors of the matrix representation of the Pauli-Fierz Hamiltonian represented within the coherent-state basis ($\hat{H}_\text{CS}$, Eq.~\ref{EQN:PFH_CS}), {\em i.e.}, by solving the eigenvalue problem
\begin{equation}\label{EQN:H-QED-CIS-1}
\begin{bmatrix}
0 & 0 & 0 & \hbar {\bf g} \\
0 & {\bf A} + {\bm \Delta}  & \hbar {\bf g}^{\dagger}  & \hbar {\bf G} \\
0 & \hbar {\bf g} & \hbar \omega & 0 \\
\hbar {\bf g}^{\dagger} & \hbar {\bf G} & 0 & {\bf A} + {\bm \Delta} + \hbar \Omega 
\end{bmatrix}
\begin{bmatrix}
{\bf c}^0_0 \\
{\bf c}^0_{ia} \\
{\bf c}^1_0 \\
{\bf c}^1_{ia}
\end{bmatrix}
=
\Omega_\text{QED-CIS-1}
\begin{bmatrix}
{\bf c}^0_0 \\
{\bf c}^0_{ia} \\
{\bf c}^1_0 \\
{\bf c}^1_{ia},
\end{bmatrix}
\end{equation}
Note that the matrix on the left-hand side of Eq.~\ref{EQN:H-QED-CIS-1} actually is the matrix representation of $\hat{H}_{\rm CS} - E_\text{QED-HF}$, where $E_\text{QED-HF}$ is the energy of the QED-HF reference state. The elements of ${\bf A}$ are similar to those encountered in canonical CIS theory,
\begin{equation}\label{EQN:CIS-A}
    A_{ia,jb} = F^{\rm C}_{ab} \delta_{ij} -  F^{\rm C}_{ij }\delta_{ab} + 2(ia|jb) - (ij|ab),
\end{equation}
with important differences being that (i) the two-electron integrals are performed over QED-HF orbitals, and (ii) ${\bf F}^{{\rm C}}$ is not diagonal in the QED-HF basis when the coupling strength is non-zero.
The dipole self energy contribution to the Hamiltonian in the subspace of spin-adapted singly-excited functions is contained in the ${\bm \Delta}$ matrix, with elements
\begin{equation}\label{EQN:DSE-ELEMENTS}
    \Delta_{ia,jb} =F^{\rm DSE}_{ab} \delta_{ij} -  F^{\rm DSE}_{ij }\delta_{ab} + 2 d_{ia} d_{jb} 
    -d_{ij} d_{ab}, 
\end{equation}
Again, we note that ${\bf F}^{{\rm DSE}}$ is not necessarily diagonal in the QED-HF basis.
The symbol $\Omega$ represents a diagonal matrix of photon energy contributions, defined by 
\begin{equation}
    \Omega_{ia,jb} = \omega \delta_{ij} \delta_{ab}.
\end{equation}
The symbols ${\bf g}$ and ${\bf G}$ arise from the bilinear coupling term in $\hat{H}_{\rm CS}$ and are defined by
\begin{equation}\label{EQN:BLC-ELEMENTS-1}
    g_{ia} = -\sqrt{\omega} d_{ia}
\end{equation}
and
\begin{equation}\label{EQ:blc_elements_2}
    G_{ia,jb} = \sqrt{\frac{\omega}{2}} \left( d_{ij} \delta_{ab} 
    -  d_{ab} \delta_{ij}  + \langle d \rangle \delta_{ij} \delta_{ab} \right)
\end{equation}
The {\bf g} term couples the reference to $|\Phi_i^a\rangle |1\rangle$, while ${\bf G}$ couples singly-excited configurations with different photon numbers, {\em i.e.}, $|\Phi_i^a\rangle |0\rangle$ and $|\Phi_i^a\rangle |1\rangle$. Note that the fact that ${\bf g}$ couples the reference to $|\Phi_i^a\rangle |1\rangle$  implies that QED-CIS-1 captures some electron-photon correlation effects. Indeed, the lowest eigenvalue, $\Omega_\text{QED-CIS-1}$, obtained from solving Eq.~\ref{EQN:H-QED-CIS-1} is nonpositive and represents an electron-photon correlation energy.

\label{sec:qed-cis}

\subsection{Cavity QED Time-Dependent Density Functional Theory (QED-TDDFT)}

\label{SEC:QED-TDDFT}

Given the popularity of time-dependent DFT (TDDFT) for the electronic structure problem, it is not surprising that multiple  generalizations of TDDFT have been proposed and applied to cavity-embedded molecular systems. Both real-time\cite{Bauer11_042107,Tokatly13_233001,Rubio14_012508,Rubio17_3026,Tokatly18_235123,Varga22_194106} and linear-response\cite{Rubio19_2757,Rubio20_508,Narang20_094116,Narang21_104109,Shao21_064107,Shao22_124104,DePrince22_9303} formulations have been put forward; here, we focus on the linear-response approaches because they more closely resemble the QED-CIS-1 method discussed above. Both real-space\cite{Rubio19_2757,Rubio20_508,Narang20_094116,Narang21_104109} and atom-centered Gaussian basis function\cite{Shao21_064107,Shao22_124104,DePrince22_9303,DePrince23_5264} representations of the electronic structure have been used within linear-response QED-TDDFT. In the latter category, Refs.~\citenum{Shao21_064107} and \citenum{Shao22_124104} have considered QED-TDDFT calculations on top of canonical Kohn-Sham reference configurations ({\em i.e.}, $|0^{\rm e}\rangle \otimes |0\rangle$, where $|0^{\rm e}\rangle$ is a Kohn-Sham determinant optimized in the absence of the cavity), while Refs.~\citenum{DePrince22_9303} and \citenum{DePrince23_5264} have considered fully relaxed QED-DFT reference functions and represented the QED-TDDFT problem in the coherent-state basis, similar to what is done in QED-CIS-1. As discussed in Ref.~\citenum{DePrince23_5264}, significant differences in excitation energies obtained from these ``unrelaxed'' and ``relaxed'' QED-TDDFT protocols can occur when considering large coupling strengths. In either case, linear-response QED-TDDFT can be implemented as a solution to a generalization of Casida's equations
\begin{equation}\label{EQN:TDDFT}
\begin{bmatrix}
{\bf A} +{\bm \Delta}  & {\bf B} + {\bm \Delta}' & \hbar {\bf g}^{\dagger} & \hbar {\bf \tilde{g}}^{\dagger} \\
{\bf B} +{\bm \Delta}'  & {\bf A} + {\bm \Delta} & \hbar {\bf g}^{\dagger} & \hbar {\bf \tilde{g}}^{\dagger} \\
\hbar {\bf g} & \hbar {\bf g} & \hbar {\bm \omega} & 0 \\
\hbar {\bf \tilde{g}} & \hbar {\bf \tilde{g}} & 0 & \hbar {\bm \omega}
\end{bmatrix}
\begin{bmatrix}
{\bf X} \\
{\bf Y} \\
{\bf M} \\
{\bf N}
\end{bmatrix}
=
\Omega^\text{QED-TDDFT}
\begin{bmatrix}
{\bf 1} & 0 & 0 & 0 \\
0 & -{\bf 1} & 0 & 0 \\
0 & 0 & {\bf 1} & 0 \\
0 & 0 & 0 & -{\bf 1} 
\end{bmatrix}
\begin{bmatrix}
{\bf X} \\
{\bf Y} \\
{\bf M} \\
{\bf N}
\end{bmatrix}
\end{equation}
Assuming a spin-adapted basis, the ${\bf A}$ matrix is the same as that given in Eq.~\ref{EQN:CIS-A}, except that the exchange term $(ij|ab)$ is replaced with appropriate derivatives of the exchange-correlation energy. For a cavity QED random phase approximation (RPA), the ${\bf B}$ matrix has elements
\begin{equation}
B_{ia,jb} =  2(ia|jb) - (ib|ja)
\end{equation}
and, for QED-TDDFT, the exchange term $(ib|ja)$ is again replaced by the appropriate derivatives of the exchange-correlation energy. The ${\bm \Delta}'$ matrix has elements
\begin{eqnarray}
\label{EQN:DELTA_PRIME}
\Delta^\prime_{ai,bj} = 2 d_{ai} d_{bj} - d_{aj} d_{ib}
\end{eqnarray}
and, lastly, ${\bf \tilde{g}} = {\bf g}$. As described, the QED-TDDFT formalism corresponds to the ``relaxed'' one developed in Ref.~\citenum{DePrince22_9303}. The ``unrelaxed'' QED-TDDFT method proposed in Ref.~\citenum{Shao21_064107} can be obtained by ignoring the effects of the cavity in the underlying ground-state Kohn-Sham problem and taking
\begin{equation}
    \Delta_{ia,jb} = \Delta'_{ia,jb} = 2 d_{ai} d_{bj}
\end{equation}

The elements of ${\bf X}$, ${\bf Y}$, ${\bf M}$, and ${\bf N}$ parametrize the QED-TDDFT excited states; the elements of ${\bf X}$ and ${\bf Y}$ correspond to the usual electronic excitation and de-excitation amplitudes encountered in conventional TDDFT, while ${\bf M}$ and ${\bf N}$ refer to photon creation and annihilation amplitudes, respectively. We see clear connections to QED-CIS-1, where the CI coefficients $c_{ai}^0$ and $c_0^1$ play roles that are similar to those of the elements of ${\bf X}$ and ${\bf M}$, respectively.  Unlike QED-CIS-1, however, the linear-response QED-TDDFT equations do not couple the QED-DFT reference to any excited configurations. Hence, this approach does not account for any explicit electron-photon correlation effects, absent any that are included via the exchange-correlation functional. Such effects were ignored in Refs. \citenum{Shao21_064107,Shao22_124104, DePrince22_9303, DePrince23_5264}; all calculations reported therein used standard density functional approximations designed for non-QED applications.

\subsection{The QED-TDDFT and QED-CIS prisms}

\begin{figure}
    \centering
    \includegraphics[width=0.45\textwidth]{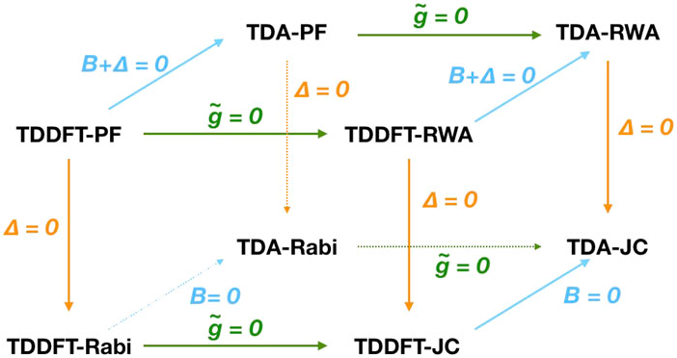}
    \caption{The QED-TDDFT prism with the front-most upper-left vertex representing full linear-response TDDFT applied to the Pauli-Fierz Hamiltonian (TDDFT-PF, herein referred to as QED-TDDFT), the back-most lower right vertex representing the most severe approximation (TDA-JC) through neglect of the electronic de-excitation matrix ${\bf B}$, the dipole self energy ${\bm \Delta}$/${\bm \Delta}'$ matrices, and the counter-rotating bilinear coupling terms ${\bf \tilde{g}}$.  Reproduced with permission from Ref.~\citenum{Shao21_064107}}
    \label{fig:qed-tddft-prism}
\end{figure}

As mentioned above, some coefficients from the QED-CIS-1 problem map directly onto amplitudes that arise in QED-TDDFT. However, QED-CIS-1 lacks analogues to the de-excitation and annihilation amplitudes (${\bf Y}$ and ${\bf N}$, respectively). That said, in Ref.~\citenum{Shao21_064107}, Shao and coworkers explored an approximation to QED-TDDFT that ignored these terms, called the Tamm-Dancoff - Rotating Wave Approximation (TDA-RWA) in that work, which has a simpler structure that is more similar to QED-CIS-1. The TDA-RWA eigenvalue problem is
\begin{equation}\label{EQN:TDA_RWA}
\begin{bmatrix}
{\bf A} +{\bm \Delta} & \hbar {\bf g}^{\dagger} \\
\hbar {\bf g} & \hbar {\bm \omega}
\end{bmatrix}
\begin{bmatrix}
{\bf X} \\
{\bf M} \\
\end{bmatrix}
=
\Omega^\text{TDA-RWA}
\begin{bmatrix}
{\bf X} \\
{\bf M}
\end{bmatrix}.
\end{equation}
The primary differences between QED-CIS-1 and TDA-RWA are (i) the different definitions of the {\bf A} matrix that we have already discussed and (ii) the fact that TDA-RWA, like QED-TDDFT,  does not account for simultaneous electronic excitations and photon creation, which would couple the QED-DFT reference to excited configurations. Other subtle differences exist, depending on whether the TDA-RWA is done in a fully relaxed way or not (as discussed in the context of QED-TDDFT above). The TDA-RWA approach is only one of eight possible approximations to QED-TDDFT that Shao and co-workers analyzed in Ref.~\citenum{Shao21_064107}; these approximations live on what those authors describe as the QED-TDDFT prism (see Figure~\ref{fig:qed-tddft-prism}). The facets of their prism include all possible combinations of including or neglecting of the ${\bf B}$ matrix, the ${\bm \Delta}$/${\bm \Delta}'$ matrices, and  ${\bf \tilde{g}}$.

An analogous family of approximations to QED-CIS-1 can be developed by neglecting ${\bm \Delta}$ or the bilinear coupling terms in Eq.~\ref{EQN:H-QED-CIS-1} or by excluding simultaneous electron excitation and photon creation terms ($|\Phi_i^a\rangle \otimes |1\rangle$) in Eq.~\ref{EQN:QED-CIS-1}. For example, excluding $|\Phi_i^a\rangle \otimes |1\rangle$ from the wave function expansion results in a QED-CIS method has the same structure as TDA-RWA:
\begin{equation}\label{EQN:QED-CIS}
\begin{bmatrix}
{\bf A} +\Delta  & \hbar {\bf g}^{\dagger}  \\
 \hbar {\bf g} & \hbar \omega
\end{bmatrix}
\begin{bmatrix}
{\bf c}^0_{ia} \\
{\bf c}^1_0 
\end{bmatrix}
=
\Omega_\text{QED-CIS}
\begin{bmatrix}
{\bf c}^0_{ia} \\
{\bf c}^1_0
\end{bmatrix}
\end{equation}
On the other hand, neglecting ${\bm \Delta}$ Eq.~\ref{EQN:H-QED-CIS-1} leads to a Jaynes-Cummings-like approximation to QED-CIS-1 (JC-CIS-1):
\begin{equation}\label{EQN:H-CISS-JC}
\begin{bmatrix}
0 & 0 & 0 & \hbar {\bf g} \\
0 & {\bf A}  & \hbar {\bf g}^{\dagger}  & \hbar {\bf G} \\
0 & \hbar {\bf g} & \hbar \omega & 0 \\
\hbar {\bf g}^{\dagger} & \hbar {\bf G} & 0 & {\bf A} + \hbar \Omega 
\end{bmatrix}
\begin{bmatrix}
{\bf c}^0_0 \\
{\bf c}^0_{ia} \\
{\bf c}^1_0 \\
{\bf c}^1_{ia}
\end{bmatrix}
=
\Omega_\text{JC-CIS-1}
\begin{bmatrix}
{\bf c}^0_0 \\
{\bf c}^0_{ia} \\
{\bf c}^1_0 \\
{\bf c}^1_{ia}
\end{bmatrix}
\end{equation}
and if we neglect ${\bm \Delta}$ from 
Eq.~\ref{EQN:QED-CIS},
we arrive at a JC-CIS methods that has the same structure as the TDA-JC method of 
Shao and co-workers~\cite{Shao21_064107}:
\begin{equation}\label{EQN:CIS-JC}
\begin{bmatrix}
{\bf A}  & \hbar {\bf g}^{\dagger}  \\
 \hbar {\bf g} & \hbar \omega
\end{bmatrix}
\begin{bmatrix}
{\bf c}^0_{ia} \\
{\bf c}^1_0 
\end{bmatrix}
=
\Omega_\text{JC-CIS}
\begin{bmatrix}
{\bf c}^0_{ia} \\
{\bf c}^1_0
\end{bmatrix}
\end{equation}

\begin{figure}[!ht]
    \centering
    \includegraphics[height=0.49\textwidth, angle=270]{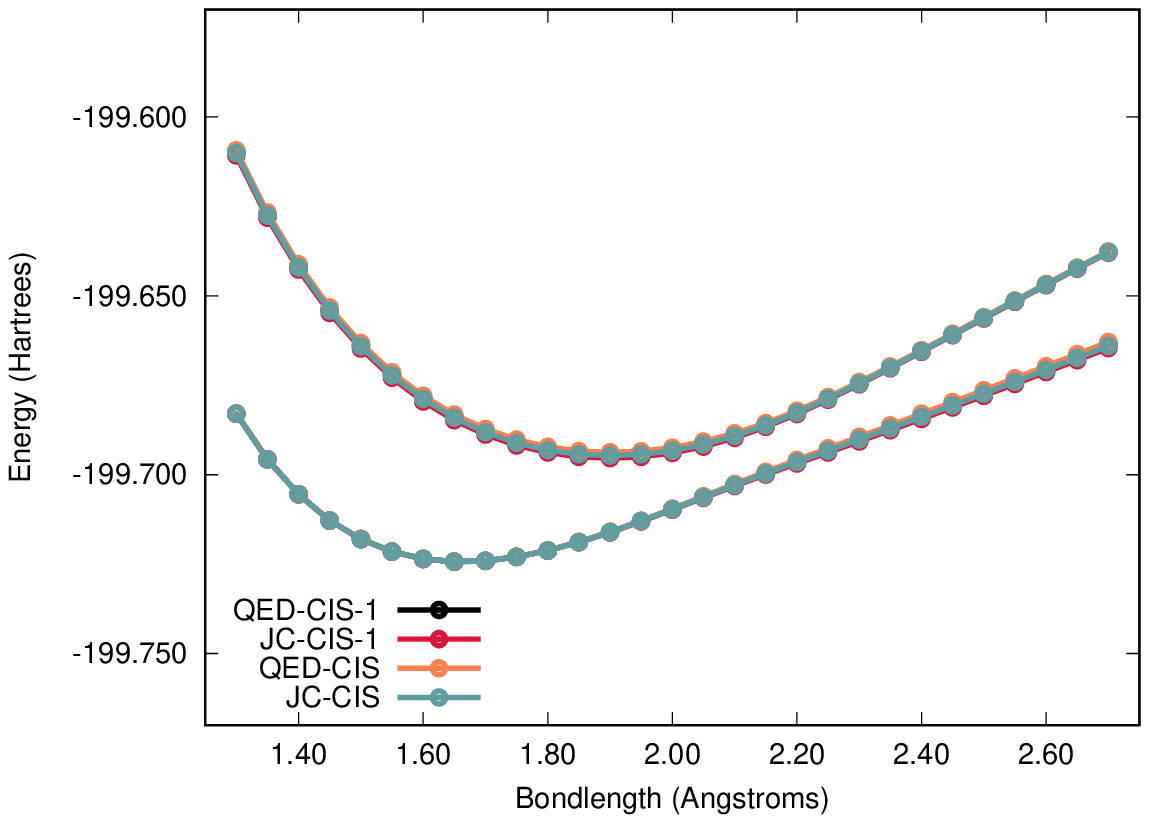}
    \includegraphics[height=0.49\textwidth, angle=270]{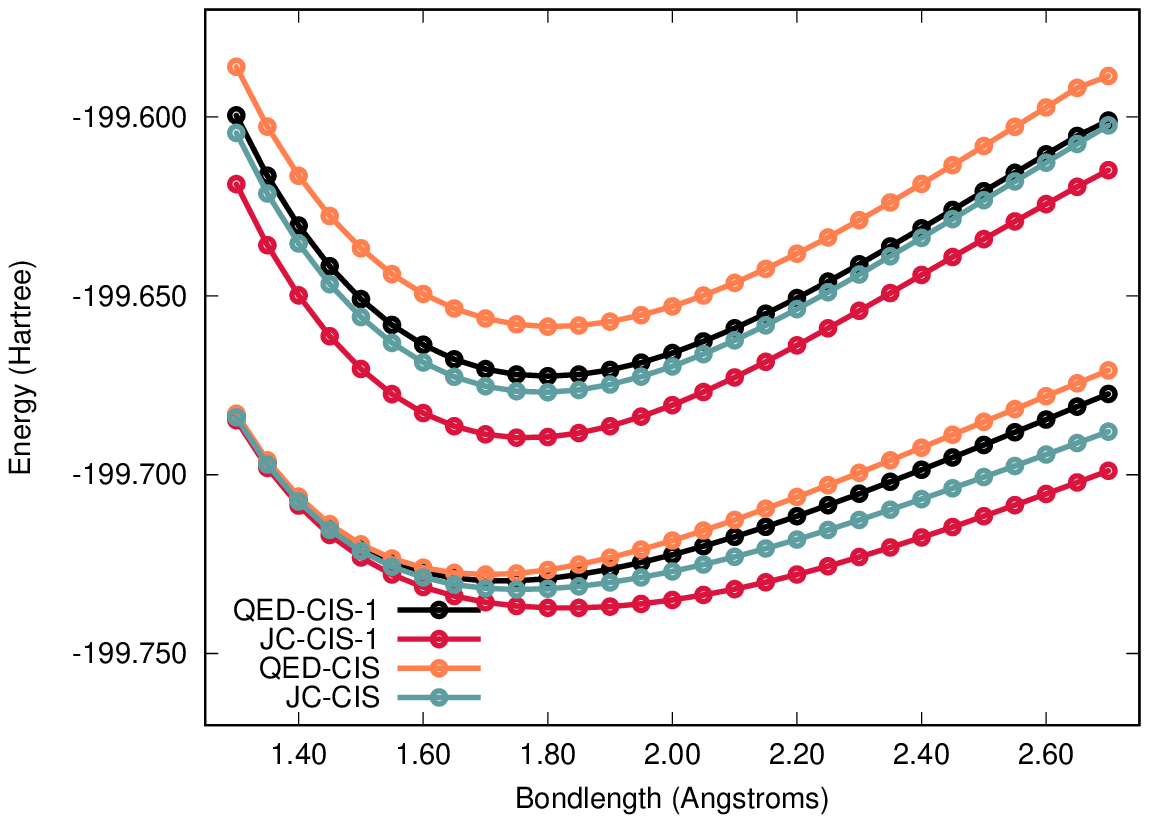}
    \caption{Potential energy curves of the lower- and upper-polariton states
    of MgH$^{+}$ coupled to a photon with energy $\omega = 4.75$ eV and fundamental
    coupling strength $|{\bm \lambda}| = 0.01$ a.u. (left),
    and $|{\bm \lambda}| = 0.05$ a.u. (right). Energies correspond to E$_\text{QED-HF}$ + $\Omega$, where $\Omega$ is the excitation energy for a given QED-CIS-1 facet.}
    \label{FIG:MGHP_PES}
\end{figure}

Ref.~\citenum{Shao21_064107} provides a detailed analysis of the behavior of different facets of the QED-TDDFT prism for several cavity-coupled molecular systems. Here, we consider how the description of an MgH$^+$ cation coupled to a single-mode cavity differs for facets of the QED-CIS-1 prism. The cavity mode frequency is chosen to be resonant with the $S_0 \to S_1$ transition in MgH$^+$ at an Mg--H distance of 2.2 \AA~(4.75 eV, as evaluated at the CIS/cc-pVDZ level of theory). The molecule is chosen to be oriented along the cavity mode polarization axis, and we consider two coupling strengths, $|{\bm \lambda}| = 0.01$ a.u. and $|{\bm \lambda}| = 0.05$ a.u. For the smaller coupling strength ($|{\bm \lambda}| = 0.01$ a.u.), all facets of the prism provide a similar description of the upper and lower polariton states (see Fig.~\ref{FIG:MGHP_PES}). On the other hand, clear differences between each model become evident for the stronger coupling strength ($|{\bm \lambda}| = 0.05$ a.u.).  Not surprisingly, energies from Jaynes-Cummings approximations (JC-CIS-1 and JC-CIS) are consistently lower than those from the Pauli-Fierz approaches (QED-CIS-1 and QED-CIS) because the Jaynes-Cummings model neglects the quadratic dipole self energy contributions, which are non-negative.  We also see that  QED-CIS-1 energies are consistent lower bounds to energies from QED-CIS; the reason is that simultaneous electron excitations and photon creation terms in QED-CIS-1 account for electron-photon correlation effects that lower the energy. For large coupling strengths, these effects can be quite large; at an Mg--H bond length of 2.2 \AA~and $|{\bm \lambda}| = 0.05$ a.u., for example, the energies of the upper- and lower-polariton states computed by QED-CIS and QED-CIS-1 energies differ by  12.4 mE$_\text{h}$ and 5.35 mE$_\text{h}$, respectively. 

As mentioned above, simultaneous electron excitations and photon creation terms in QED-CIS-1 incorporate electron-photon correlation effects into the approach and, as a result, the lowest-energy eigenvalue associated with Eq.~\ref{EQN:H-QED-CIS-1} is nonpositive and corresponds to an electron-photon correlation contribution to the ground-state energy. Table \ref{table:ciss_ground_state} quantifies these effects for a formaldehyde molecule  coupled to a single-mode cavity with two different coupling vectors, ${\bm \lambda}_z$ and ${\bm \lambda}_{yz}$, which both have magnitudes of 0.1 a.u.~and were defined in \ref{sec:cs-qed-hf}. The geometry for formaldehyde was taken from Ref.~\citenum{Foley_154103}, with the principal axis of the molecule aligned in the $z$-direction. The authors of Ref.~\citenum{Foley_154103} considered a photon mode with $\omega$ = 10.4 eV, which is approximately resonant with the first two dipole allowed transitions at the CIS/cc-pVDZ level of theory. The changes to the ground-state energy as predicted by QED-CIS-1 are given relative to the canonical RHF method and the QED-HF method in Table~\ref{table:ciss_ground_state}.  A Jupyter-notebook-based tutorial implementing the prism of QED-CIS-1 methods can be found \href{https://github.com/FoleyLab/psi4polaritonic/blob/cpr/QED-CIS-1.ipynb}{online}.\cite{Foley23_QED_CIS_TUTORIAL}  The tutorial provides a benchmark calculation on the MgH$^{+}$ ion, and it can easily be modified to study other systems.

\begin{table}[!ht]\label{table:ciss_ground_state}
\begin{tabular}{c|c|c}
\hline
 & Relative to RHF & Relative to QED-HF \\
\hline
 Polarization & $\Delta E$ (eV) &   $\Delta E$ (eV)   \\
 \hline
${\bm \lambda}_z$ & 0.811  & -0.318   \\
 \hline
${\bm \lambda}_{yz}$ & 0.771 & -0.266  \\
  \hline
\end{tabular}
\caption{Changes to the ground-state energy predicted by QED-CIS-1 relative to the canonical RHF energy as well as the QED-HF energy in atomic units. Calculations were performed with a fixed magnitude of $|\lambda| = 0.1$ a.u.~for the $\lambda_z$ and $\lambda_{yz}$ polarizations. QED-CIS-1 method calculations were performed to reflect formaldehyde coupling to a photon with $\hbar \omega = 10.4$ eV.}
\end{table}

\section{Cavity QED Coupled Cluster (QED-CC)}

\label{sec:qed-cc}

Beyond the single-particle theories discussed in the previous sections, a number of groups have considered many-body frameworks for {\em ab initio} cavity QED calculations. Many of these efforts have focused on the coupled-cluster (CC)\cite{Cizek66_4256, Paldus71_359, Bartlett09_book, Musial07_291} {\em ansatz}, which has enjoyed great success in conventional (non-QED) quantum chemistry applications. CC methods exhibit a number of desirable features that have contributed to this success, including the size-extensivity of truncated CC expansions, the size-intensivity of equation of motion (EOM)\cite{Bartlett93_7029,Bartlett12_126,Musial07_291,Krylov08_433} or linear-response\cite{Monkhorst77_421,Mukherjee79_325,Monkhorst83_1217,Jorgensen90_3333,Helgaker90_3345,Koch97_8059,Jorgensen95_7429,Jorgensen19_134109} CC excitation energies, and systematic convergence of the approach toward the full CI limit.

Two slightly different generalizations of CC theory for use with the PF Hamiltonian appeared in the literature at roughly the same time.\cite{Manby20_023262, Koch20_041043} The polaritonic coupled-cluster theory of Mordovina, Bungey, Appel, Knowles, Rubio, and Manby
\cite{Manby20_023262} considered an exponential parametrization of the ground-state polaritonic wave function that included single and double electronic transition operators, as well as photon creation operators and coupled electron transition and photon creation operators. They applied this {\em ansatz}, along with QED full CI, to the description of strong coupling between a single photon mode and a four-site Hubbard model. It should be noted that this work did not use typical boson creation operators, but, rather, nilpotent operators that lead to a linear parametrization of the photon space. On the otherhand, the QED-CCSD-1 model presented by Haugland, Enrico Ronca, Kj{\o}nstad,  Rubio, and  Koch\cite{Koch20_041043} used an exponential parametrization of similar complexity, along with more familiar (non-nilpotent) boson creation operators, and they applied this approach strong coupling problems involving an {\em ab initio} molecular Hamiltonian. The ground-state QED-CCSD-1 wave function is
\begin{equation}
    \label{EQN:QED-CC-WFN}
    |\Psi_{\rm CC}\rangle = e^{\hat{T}}|\Phi_{0}\rangle
\end{equation}
with
\begin{eqnarray}
    \label{EQN:T}
    \hat{T} &=& \sum_{ia} t_i^a \hat{a}^\dagger_a \hat{a}_i  + \frac{1}{4} \sum_{ijab} t_{ij}^{ab} \hat{a}^\dagger_a \hat{a}^\dagger_b \hat{a}_j \hat{a}_i 
    + u_0 \hat{b}^\dagger + \sum_{ia} u_i^a \hat{a}^\dagger_a \hat{a}_i \hat{b}^\dagger + \frac{1}{4} \sum_{ijab} u_{ij}^{ab} \hat{a}^\dagger_a \hat{a}^\dagger_b \hat{a}_j \hat{a}_i \hat{b}^\dagger
\end{eqnarray}
and where $|\Phi_0\rangle$ is a reference configuration of the form
\begin{equation}
    |\Phi_0\rangle = |0^{\rm e}\rangle \otimes |0\rangle
\end{equation}
In Eq.~\ref{EQN:T}, the symbols $t_i^a$, $t_{ij}^{ab}$, $u_0$, $u_i^a$, and $u_{ij}^{ab}$ represent the cluster amplitudes, and we can see that QED-CCSD-1 is an extension of the usual CCSD model\cite{Bartlett82_1910} that includes both photon creation operators and products of electronic transition and photon creation operators. 

Excited states in QED-CC theory are represented within the EOM-CC framework,\cite{Bartlett93_7029,Bartlett12_126,Musial07_291,Krylov08_433} in which we define both left- and right-hand excited states of the form
\begin{eqnarray}
    | \Psi_I \rangle = \hat{R}_I e^{\hat{T}} | \Phi_0  \rangle \\
    \langle \tilde{\Psi}_I | = \langle \Phi_0 |  \hat{L}_I e^{-\hat{T}}
\end{eqnarray}
where, the label {\em I} denotes the state. These functions satisfy left- and right-hand eigenvalue equations 
\begin{align}
\label{EQN:EOM_CC_eig_L}
    \langle \Phi_0 |  \hat{L}_I \bar{H}  &=  \langle \Phi | \hat{L}_I E_I \\
    \label{EQN:EOM_CC_eig_R}
    \bar{H} \hat{R}_I |\Phi_0 \rangle &= E_I \hat{R}_I |\Phi \rangle 
\end{align} 
involving the similarity transformed PF Hamiltonian, $\bar{H} = e^{-\hat{T}}\hat{H}e^{\hat{T}}$. Here, $\hat{H}$ is represented in the coherent-state basis. At the EOM-QED-CCSD-1 level of theory, the $\hat{R}_I$ and $\hat{L}_I$ operators are defined by
\begin{eqnarray}
\label{EQN:EOM_CC_L}
    \hat{L}_I &=&{l}_0 + \sum_{ai} {l}^i_a \hat{a}^\dagger_i\hat{a}_a + \frac{1}{4} \sum_{abij} {l}_{ab}^{ij} \hat{a}^\dagger_i \hat{a}^\dagger_j \hat{a}_b \hat{a}_a \nonumber\\
    &+&{m}_0 \hat{b} + \sum_{ai} {m}^i_a \hat{a}^\dagger_i\hat{a}_a \hat{b} + \frac{1}{4} \sum_{abij}{m}_{ab}^{ij} \hat{a}^\dagger_i \hat{a}^\dagger_j \hat{a}_b \hat{a}_a \hat{b}
\end{eqnarray}
and
\begin{eqnarray}
\label{EQN:EOM_CC_R}
    \hat{R}_I &=&{r}_0 + \sum_{ai} {r}^a_i \hat{a}^\dagger_a\hat{a}_i + \frac{1}{4} \sum_{abij} {r}_{ij}^{ab} \hat{a}^\dagger_a \hat{a}^\dagger_b \hat{a}_j \hat{a}_i \nonumber\\
    &+&{s}_0 \hat{b}^\dagger + \sum_{ai} {s}^a_i \hat{a}^\dagger_a\hat{a}_i \hat{b}^\dagger + \frac{1}{4} \sum_{abij} {s}_{ij}^{ab} \hat{a}^\dagger_a \hat{a}^\dagger_b \hat{a}_j \hat{a}_i \hat{b}^\dagger
\end{eqnarray} 
respectively, and the amplitudes appearing in Eqs.~\ref{EQN:EOM_CC_L} and \ref{EQN:EOM_CC_R} are determined by solving Eqs.~\ref{EQN:EOM_CC_eig_L} and \ref{EQN:EOM_CC_eig_R}.

Since 2020, several groups have developed implementations of similar QED-CC approaches and explored the influence of cavity effects on various ground-state properties. DePrince\cite{DePrince21_094112} used QED-CCSD-1 to demonstrate that strong coupling leads to appreciable changes in electron affinities in sodium halide compounds and that QED-HF significantly overestimates these effects. Ionization potentials were found to be less sensitive to cavity effects in these systems.
Pavošević and Flick\cite{Flick21_9100} also explored the influence of cavity effects on electron affinities using a unitary formuation of QED-CCSD-1, implemented using the variational quantum eigensolver (VQE)\cite{Obrien14_4213, Solano14_3589, Aspuru-Guzik16_023023} algorithm, on a quantum computer. They also extended the framework to include up to two photon creation operators plus single and double electronic excitaitons (termed QED-CCSD-2). These works led to a study on the features of ionization in QED environments by Riso, Haugland, Ronka and Koch\cite{Koch22_234103} that highlighted the importance of an appropriate treatment of the ionized electron.

Beyond these studies on ionization / electron attachment, a number of works have used QED-CC approaches to explore how vacuum fluctuations can be leveraged in chemical contexts. Here, it is important to note that we are referring to changes to ground states of cavity-embedded systems, without driving transitions or creating polariton states via the addition of photons to the cavity.  Pavošević, Hammes-Schiffer, Rubio, and Flick\cite{Flick22_4995} used non-unitary QED-CCSD-2 to show that strong coupling leads to non-negligible changes in proton transfer reaction barrier heights; changes as large as 20\% were reported in Ref.~\citenum{Flick22_4995}. These authors also introduced an approximation to QED-CCSD-2 in which  single electron transitions appear with up to two photon creation operators, but double electron transitions only appear with up to single photon creation operators (termed QED-CCSD-21). This QED-CCSD-21 model has a similar structure to the approach of White, Gao, Minnich, and Chan,\cite{Chan20_224112} which was developed to model electron-phonon interactions. Pavošević, Smith, and Rubio applied an approximate QED-CCSD-1 model (that ignores coupled two-electron plus photon interactions) to two cycloaddition reactions. In that work, the authors demonstrated that sufficiently strong coupling, along with precise control over the relative orientation of molecules and the cavity mode axis, could influence the major products of these reactions.  Pavošević and Rubio have also incorporated QED-CCSD-1 into an embedding protocol\cite{Rubio22_094101} that treats a subset of a cavity-embedded molecular system using QED-CC and the remainder of the system via QED-DFT or QED-HF (termed `QED-CC-in-QED-SCF''). Assuming that electron-photon correlations are limited to the embedded  region, this protocol could circumvent the high computational cost of the many-body {\em ab initio} cavity QED framework. 

Haugland, Schäfer, Ronca, Rubio and Koch used QED-CCSD-1, QED-DFT, and QED full CI to model the effects of vacuum fluctuations on nature of intermolecular interactions.\cite{Koch21_094113}  Not surprisingly, QED-HF and QED-DFT do not provide good descriptions of intermolecular interactions in a cavity, particularly for van der Waals interactions. Additional notable observations include an $R^{-3}$ contribution to van der Waals interactions (which display $R^{-6}$ dependence in the absence of a cavity), stemming from electron-photon correlations, and an apparently infinite distance over which cavity-embedded molecules remain correlated, which results from the dipole self-energy contribution to the interaction energy. It should be noted that the coupling strength employed in this study was quite large: $\lambda = 0.1$ a.u., which, assuming a single cavity mode, corresponds to an effective mode volume of $\approx 0.2$ nm$^3$. The authors correctly note that, at the mean-field level, multiple modes polarized along the same axis can be treated as a single effective mode with coupling strength, $\lambda_{\rm eff}^2 = \sum_i \lambda_i^2$. Even so, some conclusions regarding long-range correlation effects involve inter-molecule distances on the order of hundreds of \AA, which seems inconsistent with such large coupling strengths. More recently, Philbin, Haugland, Ghosh, Ronca, Chen, Narang, and Koch\cite{Koch22_arXiv:2209.07956} used machine learning (ML) techniques to learn intermolecular potentials for cavity-embedded dimers of H$_2$ molecules, which were treated using QED-CCSD-1 plus two-photon creation operators (termed QED-CCSD-12-SD1 in that work) and QED full CI with up to five photon creation operators (QED-FCI-5). Interestingly, comparisons between QED-CCSD-1 and QED-CCSD-12-SD1 revealed that two-photon transitions are crucial for recovering the correct sign on interaction energies for H$_2$ molecules separated by large distances; QED-CCSD-12-SD1 and QED-FCI-5 predict these interactions to be attractive, while QED-CCSD-1 predicts a repulsive interaction. Given machine-learned potentials, path integral molecular dynamics simulations on hundreds of cavity embedded molecules revealed that cavity-modified van der Waals interactions result in orientational order not seen in cavity-free simulations. 

In 2022, Riso,  Grazioli,  Ronca,  Giovannini, and Koch\cite{Koch22_chiral} developed a formulation of QED-CCSD-1 that models interactions between electronic degrees of freedom and the quantized photon field of a chiral cavity mode. They found that a proper description requires that the photon field be treated beyond the dipole (or even multipolar) approximation, which results in a complex-valued Hamiltonian that depends two cavity modes (for a single resonant frequency). These complications aside, Ref.~\citenum{Koch22_chiral} demonstrated that circularly polarized light can discriminate between enantiomers of chiral molecules embedded within a chiral cavity ({\em e.g.}, via changes to the energies of the ground states of the enantiomers or their rotational spectra). Moreover, the discriminating power of the cavity increases with the number of molecules.  

Cleary, a large body of work has considered the effects of strong light-matter interactions on ground states of cavity-embedded systems. Somewhat less work has considered excited-state electronic/polaritonic structure of such systems. The initial papers\cite{Manby20_023262, Koch20_041043} describing generalizations of CC theory for use with the PF Hamiltonian developed and applied QED-EOM-CC formalisms to cavity-embedded systems. In particular, Ref.~\citenum{Koch20_041043} describes how polariton formation can manipulate conical intersections; QED-CCSD-1 calculations on a cavity-coupled pyrole molecule show sufficiently strong coupling can open a gap at a conical intersection between the ${}^1{\rm B}_1$ ${}^1{\rm A}_2$ states. An exciting chemical consequence is that such modifications to the energy landscape could lead to changes in relaxation pathways or dynamics in chemical reactions. This idea has also been put forward in the context of linear response QEDFT, as well;\cite{Rubio19_2757} QEDFT simulations on cavity-embedded formaldehyde\cite{Narang20_094116} have showed that different combinations of cavity parameters can move or suppress avoided crossings between excited states. 
While we have limited this discussion to consider descriptions of purely electronic strong coupling, we recognize that Vidal, Manby, and Knowles\cite{Knowles22_204119} have used similar QED-EOM-CC approaches to explore how coupling to a cavity mode can affect vibronic structure.

Liebenthal and DePrince\cite{DePrince22_054105} extended QED-EOM-CC theory to consider non-particle-conserving excitation operators. Specifically, they developed a QED-EOM-CCSD-1 model for electron attachmentment (EA), which is a cavity QED generalization of the EOM-EA-CC approach\cite{Bartlett95_3629} from electronic structure theory. One of the key findings in Ref.~\citenum{DePrince22_054105} was that, in order to recover electron affinities obtained from separate QED-CCSD-1 calculations on different charge states,\cite{DePrince21_094112} QED-EOM-EA-CCSD-1 calculations starting from an $N$-electron reference must employ the coherent-state basis defined for the ($N+1$)-electron state. This finding suggests that the coherent-state basis should be chosen with care in any QED-EOM-CC model that samples non-particle or spin-conserving sectors of Fock space. This work also revealed defects in the similarity-transformed PF Hamiltonian ({\em i.e.}, complex eigenvalues) at a same-symmetry conical intersection in magnesium fluoride (MgF), involving the lower-polariton state. Such defects can emerge in standard EOM-CC theories that make use of truncated cluster expansions; the MgF example highlights that this issue persists in the cavity QED generalization of EOM-CC. 

We note that most QED-CC studies are formulated within the coherent-state basis introduced in Sec.~\ref{SEC:CoherentStateBasis}. The primary reason for this choice is that it guarantees that the correlated calculation will be strictly origin invariant, even for charged species. Liebenthal, Vu, and DePrince\cite{DePrince23_5264} studied the numerical consequences of this choice by comparing QED-CCSD-1 and QED-EOM-CCSD-1 calculations in the coherent-state basis, using a QED-HF reference (termed ``relaxed''), to calculations performed in the canonical Hartree-Fock basis, using a Hartree-Fock wave function that was not perturbed by cavity interactions (termed ``unrelaxed''). For the unrelaxed case, they found that the presence of exponentiated single electron transitions ($e^{\hat{T}_1}$) do a good job of accounting for orbital relaxation effects from QED-HF, while exponentiated boson creation operators ($e^{u_0\hat{b}^\dagger}$) can mimic the effects of the coherent-state transformation itself. For example, ground-state unrelaxed QED-CCSD-1 energies on charged species acquire only modest origin dependence; for a cavity-bound HF${^+}$ cation, described by a cc-pVDZ basis set and a large coupling strength of $\lambda = 0.05$ a.u., that work showed that the energy changes by less than 1$\times 10^{-3}$ $E_{\rm h}$ when shifting the molecule 10 \AA~from the origin.  Moreover, for the most part, excitation energies from relaxed and unrelaxed QED-EOM-CCSD-1 are similar, particularly for experimentally feasible coupling strengths ({\em i.e.}, $\lambda < 0.05$). These results stand in stark contrast to results obtained from  unrelaxed and relaxed formulations of QED-DFT and QED-TDDFT. First, unrelaxed QED-DFT acquires a substantial origin dependence in the energy (stemming from the dipole self energy contribution). Second, relaxed and unrelaxed QED-TDDFT yield significantly different spectra, with relaxed QED-TDDFT generally doing a better job of reproducing some trends from relaxed QED-EOM-CCSD-1. These observations are important, given that multiple formulations of of QED-TDDFT can be found in the literature, and not all of them account for cavity self-consistently in the ground state.\cite{Shao21_064107, Shao22_124104} 

Fregoni, Haugland, Pipolo, Giovannini, Koch, and Corni have applied QED-EOM-CCSD-1 to interactions between a molecular system and a plasmonic nano/picocavity.\cite{Corni21_6664} Their protocol is similar to that discussed throughout this Section, except for the precise form of the Hamiltonian. First, a polarized continuum model for nanoparticles\cite{Corni19_315} is applied to describe the plasmon mode. Second, the dipole self-energy contribution is not included in the Hamiltonian for the coupled system. The argument for neglecting the dipole self energy is that the collective electronic oscillations comprising the plasmon excitation interact with the molecule through longitudinal Coulomb interactions, and this interaction
dominates over the coupling between the molecule transverse component of the vector potential.~\cite{Feist19_021057, Fesit21_477}  It should also be noted that in the case of
strong coupling to a cavity mode with a significant material contribution to the excitation (such as a plasmonic mode), Eq.~\ref{EQN:PdotA} should be augmented to include coupling between the charged particles of the molecular subsystem and the electric scalar potential $\phi(x)$ associated with the plasmon excitation:  $\hat{H}_{{\rm p \cdot A}} = \sum_i^N \frac{1}{2m_i}\left(\hat{p}_i - z_i \hat{{\bf A}}_{\perp} \right)^2 + z_i \phi(x_i) + \hat{V}({\bf \hat{x}}) + \hbar \omega_{\rm cav} \bhd \bh$.  We note that the dipole self energy term (even if very small) still emerges upon PZW transformation of this Hamiltonian, particularly through transformation of the energy of the cavity mode $\hbar \omega_{\rm cav} \bhd \bh$ (see Eq.~\ref{EQN:PZW_HP}). Third, the bilinear coupling term takes a slightly different form.  Despite these differences, the QED-EOM-CCSD-1 wave function {\em ansatz} is the same as that discussed herein. Building upon this work,
Romanelli, Riso, Haugland, Ronca, Corni, and Koch\cite{Koch23_2302.05381} have developed a QED-CC model that folds in the effects of multiple plasmonic modes into a single effective mode. Other models for plasmon-molecule interactions that make use of quantized radiation fields and parametrized plasmon modes have been proposed as well.\cite{DePrince15_214104}

{\color{black}Lastly, two many-body perturbation theory approaches to cavity QED have recently emerged. First, a} cavity QED extension of second-order {\color{black}M{\o}ller-Plesset} perturbation theory (MP2) and the algebraic diagrammatic construction (ADC) has been developed by Bauer and Dreuw.\cite{Dreuw23_124128} QED-MP2 is an approximation to QED-CCSD-1, and, like conventional ADC, QED-ADC can be thought of a Hermitian approximation to QED-EOM-CCSD-1.  The data presented in Ref.~\citenum{Dreuw23_124128} suggest that the QED-MP2 correlation energy is much more sensitive to the frequency of the cavity mode than the correlation energy from QED-CCSD-1. This sensitivity is increased if the QED-MP2 calculations are performed on top of Hartree-Fock reference wave functions evaluated in the absence of the cavity. Hence, it appears that, like QED-DFT and QED-TDDFT, the QED-MP2 {\em ansatz} is not as robust as QED-CCSD-1 to the description of cavity effects at the mean-field level. {\color{black}On the other hand, the Rayleigh-Schr\"{o}dinger perturbation theory\cite{Narang23_arXiv:2307.14822} presented by Haugland, Philbin, Ghosh, Chen, Koch and Narang does an excellent job of reproducing ground-state energies from full QED-CC over a wide range of cavity frequencies and coupling strengths. This perturbation theory is general and can be implemented for any electronic structure theory for which linear-response theory has been formulated. }

\section{Transformation of Operators}\label{sec:transformation_of_operators}
In the preceeding sections, we have obtained (approximate) eigenstates of $\hat{H}_{\rm CS}$, where
$\hat{H}_{\rm CS}$ results from a unitary transformation of our original Hamiltonian in Eq.~\ref{EQN:PdotA}.  In the following, we discuss relationships that hold between the
exact eigenstates of $\hat{H}_{\rm  CS}$ (which could be obtained, for example, through full configuration interaction in a complete single-particle basis) and 
$\hat{H}_{\rm p \cdot A}$.  Although it is generally not possible to obtain the exact eigenfunctions of $\hat{H}_{\rm  CS}$ or $\hat{H}_{\rm p \cdot A}$, we will work out
practical relationships for the photonic character and the dipole operator and apply them to expectation values taken with approximate eigenfunctions obtained from the QED-CIS-1 method.

The exact eigenvalues of an operator are preserved under unitary rotations, while the eigenfunctions
of $\hat{H}_{\rm CS}$ are related to the eigenfunctions of $\hat{H}_{\rm  p \cdot A}$ by a unitary transformation.  In particular, we have:
\begin{align}
    \hat{H}_{\rm  p \cdot A}  \longrightarrow  \hat{H}_{\rm CS} \; \; &{\rm via} \; \; \hat{U}\hat{H}_{\rm  p \cdot A}\hat{U}^{\dagger} \\
    |\Psi_I\rangle  \longrightarrow  |\Psi^{'}_{I}\rangle \; \; &{\rm via} \; \; \hat{U}|\Psi_I\rangle \\
    \hat{H}_{\rm  p \cdot A} |\Psi_I\rangle &= E_I |\Psi_I\rangle \\
    \hat{H}_{\rm  CS} |\Psi^{'}_{I}\rangle &= E_I |\Psi^{'}_{I}\rangle.
\end{align}
Therefore, in order for expectation values computed with these transformed eigenstates to have correspondance with the expectation values computed with the eigenstates of 
$\hat{H}_{\rm p \cdot A}$, we must transform the operators as follows:
\begin{align}
    \langle \Psi_I | \hat{O} | \Psi_I \rangle &=  \langle \Psi_I^{'} | \hat{O}^{'} | \Psi_I^{'} \rangle \\
    &= \langle \Psi_I | \hat{U}^{\dagger} \hat{O}^{'} \hat{U}| \Psi_I \rangle \\
    &= \langle \Psi_I | \hat{U}^{\dagger} \hat{U} \hat{O} \hat{U}^{\dagger} \hat{U}| \Psi_I \rangle.
\end{align}
Thus we see the transformation for operators to use with our
transformed eigenstates is also $\hat{O}^{'} = \hat{U} \hat{O} \hat{U}^{\dagger}$.
Specifically, following transformation of the Hamiltonian from the miminal coupling Hamiltonian in Eq.~\ref{EQN:PdotA} to the Pauli-Fierz Hamiltonian in the length gauge and to the coherent state basis, we must apply the same transformations to operators for the purposes of computing expectation values with the eigenfunctions of Eq.~\ref{EQN:PFH_CS}.  
Following transformation of the Hamiltonian from the miminal coupling Hamiltonian in Eq.~\ref{EQN:PdotA} to the Pauli-Fierz Hamiltonian in the length gauge and to the coherent state basis, we apply the same transformations to operators for the purposes of computing expectation values with the eigenfunctions of Eq.~\ref{EQN:PFH_CS}.

Some operators will commute with the operators that provide these transformations ($\hat{U}_{{\rm PZW}}$,  $\hat{U}_{\phi}$, and $\hat{U}_{{\rm CS}}$) and will be unchanged, while others will be transformed.  It is common to compute the photonic character of a polaritonic state, and so here we investigate the behaviour of the photon number operator, $\hat{N}_{\rm p} = \hat{b}^{\dagger} \hat{b}$ for a single photon mode.  Furthermore, the dipole moment expectation value of the polariton system can be of interest~\cite{Foley_154103}, so we will also investigate the behaviour of the dipole moment operator $\hat{{\bm \mu}}.$

For a single photonic mode:
\begin{equation}
\hat{U}_{{\rm PZW}} \bhd \bh \hat{U}^{\dagger}_{{\rm PZW}} = \bhd \bh + \frac{i}{\hbar} \sqrt{\frac{1}{2\omega_{\rm cav}}} {\bm \lambda} \cdot \hat{\bm \mu} (\bhd - \bh) + \frac{1}{\hbar^2}\frac{1}{2\omega_{\rm cav}}({\bm \lambda} \cdot \hat{\bm \mu})^2,
\end{equation}
\begin{equation}
\hat{U}_{\phi} \hat{U}_{{\rm PZW}} \bhd \bh \hat{U}^{\dagger}_{{\rm PZW}} \hat{U}^{\dagger}_{\phi} = \bhd \bh - \frac{1}{\hbar} \sqrt{\frac{1}{2\omega_{\rm cav}}} {\bm \lambda} \cdot \hat{\bm \mu} (\bhd + \bh) + \frac{1}{\hbar^2}\frac{1}{2\omega_{\rm cav}}({\bm \lambda} \cdot \hat{\bm \mu})^2,
\end{equation}
and 
\begin{equation}
 \hat{N}_{CS} = \bhd \bh - \frac{1}{\hbar} \sqrt{\frac{1}{2\omega_{\rm cav}}} [{\bm{\lambda}} \cdot ({\bm{\hat{\mu}}} - \langle {\bm{\hat{\mu}}} \rangle )] (\bhd + \bh) + \frac{1}{\hbar^2}\frac{1}{2\omega_{\rm cav}}[{\bm{\lambda}} \cdot ({\bm{\hat{\mu}}} - \langle {\bm{\hat{\mu}}} \rangle )]^2,
\end{equation}
where $\hat{N}_{CS} = \hat{U}_{\rm CS}\hat{U}_{\phi} \hat{U}_{{\rm PZW}} \bhd \bh \hat{U}^{\dagger}_{{\rm PZW}} \hat{U}^{\dagger}_{\phi} \hat{U}^{\dagger}_{\rm CS}$.

On the other hand, The PZW transformation of the dipole operator can be shown to preserve the expectation values because the dipole operator can be shown to commute with $\hat{\bm \mu} \cdot \hat{\bf A}$ since $\hat{\bf A}$ operators only on photon degrees of freedom, and $\hat{\bm \mu}$ must commute with itself.  Similarly, since the phase and coherent state transformations involve only photon operators and $\hat{\bm \mu}$ involves only electron operators, the dipole operator is unchanged by these transformations, and we have 
\begin{equation}
    \hat{U}_{CS} \hat{U}_{\phi} \hat{U}_{PZW} \hat{\bm \mu}\hat{U}^{\dagger}_{PZW}
 \hat{U}^{\dagger}_{\phi} \hat{U}^{\dagger}_{CS} = \hat{\bm \mu}.
\end{equation}

Of course we are not typically able to obtain the exact eigenfunctions for $\hat{H}_{\rm CS}$; for example we will perform some truncation in the single-particle basis and/or in the
many-particle basis.  We will derive explicit expressions in the case that we have truncated the many-particle basis consistent with QED-CIS-1; these expressions are independent of the level of truncation of the single-particle basis.

Recalling the form of the QED-CIS-1 wavefunction (~\ref{EQN:QED-CIS-1}),
we will examine the explicit expressions for the photonic occupation of a given
electronic state $\Psi_I$ that can be defined as
\begin{align}\label{EQN:PHOTON_OCC}
\langle N_{CS} \rangle &= \langle \Psi_I | \hat{N}_{CS} | \Psi_I \rangle \nonumber \\
&= \langle \Psi_I | \bhd \bh | \Psi_I \rangle 
- \frac{1}{\sqrt{2\omega_{cav}}} \langle \Psi_I | {\bm{\lambda}} \cdot ({\bm{\hat{\mu}}} - \langle {\bm{\hat{\mu}}} \rangle )(\bhd + \bh) | \Psi_I \rangle 
+ \frac{1}{2\omega_{cav}} \langle \Psi_I | {\bm{\lambda}} \cdot ({\bm{\hat{\mu}}} - \langle {\bm{\hat{\mu}}} \rangle )^2 | \Psi_I \rangle.
\end{align}
The first expectation value can be computed as follows:
\begin{equation}\label{EQN:ZO_PHOTON_OCC}
    \langle \Psi_I | \bhd \bh | \Psi_I \rangle = |c_0^1|^2 + \sum_{ia} |c_{ia}^1|^2.
\end{equation}
The second expectation value can be computed as follows:
\begin{align}\label{EQN:FO_PHOTON_OCC}
-\frac{1}{\sqrt{2\omega_{cav}}} \langle \Psi_I | {\bm \lambda} \cdot ({\bm{\hat{ \mu}_{\rm e}}} - 
\langle {\bm \mu} \rangle_e)(\bhd + \bh) | \Psi_I \rangle = -\frac{1}{\sqrt{2\omega_{cav}}}{\bf c}^{\rm T} {\bf H}_{{\rm blc}} {\bf c},
\end{align}
where ${\bf c}$ denotes the QED-CIS-1 eigenvector for state $I$ and ${\bf H}_{{\rm blc}}$ is the contribution of the Hamiltonian
matrix in Eq.~\ref{EQN:H-QED-CIS-1} that contains only the elements given in Eqs.~\ref{EQN:BLC-ELEMENTS-1} and \ref{EQ:blc_elements_2}.
The third expectation value can be computed as 
\begin{equation}\label{EQN:SO_PHOTON_OCC}
    \frac{1}{2\omega_{cav}} \langle \Psi_I | ({\bm \lambda} \cdot({\bm {\hat{\mu}_{\rm e}}} -\langle {\bm \mu}_{\rm e} \rangle))^2 | \Psi_I \rangle = 
    \frac{1}{2\omega_{cav}} {\bf c}^{\rm T} {\bf H}_{{\rm dse}} {\bf c},
\end{equation}
where ${\bf H}_{{\rm blc}}$ is the contribution of the Hamiltonian
matrix in Eq.~\ref{EQN:H-QED-CIS-1} that contains only the elements given in Eqs.~\ref{EQN:DSE-ELEMENTS}.

\begin{figure}[!h]
    \centering
    \includegraphics[height=0.50\textwidth, angle=270]{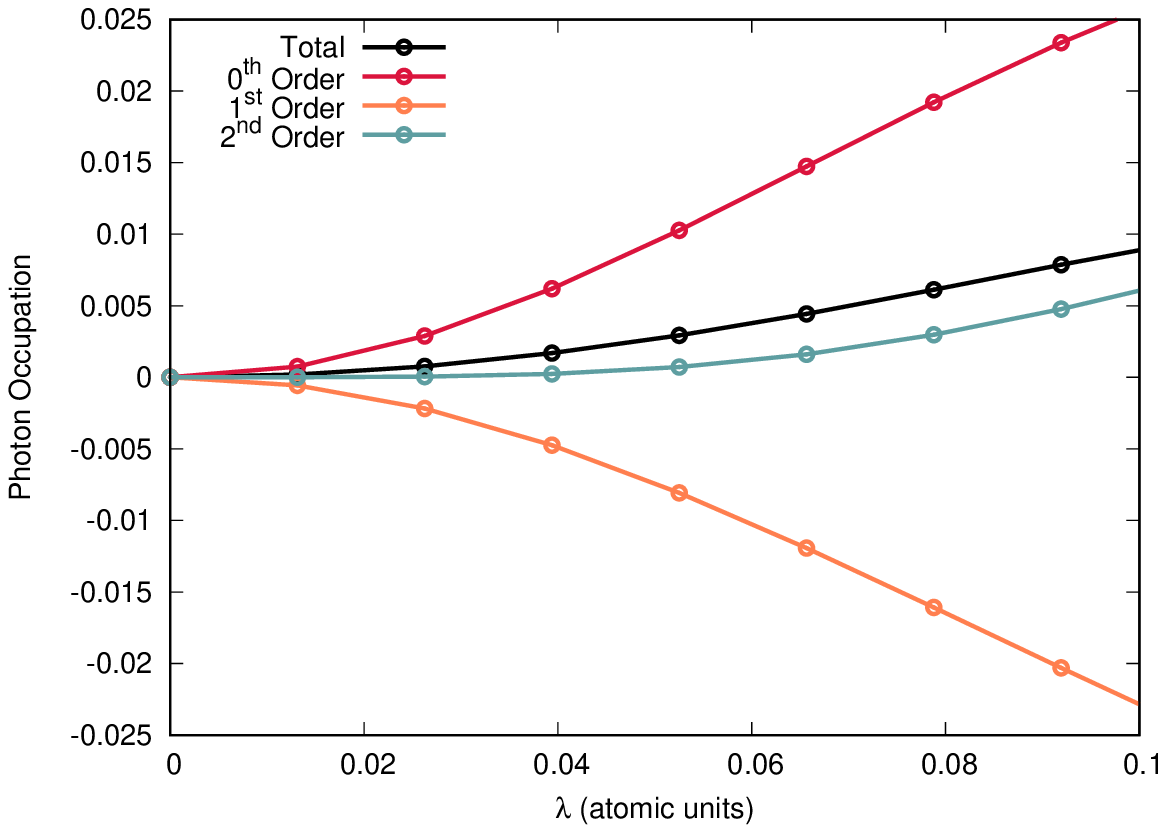}
    \caption{Photon occupation of the ground-state of the ${\rm MgH^+}$ ion as a function of the fundamental coupling strength $\lambda$.  The total photon occupation is computed using Eq.~\ref{EQN:PHOTON_OCC}, the $0^{{\rm th}}$ Order contribution comes from 
    Eq.~\ref{EQN:ZO_PHOTON_OCC}, the $1^{{\rm st}}$ Order contribution comes from 
    Eq.~\ref{EQN:FO_PHOTON_OCC}, and the $2^{{\rm nd}}$ Order contribution comes from 
    Eq.~\ref{EQN:SO_PHOTON_OCC}.  This calculation is performed at the QED-CIS-1/cc-pVDZ level of theory with a photon frequency of $\hbar \omega_{\rm cav} = 4.75 {\rm eV}$.  The bondlength is fixed at 2.2 Angstroms. }
    \label{FIG:PHOTON_OCCUPATION}
\end{figure}

We plot these various contributions and the total photon occupation of the QED-CIS-1 ground-state of the ${\rm MgH^+}$ ion as a function of the fundamental coupling strength $\lambda = \sqrt{\frac{\hbar}{\epsilon_0 V}}$ from a photon polarized purely along the principle
axis of the molecule in
Figure~\ref{FIG:PHOTON_OCCUPATION}.  Here we denote the $0^{\rm th}$ order contribution
as arising from Eq.~\ref{EQN:ZO_PHOTON_OCC}, the $1^{\rm st}$ order contribution
as arising from Eq.~\ref{EQN:FO_PHOTON_OCC}, the 
$2^{\rm nd}$ order contribution
as arising from Eq.~\ref{EQN:SO_PHOTON_OCC}, and the Total as arising from the sum of these three terms, e.g. Eq.~\ref{EQN:PHOTON_OCC}.

\section{Concluding Remarks}
Despite the impressive surge of theoretical and experimental advances in polariton chemistry and molecular polaritonics, many challenges and opportunities remain to advance the field towards its full promise.  While
it may seem daunting to span the chasm that exists between the majority of polariton experiments (done 
in the regime of 10$^6$ to 10$^9$ molecules within the cavity mode volume) to the regime accessible by even large-scale atomistic methods~\cite{Li_PNAS_2022} (~100s of molecules), we assert that all advances in the theoretical treatment of cavity-molecule interactions provide value towards the goal of understanding and controlling polariton chemistry.  In particular, 
single- and few-molecule strong coupling has been experimentally realized with several different cavity platforms,\cite{Baumberg16_127, Sandoghdar_PRL_2021} and, as the limits of this regime are expanded, there is an urgent need for rigorous and non-perturbative quantum mechanical methods that can accurately capture modifications to ground- and excited-state properties and emergent phenomena.  The techniques described in
this review provide such a rigorous foundation, although we should note that there are additional advances 
required for plasmonic nanocavities, such as rigorous inclusion of longitudinal scalar potential coupling 
to capture the material contribution of plasmon excitation, and inclusion of the modified chemical environment 
that molecules experience in the vicinity of plasmonic particles in the dark.\cite{Feist22_ACSPhoton}  Some of these effects are
more naturally included in the real-space Coulomb gauge formulations described in Refs.~\citenum{Rubio22_Arxiv, Varga_JCP_2022}, 
which then leaves us with an intriguing theoretical challenge for formulations based on Gaussian basis sets and in the length gauge, or Coulomb gauge formulations with Gaussian basis sets, as reported by Koch and co-workers.\cite{Koch22_chiral}  Moreover, theoretical approaches (quantum and classical) can be deployed to approach collective strong coupling 
from the bottom up, which may provide valuable insights into some of the phenomena
that are observed in this regime.  In this case, the availability of 
rigorous methods to benchmark lower-scaling methods (e.g. density functional based approaches, parameterized and semi-empirical approaches, and classical force fields) will be paramount.  We hope that this tutorial review will serve to orient
researchers towards these varied areas of development, as well as to provide the foundation for further development of \emph{ab initio} QED approaches and
the sound deployment of these methods.

\vspace{0.5cm}

\vspace{0.5cm}

{\bf Author Information} 

Present Address

\textsuperscript{\textbar \textbar} Department of Chemistry, 
             Texas A\&M University, 
             College Station, TX 77843
\vspace{0.5cm}
\vspace{0.5cm}

{\bf Acknowledgments} This material is based upon work supported by the National Science Foundation under Grant No. CHE-2100984.  J.J.F Acknowledges support from the Research Corporation for Scientific Advancement Cottrell Scholar Award.  JJF and J.M. and the NSF CAREER Award CHE-2043215. J.J.F. acknowledges support from the Center for MAny-Body Methods, Spectroscopies, and Dynamics for Molecular POLaritonic Systems (MAPOL) under subcontract from FWP 79715, which is funded as part of the Computational Chemical Sciences (CCS) program by the U.S. Department of Energy, Office of Science, Office of Basic Energy Sciences, Division of Chemical Sciences, Geosciences and Biosciences at Pacific Northwest National Laboratory (PNNL). PNNL is a multi-program national laboratory operated by Battelle Memorial Institute for the United States Department of Energy under DOE contract number DE-AC05-76RL1830.   
\bibliography{Journal_Short_Name.bib,main.bib}

\begin{thebibliography}{115}%
\makeatletter
\providecommand \@ifxundefined [1]{%
 \@ifx{#1\undefined}
}%
\providecommand \@ifnum [1]{%
 \ifnum #1\expandafter \@firstoftwo
 \else \expandafter \@secondoftwo
 \fi
}%
\providecommand \@ifx [1]{%
 \ifx #1\expandafter \@firstoftwo
 \else \expandafter \@secondoftwo
 \fi
}%
\providecommand \natexlab [1]{#1}%
\providecommand \enquote  [1]{``#1''}%
\providecommand \bibnamefont  [1]{#1}%
\providecommand \bibfnamefont [1]{#1}%
\providecommand \citenamefont [1]{#1}%
\providecommand \href@noop [0]{\@secondoftwo}%
\providecommand \href [0]{\begingroup \@sanitize@url \@href}%
\providecommand \@href[1]{\@@startlink{#1}\@@href}%
\providecommand \@@href[1]{\endgroup#1\@@endlink}%
\providecommand \@sanitize@url [0]{\catcode `\\12\catcode `\$12\catcode
  `\&12\catcode `\#12\catcode `\^12\catcode `\_12\catcode `\%12\relax}%
\providecommand \@@startlink[1]{}%
\providecommand \@@endlink[0]{}%
\providecommand \url  [0]{\begingroup\@sanitize@url \@url }%
\providecommand \@url [1]{\endgroup\@href {#1}{\urlprefix }}%
\providecommand \urlprefix  [0]{URL }%
\providecommand \Eprint [0]{\href }%
\providecommand \doibase [0]{http://dx.doi.org/}%
\providecommand \selectlanguage [0]{\@gobble}%
\providecommand \bibinfo  [0]{\@secondoftwo}%
\providecommand \bibfield  [0]{\@secondoftwo}%
\providecommand \translation [1]{[#1]}%
\providecommand \BibitemOpen [0]{}%
\providecommand \bibitemStop [0]{}%
\providecommand \bibitemNoStop [0]{.\EOS\space}%
\providecommand \EOS [0]{\spacefactor3000\relax}%
\providecommand \BibitemShut  [1]{\csname bibitem#1\endcsname}%
\let\auto@bib@innerbib\@empty
\bibitem [{\citenamefont {Frisk~Kockum}\ \emph {et~al.}(2019)\citenamefont
  {Frisk~Kockum}, \citenamefont {Miranowicz}, \citenamefont {De~Liberato},
  \citenamefont {Savasta},\ and\ \citenamefont {Nori}}]{Nori19_19}%
  \BibitemOpen
  \bibfield  {author} {\bibinfo {author} {\bibfnamefont {A.}~\bibnamefont
  {Frisk~Kockum}}, \bibinfo {author} {\bibfnamefont {A.}~\bibnamefont
  {Miranowicz}}, \bibinfo {author} {\bibfnamefont {S.}~\bibnamefont
  {De~Liberato}}, \bibinfo {author} {\bibfnamefont {S.}~\bibnamefont
  {Savasta}}, \ and\ \bibinfo {author} {\bibfnamefont {F.}~\bibnamefont
  {Nori}},\ }\href {\doibase 10.1038/s42254-018-0006-2} {\bibfield  {journal}
  {\bibinfo  {journal} {Nat. Rev. Phys.}\ }\textbf {\bibinfo {volume} {1}},\
  \bibinfo {pages} {19} (\bibinfo {year} {2019})}\BibitemShut {NoStop}%
\bibitem [{\citenamefont {Flick}, \citenamefont {Rivera},\ and\ \citenamefont
  {Narang}(2018)}]{Narang18_1479}%
  \BibitemOpen
  \bibfield  {author} {\bibinfo {author} {\bibfnamefont {J.}~\bibnamefont
  {Flick}}, \bibinfo {author} {\bibfnamefont {N.}~\bibnamefont {Rivera}}, \
  and\ \bibinfo {author} {\bibfnamefont {P.}~\bibnamefont {Narang}},\ }\href
  {\doibase https://doi.org/10.1515/nanoph-2018-0067} {\bibfield  {journal}
  {\bibinfo  {journal} {Nanophotonics}\ }\textbf {\bibinfo {volume} {7}},\
  \bibinfo {pages} {1479 } (\bibinfo {year} {2018})}\BibitemShut {NoStop}%
\bibitem [{\citenamefont {Törmä}\ and\ \citenamefont
  {Barnes}(2014)}]{Barnes14_013901}%
  \BibitemOpen
  \bibfield  {author} {\bibinfo {author} {\bibfnamefont {P.}~\bibnamefont
  {Törmä}}\ and\ \bibinfo {author} {\bibfnamefont {W.~L.}\ \bibnamefont
  {Barnes}},\ }\href {\doibase 10.1088/0034-4885/78/1/013901} {\bibfield
  {journal} {\bibinfo  {journal} {Reports on Progress in Physics}\ }\textbf
  {\bibinfo {volume} {78}},\ \bibinfo {pages} {013901} (\bibinfo {year}
  {2014})}\BibitemShut {NoStop}%
\bibitem [{\citenamefont {Lidzey}\ \emph {et~al.}(1998)\citenamefont {Lidzey},
  \citenamefont {Bradley}, \citenamefont {Skolnick}, \citenamefont {Virgili},
  \citenamefont {Walker},\ and\ \citenamefont {Whittaker}}]{Whittaker98_6697}%
  \BibitemOpen
  \bibfield  {author} {\bibinfo {author} {\bibfnamefont {D.~G.}\ \bibnamefont
  {Lidzey}}, \bibinfo {author} {\bibfnamefont {D.~D.~C.}\ \bibnamefont
  {Bradley}}, \bibinfo {author} {\bibfnamefont {M.~S.}\ \bibnamefont
  {Skolnick}}, \bibinfo {author} {\bibfnamefont {T.}~\bibnamefont {Virgili}},
  \bibinfo {author} {\bibfnamefont {S.}~\bibnamefont {Walker}}, \ and\ \bibinfo
  {author} {\bibfnamefont {D.~M.}\ \bibnamefont {Whittaker}},\ }\href {\doibase
  10.1038/25692} {\bibfield  {journal} {\bibinfo  {journal} {Nature}\ }\textbf
  {\bibinfo {volume} {395}},\ \bibinfo {pages} {53} (\bibinfo {year}
  {1998})}\BibitemShut {NoStop}%
\bibitem [{\citenamefont {Bellessa}\ \emph {et~al.}(2004)\citenamefont
  {Bellessa}, \citenamefont {Bonnand}, \citenamefont {Plenet},\ and\
  \citenamefont {Mugnier}}]{Mugnier04_036404}%
  \BibitemOpen
  \bibfield  {author} {\bibinfo {author} {\bibfnamefont {J.}~\bibnamefont
  {Bellessa}}, \bibinfo {author} {\bibfnamefont {C.}~\bibnamefont {Bonnand}},
  \bibinfo {author} {\bibfnamefont {J.~C.}\ \bibnamefont {Plenet}}, \ and\
  \bibinfo {author} {\bibfnamefont {J.}~\bibnamefont {Mugnier}},\ }\href
  {\doibase 10.1103/PhysRevLett.93.036404} {\bibfield  {journal} {\bibinfo
  {journal} {Phys. Rev. Lett.}\ }\textbf {\bibinfo {volume} {93}},\ \bibinfo
  {pages} {036404} (\bibinfo {year} {2004})}\BibitemShut {NoStop}%
\bibitem [{\citenamefont {Hutchison}\ \emph {et~al.}(2012)\citenamefont
  {Hutchison}, \citenamefont {Schwartz}, \citenamefont {Genet}, \citenamefont
  {Devaux},\ and\ \citenamefont {Ebbesen}}]{Ebbesen12_1592}%
  \BibitemOpen
  \bibfield  {author} {\bibinfo {author} {\bibfnamefont {J.~A.}\ \bibnamefont
  {Hutchison}}, \bibinfo {author} {\bibfnamefont {T.}~\bibnamefont {Schwartz}},
  \bibinfo {author} {\bibfnamefont {C.}~\bibnamefont {Genet}}, \bibinfo
  {author} {\bibfnamefont {E.}~\bibnamefont {Devaux}}, \ and\ \bibinfo {author}
  {\bibfnamefont {T.~W.}\ \bibnamefont {Ebbesen}},\ }\href {\doibase
  10.1002/anie.201107033} {\bibfield  {journal} {\bibinfo  {journal} {Angew.
  Chem., Int. Ed.}\ }\textbf {\bibinfo {volume} {51}},\ \bibinfo {pages} {1592}
  (\bibinfo {year} {2012})}\BibitemShut {NoStop}%
\bibitem [{\citenamefont {Coles}\ \emph {et~al.}(2014)\citenamefont {Coles},
  \citenamefont {Yang}, \citenamefont {Wang}, \citenamefont {Grant},
  \citenamefont {Taylor}, \citenamefont {Saikin}, \citenamefont {Aspuru-Guzik},
  \citenamefont {Lidzey}, \citenamefont {Tang},\ and\ \citenamefont
  {Smith}}]{Smith14_5561}%
  \BibitemOpen
  \bibfield  {author} {\bibinfo {author} {\bibfnamefont {D.~M.}\ \bibnamefont
  {Coles}}, \bibinfo {author} {\bibfnamefont {Y.}~\bibnamefont {Yang}},
  \bibinfo {author} {\bibfnamefont {Y.}~\bibnamefont {Wang}}, \bibinfo {author}
  {\bibfnamefont {R.~T.}\ \bibnamefont {Grant}}, \bibinfo {author}
  {\bibfnamefont {R.~A.}\ \bibnamefont {Taylor}}, \bibinfo {author}
  {\bibfnamefont {S.~K.}\ \bibnamefont {Saikin}}, \bibinfo {author}
  {\bibfnamefont {A.}~\bibnamefont {Aspuru-Guzik}}, \bibinfo {author}
  {\bibfnamefont {D.~G.}\ \bibnamefont {Lidzey}}, \bibinfo {author}
  {\bibfnamefont {J.~K.-H.}\ \bibnamefont {Tang}}, \ and\ \bibinfo {author}
  {\bibfnamefont {J.~M.}\ \bibnamefont {Smith}},\ }\href {\doibase
  10.1038/ncomms6561} {\bibfield  {journal} {\bibinfo  {journal} {Nat.
  Commun.}\ }\textbf {\bibinfo {volume} {5}},\ \bibinfo {pages} {5561}
  (\bibinfo {year} {2014})}\BibitemShut {NoStop}%
\bibitem [{\citenamefont {Orgiu}\ \emph {et~al.}(2015)\citenamefont {Orgiu},
  \citenamefont {George}, \citenamefont {Hutchison}, \citenamefont {Devaux},
  \citenamefont {Dayen}, \citenamefont {Doudin}, \citenamefont {Stellacci},
  \citenamefont {Genet}, \citenamefont {Schachenmayer}, \citenamefont {Genes},
  \citenamefont {Pupillo}, \citenamefont {Samor{\`i}},\ and\ \citenamefont
  {Ebbesen}}]{Ebbesen15_1123}%
  \BibitemOpen
  \bibfield  {author} {\bibinfo {author} {\bibfnamefont {E.}~\bibnamefont
  {Orgiu}}, \bibinfo {author} {\bibfnamefont {J.}~\bibnamefont {George}},
  \bibinfo {author} {\bibfnamefont {J.~A.}\ \bibnamefont {Hutchison}}, \bibinfo
  {author} {\bibfnamefont {E.}~\bibnamefont {Devaux}}, \bibinfo {author}
  {\bibfnamefont {J.~F.}\ \bibnamefont {Dayen}}, \bibinfo {author}
  {\bibfnamefont {B.}~\bibnamefont {Doudin}}, \bibinfo {author} {\bibfnamefont
  {F.}~\bibnamefont {Stellacci}}, \bibinfo {author} {\bibfnamefont
  {C.}~\bibnamefont {Genet}}, \bibinfo {author} {\bibfnamefont
  {J.}~\bibnamefont {Schachenmayer}}, \bibinfo {author} {\bibfnamefont
  {C.}~\bibnamefont {Genes}}, \bibinfo {author} {\bibfnamefont
  {G.}~\bibnamefont {Pupillo}}, \bibinfo {author} {\bibfnamefont
  {P.}~\bibnamefont {Samor{\`i}}}, \ and\ \bibinfo {author} {\bibfnamefont
  {T.~W.}\ \bibnamefont {Ebbesen}},\ }\href {\doibase 10.1038/nmat4392}
  {\bibfield  {journal} {\bibinfo  {journal} {Nat. Mater.}\ }\textbf {\bibinfo
  {volume} {14}},\ \bibinfo {pages} {1123} (\bibinfo {year}
  {2015})}\BibitemShut {NoStop}%
\bibitem [{\citenamefont {Chikkaraddy}\ \emph {et~al.}(2016)\citenamefont
  {Chikkaraddy}, \citenamefont {de~Nijs}, \citenamefont {Benz}, \citenamefont
  {Barrow}, \citenamefont {Scherman}, \citenamefont {Rosta}, \citenamefont
  {Demetriadou}, \citenamefont {Fox}, \citenamefont {Hess},\ and\ \citenamefont
  {Baumberg}}]{Baumberg16_127}%
  \BibitemOpen
  \bibfield  {author} {\bibinfo {author} {\bibfnamefont {R.}~\bibnamefont
  {Chikkaraddy}}, \bibinfo {author} {\bibfnamefont {B.}~\bibnamefont
  {de~Nijs}}, \bibinfo {author} {\bibfnamefont {F.}~\bibnamefont {Benz}},
  \bibinfo {author} {\bibfnamefont {S.~J.}\ \bibnamefont {Barrow}}, \bibinfo
  {author} {\bibfnamefont {O.~A.}\ \bibnamefont {Scherman}}, \bibinfo {author}
  {\bibfnamefont {E.}~\bibnamefont {Rosta}}, \bibinfo {author} {\bibfnamefont
  {A.}~\bibnamefont {Demetriadou}}, \bibinfo {author} {\bibfnamefont
  {P.}~\bibnamefont {Fox}}, \bibinfo {author} {\bibfnamefont {O.}~\bibnamefont
  {Hess}}, \ and\ \bibinfo {author} {\bibfnamefont {J.~J.}\ \bibnamefont
  {Baumberg}},\ }\href {\doibase 10.1038/nature17974} {\bibfield  {journal}
  {\bibinfo  {journal} {Nature}\ }\textbf {\bibinfo {volume} {535}},\ \bibinfo
  {pages} {127} (\bibinfo {year} {2016})}\BibitemShut {NoStop}%
\bibitem [{\citenamefont {Ebbesen}(2016)}]{Ebbesen16_2403}%
  \BibitemOpen
  \bibfield  {author} {\bibinfo {author} {\bibfnamefont {T.~W.}\ \bibnamefont
  {Ebbesen}},\ }\href {\doibase 10.1021/acs.accounts.6b00295} {\bibfield
  {journal} {\bibinfo  {journal} {Acc. Chem. Res.}\ }\textbf {\bibinfo {volume}
  {49}},\ \bibinfo {pages} {2403} (\bibinfo {year} {2016})}\BibitemShut
  {NoStop}%
\bibitem [{\citenamefont {Sukharev}\ and\ \citenamefont
  {Nitzan}(2017)}]{Nitzan17_443003}%
  \BibitemOpen
  \bibfield  {author} {\bibinfo {author} {\bibfnamefont {M.}~\bibnamefont
  {Sukharev}}\ and\ \bibinfo {author} {\bibfnamefont {A.}~\bibnamefont
  {Nitzan}},\ }\href {\doibase 10.1088/1361-648x/aa85ef} {\bibfield  {journal}
  {\bibinfo  {journal} {Journal of Physics: Condensed Matter}\ }\textbf
  {\bibinfo {volume} {29}},\ \bibinfo {pages} {443003} (\bibinfo {year}
  {2017})}\BibitemShut {NoStop}%
\bibitem [{\citenamefont {Zhong}\ \emph {et~al.}(2017)\citenamefont {Zhong},
  \citenamefont {Chervy}, \citenamefont {Zhang}, \citenamefont {Thomas},
  \citenamefont {George}, \citenamefont {Genet}, \citenamefont {Hutchison},\
  and\ \citenamefont {Ebbesen}}]{Ebbesen17_9034}%
  \BibitemOpen
  \bibfield  {author} {\bibinfo {author} {\bibfnamefont {X.}~\bibnamefont
  {Zhong}}, \bibinfo {author} {\bibfnamefont {T.}~\bibnamefont {Chervy}},
  \bibinfo {author} {\bibfnamefont {L.}~\bibnamefont {Zhang}}, \bibinfo
  {author} {\bibfnamefont {A.}~\bibnamefont {Thomas}}, \bibinfo {author}
  {\bibfnamefont {J.}~\bibnamefont {George}}, \bibinfo {author} {\bibfnamefont
  {C.}~\bibnamefont {Genet}}, \bibinfo {author} {\bibfnamefont {J.~A.}\
  \bibnamefont {Hutchison}}, \ and\ \bibinfo {author} {\bibfnamefont {T.~W.}\
  \bibnamefont {Ebbesen}},\ }\href {\doibase 10.1002/anie.201703539} {\bibfield
   {journal} {\bibinfo  {journal} {Angew. Chem., Int. Ed.}\ }\textbf {\bibinfo
  {volume} {56}},\ \bibinfo {pages} {9034} (\bibinfo {year}
  {2017})}\BibitemShut {NoStop}%
\bibitem [{\citenamefont {Chevrier}\ \emph {et~al.}(2019)\citenamefont
  {Chevrier}, \citenamefont {Benoit}, \citenamefont {Symonds}, \citenamefont
  {Saikin}, \citenamefont {Yuen-Zhou},\ and\ \citenamefont
  {Bellessa}}]{Bellessa19_173902}%
  \BibitemOpen
  \bibfield  {author} {\bibinfo {author} {\bibfnamefont {K.}~\bibnamefont
  {Chevrier}}, \bibinfo {author} {\bibfnamefont {J.~M.}\ \bibnamefont
  {Benoit}}, \bibinfo {author} {\bibfnamefont {C.}~\bibnamefont {Symonds}},
  \bibinfo {author} {\bibfnamefont {S.~K.}\ \bibnamefont {Saikin}}, \bibinfo
  {author} {\bibfnamefont {J.}~\bibnamefont {Yuen-Zhou}}, \ and\ \bibinfo
  {author} {\bibfnamefont {J.}~\bibnamefont {Bellessa}},\ }\href {\doibase
  10.1103/PhysRevLett.122.173902} {\bibfield  {journal} {\bibinfo  {journal}
  {Phys. Rev. Lett.}\ }\textbf {\bibinfo {volume} {122}},\ \bibinfo {pages}
  {173902} (\bibinfo {year} {2019})}\BibitemShut {NoStop}%
\bibitem [{\citenamefont {K{\'e}na-Cohen}\ and\ \citenamefont
  {Forrest}(2010)}]{Forrest20_371}%
  \BibitemOpen
  \bibfield  {author} {\bibinfo {author} {\bibfnamefont {S.}~\bibnamefont
  {K{\'e}na-Cohen}}\ and\ \bibinfo {author} {\bibfnamefont {S.~R.}\
  \bibnamefont {Forrest}},\ }\href {\doibase 10.1038/nphoton.2010.86}
  {\bibfield  {journal} {\bibinfo  {journal} {Nat. Photon.}\ }\textbf {\bibinfo
  {volume} {4}},\ \bibinfo {pages} {371} (\bibinfo {year} {2010})}\BibitemShut
  {NoStop}%
\bibitem [{\citenamefont {Wright}, \citenamefont {Nelson},\ and\ \citenamefont
  {Weichman}(2023)}]{Weichman23_JACS}%
  \BibitemOpen
  \bibfield  {author} {\bibinfo {author} {\bibfnamefont {A.~D.}\ \bibnamefont
  {Wright}}, \bibinfo {author} {\bibfnamefont {J.~C.}\ \bibnamefont {Nelson}},
  \ and\ \bibinfo {author} {\bibfnamefont {M.~L.}\ \bibnamefont {Weichman}},\
  }\href {\doibase 10.1021/jacs.3c00126} {\bibfield  {journal} {\bibinfo
  {journal} {J. Am. Chem. Soc.}\ }\textbf {\bibinfo {volume} {145}},\ \bibinfo
  {pages} {5982–5987} (\bibinfo {year} {2023})}\BibitemShut {NoStop}%
\bibitem [{\citenamefont {Lather}\ \emph {et~al.}(2019)\citenamefont {Lather},
  \citenamefont {Bhatt}, \citenamefont {Thomas}, \citenamefont {Ebbesen},\ and\
  \citenamefont {George}}]{George19_10635}%
  \BibitemOpen
  \bibfield  {author} {\bibinfo {author} {\bibfnamefont {J.}~\bibnamefont
  {Lather}}, \bibinfo {author} {\bibfnamefont {P.}~\bibnamefont {Bhatt}},
  \bibinfo {author} {\bibfnamefont {A.}~\bibnamefont {Thomas}}, \bibinfo
  {author} {\bibfnamefont {T.~W.}\ \bibnamefont {Ebbesen}}, \ and\ \bibinfo
  {author} {\bibfnamefont {J.}~\bibnamefont {George}},\ }\href {\doibase
  10.1002/anie.201905407} {\bibfield  {journal} {\bibinfo  {journal} {Angew.
  Chem., Int. Ed.}\ }\textbf {\bibinfo {volume} {58}},\ \bibinfo {pages}
  {10635} (\bibinfo {year} {2019})}\BibitemShut {NoStop}%
\bibitem [{\citenamefont {Climent}\ \emph {et~al.}(2019)\citenamefont
  {Climent}, \citenamefont {Galego}, \citenamefont {Garcia-Vidal},\ and\
  \citenamefont {Feist}}]{Feist19_8698}%
  \BibitemOpen
  \bibfield  {author} {\bibinfo {author} {\bibfnamefont {C.}~\bibnamefont
  {Climent}}, \bibinfo {author} {\bibfnamefont {J.}~\bibnamefont {Galego}},
  \bibinfo {author} {\bibfnamefont {F.~J.}\ \bibnamefont {Garcia-Vidal}}, \
  and\ \bibinfo {author} {\bibfnamefont {J.}~\bibnamefont {Feist}},\ }\href
  {\doibase 10.1002/anie.201901926} {\bibfield  {journal} {\bibinfo  {journal}
  {Angew. Chem., Int. Ed.}\ }\textbf {\bibinfo {volume} {58}},\ \bibinfo
  {pages} {8698} (\bibinfo {year} {2019})}\BibitemShut {NoStop}%
\bibitem [{\citenamefont {Dunkelberger}\ \emph {et~al.}(2022)\citenamefont
  {Dunkelberger}, \citenamefont {Simpkins}, \citenamefont {Vurgaftman},\ and\
  \citenamefont {Owrutsky}}]{Simpkins_ARPC_2023}%
  \BibitemOpen
  \bibfield  {author} {\bibinfo {author} {\bibfnamefont {A.~D.}\ \bibnamefont
  {Dunkelberger}}, \bibinfo {author} {\bibfnamefont {B.~S.}\ \bibnamefont
  {Simpkins}}, \bibinfo {author} {\bibfnamefont {I.}~\bibnamefont
  {Vurgaftman}}, \ and\ \bibinfo {author} {\bibfnamefont {J.~C.}\ \bibnamefont
  {Owrutsky}},\ }\href {\doibase 10.1146/annurev-physchem-082620-014627}
  {\bibfield  {journal} {\bibinfo  {journal} {ARPC}\ }\textbf {\bibinfo
  {volume} {73}},\ \bibinfo {pages} {429} (\bibinfo {year} {2022})},\ \bibinfo
  {note} {pMID: 35081324},\ \Eprint
  {http://arxiv.org/abs/https://doi.org/10.1146/annurev-physchem-082620-014627}
  {https://doi.org/10.1146/annurev-physchem-082620-014627} \BibitemShut
  {NoStop}%
\bibitem [{\citenamefont {Cheng}\ \emph {et~al.}(2022)\citenamefont {Cheng},
  \citenamefont {Krainova}, \citenamefont {Brigeman}, \citenamefont {Khanna},
  \citenamefont {Shedge}, \citenamefont {Isborn}, \citenamefont {Yuen-Zhou},\
  and\ \citenamefont {Giebink}}]{CG_NatureCommun_2022}%
  \BibitemOpen
  \bibfield  {author} {\bibinfo {author} {\bibfnamefont {C.-Y.}\ \bibnamefont
  {Cheng}}, \bibinfo {author} {\bibfnamefont {N.}~\bibnamefont {Krainova}},
  \bibinfo {author} {\bibfnamefont {A.~N.}\ \bibnamefont {Brigeman}}, \bibinfo
  {author} {\bibfnamefont {A.}~\bibnamefont {Khanna}}, \bibinfo {author}
  {\bibfnamefont {S.}~\bibnamefont {Shedge}}, \bibinfo {author} {\bibfnamefont
  {C.}~\bibnamefont {Isborn}}, \bibinfo {author} {\bibfnamefont
  {J.}~\bibnamefont {Yuen-Zhou}}, \ and\ \bibinfo {author} {\bibfnamefont
  {N.~C.}\ \bibnamefont {Giebink}},\ }\href {\doibase
  10.1038/s41467-022-35589-4} {\bibfield  {journal} {\bibinfo  {journal} {Nat.
  Commun.}\ }\textbf {\bibinfo {volume} {13}},\ \bibinfo {pages} {7937}
  (\bibinfo {year} {2022})}\BibitemShut {NoStop}%
\bibitem [{\citenamefont {Yadav}\ \emph {et~al.}(2020)\citenamefont {Yadav},
  \citenamefont {Otten}, \citenamefont {Wang}, \citenamefont {Cortes},
  \citenamefont {Gosztola}, \citenamefont {Wiederrecht}, \citenamefont {Gray},
  \citenamefont {Odom},\ and\ \citenamefont {Basu}}]{Basu20_5043}%
  \BibitemOpen
  \bibfield  {author} {\bibinfo {author} {\bibfnamefont {R.~K.}\ \bibnamefont
  {Yadav}}, \bibinfo {author} {\bibfnamefont {M.}~\bibnamefont {Otten}},
  \bibinfo {author} {\bibfnamefont {W.}~\bibnamefont {Wang}}, \bibinfo {author}
  {\bibfnamefont {C.~L.}\ \bibnamefont {Cortes}}, \bibinfo {author}
  {\bibfnamefont {D.~J.}\ \bibnamefont {Gosztola}}, \bibinfo {author}
  {\bibfnamefont {G.~P.}\ \bibnamefont {Wiederrecht}}, \bibinfo {author}
  {\bibfnamefont {S.~K.}\ \bibnamefont {Gray}}, \bibinfo {author}
  {\bibfnamefont {T.~W.}\ \bibnamefont {Odom}}, \ and\ \bibinfo {author}
  {\bibfnamefont {J.~K.}\ \bibnamefont {Basu}},\ }\href {\doibase
  10.1021/acs.nanolett.0c01236} {\bibfield  {journal} {\bibinfo  {journal}
  {Nano Lett.}\ }\textbf {\bibinfo {volume} {20}},\ \bibinfo {pages} {5043}
  (\bibinfo {year} {2020})}\BibitemShut {NoStop}%
\bibitem [{\citenamefont {Pandya}\ \emph {et~al.}(2022)\citenamefont {Pandya},
  \citenamefont {Ashoka}, \citenamefont {Georgiou}, \citenamefont {Sung},
  \citenamefont {Jayaprakash}, \citenamefont {Renken}, \citenamefont {Gai},
  \citenamefont {Shen}, \citenamefont {Rao},\ and\ \citenamefont
  {Musser}}]{Musser_AS_2022}%
  \BibitemOpen
  \bibfield  {author} {\bibinfo {author} {\bibfnamefont {R.}~\bibnamefont
  {Pandya}}, \bibinfo {author} {\bibfnamefont {A.}~\bibnamefont {Ashoka}},
  \bibinfo {author} {\bibfnamefont {K.}~\bibnamefont {Georgiou}}, \bibinfo
  {author} {\bibfnamefont {J.}~\bibnamefont {Sung}}, \bibinfo {author}
  {\bibfnamefont {R.}~\bibnamefont {Jayaprakash}}, \bibinfo {author}
  {\bibfnamefont {S.}~\bibnamefont {Renken}}, \bibinfo {author} {\bibfnamefont
  {L.}~\bibnamefont {Gai}}, \bibinfo {author} {\bibfnamefont {Z.}~\bibnamefont
  {Shen}}, \bibinfo {author} {\bibfnamefont {A.}~\bibnamefont {Rao}}, \ and\
  \bibinfo {author} {\bibfnamefont {A.~J.}\ \bibnamefont {Musser}},\ }\href
  {\doibase https://doi.org/10.1002/advs.202105569} {\bibfield  {journal}
  {\bibinfo  {journal} {Advanced Science}\ }\textbf {\bibinfo {volume} {9}},\
  \bibinfo {pages} {2105569} (\bibinfo {year} {2022})},\ \Eprint
  {http://arxiv.org/abs/https://onlinelibrary.wiley.com/doi/pdf/10.1002/advs.202105569}
  {https://onlinelibrary.wiley.com/doi/pdf/10.1002/advs.202105569} \BibitemShut
  {NoStop}%
\bibitem [{\citenamefont {Munkhbat}\ \emph {et~al.}(2018)\citenamefont
  {Munkhbat}, \citenamefont {Wers{\"a}ll}, \citenamefont {Baranov},
  \citenamefont {Antosiewicz},\ and\ \citenamefont
  {Shegai}}]{Shegai18_eaas9552}%
  \BibitemOpen
  \bibfield  {author} {\bibinfo {author} {\bibfnamefont {B.}~\bibnamefont
  {Munkhbat}}, \bibinfo {author} {\bibfnamefont {M.}~\bibnamefont
  {Wers{\"a}ll}}, \bibinfo {author} {\bibfnamefont {D.~G.}\ \bibnamefont
  {Baranov}}, \bibinfo {author} {\bibfnamefont {T.~J.}\ \bibnamefont
  {Antosiewicz}}, \ and\ \bibinfo {author} {\bibfnamefont {T.}~\bibnamefont
  {Shegai}},\ }\href {\doibase 10.1126/sciadv.aas9552} {\bibfield  {journal}
  {\bibinfo  {journal} {Science Advances}\ }\textbf {\bibinfo {volume} {4}},\
  \bibinfo {pages} {eaas9552} (\bibinfo {year} {2018})}\BibitemShut {NoStop}%
\bibitem [{\citenamefont {Brawley}\ \emph {et~al.}(2021)\citenamefont
  {Brawley}, \citenamefont {Storm}, \citenamefont {Mora}, \citenamefont
  {Pelton},\ and\ \citenamefont {Sheldon}}]{Sheldon_JCP_2021}%
  \BibitemOpen
  \bibfield  {author} {\bibinfo {author} {\bibfnamefont {Z.~T.}\ \bibnamefont
  {Brawley}}, \bibinfo {author} {\bibfnamefont {S.~D.}\ \bibnamefont {Storm}},
  \bibinfo {author} {\bibfnamefont {D.~A.~C.}\ \bibnamefont {Mora}}, \bibinfo
  {author} {\bibfnamefont {M.}~\bibnamefont {Pelton}}, \ and\ \bibinfo {author}
  {\bibfnamefont {M.}~\bibnamefont {Sheldon}},\ }\href@noop {} {\bibfield
  {journal} {\bibinfo  {journal} {J. Chem. Phys.}\ }\textbf {\bibinfo {volume}
  {154}},\ \bibinfo {pages} {104305} (\bibinfo {year} {2021})}\BibitemShut
  {NoStop}%
\bibitem [{\citenamefont {Ishii}\ \emph {et~al.}(2021)\citenamefont {Ishii},
  \citenamefont {Bencheikh}, \citenamefont {Forget}, \citenamefont {Chénais},
  \citenamefont {Heinrich}, \citenamefont {Kreher}, \citenamefont
  {Sosa~Vargas}, \citenamefont {Miyata}, \citenamefont {Onda}, \citenamefont
  {Fujihara}, \citenamefont {Kéna-Cohen}, \citenamefont {Mathevet},\ and\
  \citenamefont {Adachi}}]{SKC_AOM_2021}%
  \BibitemOpen
  \bibfield  {author} {\bibinfo {author} {\bibfnamefont {T.}~\bibnamefont
  {Ishii}}, \bibinfo {author} {\bibfnamefont {F.}~\bibnamefont {Bencheikh}},
  \bibinfo {author} {\bibfnamefont {S.}~\bibnamefont {Forget}}, \bibinfo
  {author} {\bibfnamefont {S.}~\bibnamefont {Chénais}}, \bibinfo {author}
  {\bibfnamefont {B.}~\bibnamefont {Heinrich}}, \bibinfo {author}
  {\bibfnamefont {D.}~\bibnamefont {Kreher}}, \bibinfo {author} {\bibfnamefont
  {L.}~\bibnamefont {Sosa~Vargas}}, \bibinfo {author} {\bibfnamefont
  {K.}~\bibnamefont {Miyata}}, \bibinfo {author} {\bibfnamefont
  {K.}~\bibnamefont {Onda}}, \bibinfo {author} {\bibfnamefont {T.}~\bibnamefont
  {Fujihara}}, \bibinfo {author} {\bibfnamefont {S.}~\bibnamefont
  {Kéna-Cohen}}, \bibinfo {author} {\bibfnamefont {F.}~\bibnamefont
  {Mathevet}}, \ and\ \bibinfo {author} {\bibfnamefont {C.}~\bibnamefont
  {Adachi}},\ }\href {\doibase https://doi.org/10.1002/adom.202101048}
  {\bibfield  {journal} {\bibinfo  {journal} {Advanced Optical Materials}\
  }\textbf {\bibinfo {volume} {9}},\ \bibinfo {pages} {2101048} (\bibinfo
  {year} {2021})},\ \Eprint
  {http://arxiv.org/abs/https://onlinelibrary.wiley.com/doi/pdf/10.1002/adom.202101048}
  {https://onlinelibrary.wiley.com/doi/pdf/10.1002/adom.202101048} \BibitemShut
  {NoStop}%
\bibitem [{\citenamefont {Hu}\ \emph {et~al.}(2022)\citenamefont {Hu},
  \citenamefont {Gustin}, \citenamefont {Krauss},\ and\ \citenamefont
  {Franco}}]{Franco_JPCL_2022}%
  \BibitemOpen
  \bibfield  {author} {\bibinfo {author} {\bibfnamefont {W.}~\bibnamefont
  {Hu}}, \bibinfo {author} {\bibfnamefont {I.}~\bibnamefont {Gustin}}, \bibinfo
  {author} {\bibfnamefont {T.~D.}\ \bibnamefont {Krauss}}, \ and\ \bibinfo
  {author} {\bibfnamefont {I.}~\bibnamefont {Franco}},\ }\href {\doibase
  10.1021/acs.jpclett.2c02877} {\bibfield  {journal} {\bibinfo  {journal} {J.
  Phys. Chem. Lett.}\ }\textbf {\bibinfo {volume} {13}},\ \bibinfo {pages}
  {11503} (\bibinfo {year} {2022})},\ \bibinfo {note} {pMID: 36469838},\
  \Eprint {http://arxiv.org/abs/https://doi.org/10.1021/acs.jpclett.2c02877}
  {https://doi.org/10.1021/acs.jpclett.2c02877} \BibitemShut {NoStop}%
\bibitem [{\citenamefont {Mandal}\ \emph {et~al.}(2022)\citenamefont {Mandal},
  \citenamefont {Taylor}, \citenamefont {Weight}, \citenamefont {Koessler},
  \citenamefont {Li},\ and\ \citenamefont {Huo}}]{Huo22_Chemrxiv}%
  \BibitemOpen
  \bibfield  {author} {\bibinfo {author} {\bibfnamefont {A.}~\bibnamefont
  {Mandal}}, \bibinfo {author} {\bibfnamefont {M.}~\bibnamefont {Taylor}},
  \bibinfo {author} {\bibfnamefont {B.}~\bibnamefont {Weight}}, \bibinfo
  {author} {\bibfnamefont {E.}~\bibnamefont {Koessler}}, \bibinfo {author}
  {\bibfnamefont {X.}~\bibnamefont {Li}}, \ and\ \bibinfo {author}
  {\bibfnamefont {P.}~\bibnamefont {Huo}},\ }\href {\doibase
  10.26434/chemrxiv-2022-g9lr7} {\bibfield  {journal} {\bibinfo  {journal}
  {ChemRxiv}\ } (\bibinfo {year} {2022}),\
  10.26434/chemrxiv-2022-g9lr7}\BibitemShut {NoStop}%
\bibitem [{\citenamefont {Fregoni}, \citenamefont {Garcia-Vidal},\ and\
  \citenamefont {Feist}(2022)}]{Feist22_ACSPhoton}%
  \BibitemOpen
  \bibfield  {author} {\bibinfo {author} {\bibfnamefont {J.}~\bibnamefont
  {Fregoni}}, \bibinfo {author} {\bibfnamefont {F.~J.}\ \bibnamefont
  {Garcia-Vidal}}, \ and\ \bibinfo {author} {\bibfnamefont {J.}~\bibnamefont
  {Feist}},\ }\href {\doibase 10.1021/acsphotonics.1c01749} {\bibfield
  {journal} {\bibinfo  {journal} {ACS Photonics}\ }\textbf {\bibinfo {volume}
  {9}},\ \bibinfo {pages} {1096} (\bibinfo {year} {2022})}\BibitemShut
  {NoStop}%
\bibitem [{\citenamefont {Ruggentharler}, \citenamefont {Sidler},\ and\
  \citenamefont {Rubio}(2022)}]{Rubio22_Arxiv}%
  \BibitemOpen
  \bibfield  {author} {\bibinfo {author} {\bibfnamefont {M.}~\bibnamefont
  {Ruggentharler}}, \bibinfo {author} {\bibfnamefont {D.}~\bibnamefont
  {Sidler}}, \ and\ \bibinfo {author} {\bibfnamefont {A.}~\bibnamefont
  {Rubio}},\ }\href {\doibase 10.48550/arXiv.2211.04241} {\bibfield  {journal}
  {\bibinfo  {journal} {arXiv}\ } (\bibinfo {year} {2022}),\
  10.48550/arXiv.2211.04241}\BibitemShut {NoStop}%
\bibitem [{\citenamefont {Weight}, \citenamefont {Krauss},\ and\ \citenamefont
  {Huo}(2023)}]{Huo23_Chemrxiv}%
  \BibitemOpen
  \bibfield  {author} {\bibinfo {author} {\bibfnamefont {B.}~\bibnamefont
  {Weight}}, \bibinfo {author} {\bibfnamefont {T.}~\bibnamefont {Krauss}}, \
  and\ \bibinfo {author} {\bibfnamefont {P.}~\bibnamefont {Huo}},\ }\href
  {\doibase 10.26434/chemrxiv-2023-2hv8l} {\bibfield  {journal} {\bibinfo
  {journal} {ChemRxvi}\ } (\bibinfo {year} {2023}),\
  10.26434/chemrxiv-2023-2hv8l}\BibitemShut {NoStop}%
\bibitem [{\citenamefont {Hu}\ and\ \citenamefont {Huo}(2023)}]{Huo23_JCTC}%
  \BibitemOpen
  \bibfield  {author} {\bibinfo {author} {\bibfnamefont {D.}~\bibnamefont
  {Hu}}\ and\ \bibinfo {author} {\bibfnamefont {P.}~\bibnamefont {Huo}},\
  }\href {\doibase 10.1021/acs.jctc.3c00137} {\bibfield  {journal} {\bibinfo
  {journal} {J. Chem. Theory Comput.}\ }\textbf {\bibinfo {volume} {19}},\
  \bibinfo {pages} {2353–2368} (\bibinfo {year} {2023})}\BibitemShut
  {NoStop}%
\bibitem [{\citenamefont {Ciuti}\ and\ \citenamefont
  {Carusotto}(2006)}]{Carusotto06_033811}%
  \BibitemOpen
  \bibfield  {author} {\bibinfo {author} {\bibfnamefont {C.}~\bibnamefont
  {Ciuti}}\ and\ \bibinfo {author} {\bibfnamefont {I.}~\bibnamefont
  {Carusotto}},\ }\href {\doibase 10.1103/PhysRevA.74.033811} {\bibfield
  {journal} {\bibinfo  {journal} {Phys. Rev. A}\ }\textbf {\bibinfo {volume}
  {74}},\ \bibinfo {pages} {033811} (\bibinfo {year} {2006})}\BibitemShut
  {NoStop}%
\bibitem [{\citenamefont {Reitz}, \citenamefont {Sommer},\ and\ \citenamefont
  {Genes}(2019)}]{Genes19_203602}%
  \BibitemOpen
  \bibfield  {author} {\bibinfo {author} {\bibfnamefont {M.}~\bibnamefont
  {Reitz}}, \bibinfo {author} {\bibfnamefont {C.}~\bibnamefont {Sommer}}, \
  and\ \bibinfo {author} {\bibfnamefont {C.}~\bibnamefont {Genes}},\ }\href
  {\doibase 10.1103/PhysRevLett.122.203602} {\bibfield  {journal} {\bibinfo
  {journal} {Phys. Rev. Lett.}\ }\textbf {\bibinfo {volume} {122}},\ \bibinfo
  {pages} {203602} (\bibinfo {year} {2019})}\BibitemShut {NoStop}%
\bibitem [{\citenamefont {Shah}\ \emph {et~al.}(2013)\citenamefont {Shah},
  \citenamefont {Scherer}, \citenamefont {Pelton},\ and\ \citenamefont
  {Gray}}]{Gray13_075411}%
  \BibitemOpen
  \bibfield  {author} {\bibinfo {author} {\bibfnamefont {R.~A.}\ \bibnamefont
  {Shah}}, \bibinfo {author} {\bibfnamefont {N.~F.}\ \bibnamefont {Scherer}},
  \bibinfo {author} {\bibfnamefont {M.}~\bibnamefont {Pelton}}, \ and\ \bibinfo
  {author} {\bibfnamefont {S.~K.}\ \bibnamefont {Gray}},\ }\href {\doibase
  10.1103/PhysRevB.88.075411} {\bibfield  {journal} {\bibinfo  {journal} {Phys.
  Rev. B}\ }\textbf {\bibinfo {volume} {88}},\ \bibinfo {pages} {075411}
  (\bibinfo {year} {2013})}\BibitemShut {NoStop}%
\bibitem [{\citenamefont {Mandal}\ and\ \citenamefont
  {Huo}(2019)}]{Huo19_5519}%
  \BibitemOpen
  \bibfield  {author} {\bibinfo {author} {\bibfnamefont {A.}~\bibnamefont
  {Mandal}}\ and\ \bibinfo {author} {\bibfnamefont {P.}~\bibnamefont {Huo}},\
  }\href {\doibase 10.1021/acs.jpclett.9b01599} {\bibfield  {journal} {\bibinfo
   {journal} {J. Phys. Chem. Lett.}\ }\textbf {\bibinfo {volume} {10}},\
  \bibinfo {pages} {5519} (\bibinfo {year} {2019})}\BibitemShut {NoStop}%
\bibitem [{\citenamefont {Mandal}, \citenamefont {Krauss},\ and\ \citenamefont
  {Huo}(2020)}]{Huo20_6321}%
  \BibitemOpen
  \bibfield  {author} {\bibinfo {author} {\bibfnamefont {A.}~\bibnamefont
  {Mandal}}, \bibinfo {author} {\bibfnamefont {T.~D.}\ \bibnamefont {Krauss}},
  \ and\ \bibinfo {author} {\bibfnamefont {P.}~\bibnamefont {Huo}},\ }\href
  {\doibase 10.1021/acs.jpcb.0c03227} {\bibfield  {journal} {\bibinfo
  {journal} {J. Phys. Chem. B}\ }\textbf {\bibinfo {volume} {124}},\ \bibinfo
  {pages} {6321} (\bibinfo {year} {2020})}\BibitemShut {NoStop}%
\bibitem [{\citenamefont {Ruggenthaler}, \citenamefont {Mackenroth},\ and\
  \citenamefont {Bauer}(2011)}]{Bauer11_042107}%
  \BibitemOpen
  \bibfield  {author} {\bibinfo {author} {\bibfnamefont {M.}~\bibnamefont
  {Ruggenthaler}}, \bibinfo {author} {\bibfnamefont {F.}~\bibnamefont
  {Mackenroth}}, \ and\ \bibinfo {author} {\bibfnamefont {D.}~\bibnamefont
  {Bauer}},\ }\href {\doibase 10.1103/PhysRevA.84.042107} {\bibfield  {journal}
  {\bibinfo  {journal} {Phys. Rev. A}\ }\textbf {\bibinfo {volume} {84}},\
  \bibinfo {pages} {042107} (\bibinfo {year} {2011})}\BibitemShut {NoStop}%
\bibitem [{\citenamefont {Tokatly}(2013)}]{Tokatly13_233001}%
  \BibitemOpen
  \bibfield  {author} {\bibinfo {author} {\bibfnamefont {I.~V.}\ \bibnamefont
  {Tokatly}},\ }\href {\doibase 10.1103/PhysRevLett.110.233001} {\bibfield
  {journal} {\bibinfo  {journal} {Phys. Rev. Lett.}\ }\textbf {\bibinfo
  {volume} {110}},\ \bibinfo {pages} {233001} (\bibinfo {year}
  {2013})}\BibitemShut {NoStop}%
\bibitem [{\citenamefont {Ruggenthaler}\ \emph {et~al.}(2014)\citenamefont
  {Ruggenthaler}, \citenamefont {Flick}, \citenamefont {Pellegrini},
  \citenamefont {Appel}, \citenamefont {Tokatly},\ and\ \citenamefont
  {Rubio}}]{Rubio14_012508}%
  \BibitemOpen
  \bibfield  {author} {\bibinfo {author} {\bibfnamefont {M.}~\bibnamefont
  {Ruggenthaler}}, \bibinfo {author} {\bibfnamefont {J.}~\bibnamefont {Flick}},
  \bibinfo {author} {\bibfnamefont {C.}~\bibnamefont {Pellegrini}}, \bibinfo
  {author} {\bibfnamefont {H.}~\bibnamefont {Appel}}, \bibinfo {author}
  {\bibfnamefont {I.~V.}\ \bibnamefont {Tokatly}}, \ and\ \bibinfo {author}
  {\bibfnamefont {A.}~\bibnamefont {Rubio}},\ }\href {\doibase
  10.1103/PhysRevA.90.012508} {\bibfield  {journal} {\bibinfo  {journal} {Phys.
  Rev. A}\ }\textbf {\bibinfo {volume} {90}},\ \bibinfo {pages} {012508}
  (\bibinfo {year} {2014})}\BibitemShut {NoStop}%
\bibitem [{\citenamefont {Pellegrini}\ \emph {et~al.}(2015)\citenamefont
  {Pellegrini}, \citenamefont {Flick}, \citenamefont {Tokatly}, \citenamefont
  {Appel},\ and\ \citenamefont {Rubio}}]{Rubio15_093001}%
  \BibitemOpen
  \bibfield  {author} {\bibinfo {author} {\bibfnamefont {C.}~\bibnamefont
  {Pellegrini}}, \bibinfo {author} {\bibfnamefont {J.}~\bibnamefont {Flick}},
  \bibinfo {author} {\bibfnamefont {I.~V.}\ \bibnamefont {Tokatly}}, \bibinfo
  {author} {\bibfnamefont {H.}~\bibnamefont {Appel}}, \ and\ \bibinfo {author}
  {\bibfnamefont {A.}~\bibnamefont {Rubio}},\ }\href {\doibase
  10.1103/PhysRevLett.115.093001} {\bibfield  {journal} {\bibinfo  {journal}
  {Phys. Rev. Lett.}\ }\textbf {\bibinfo {volume} {115}},\ \bibinfo {pages}
  {093001} (\bibinfo {year} {2015})}\BibitemShut {NoStop}%
\bibitem [{\citenamefont {Flick}\ \emph {et~al.}(2018)\citenamefont {Flick},
  \citenamefont {Schäfer}, \citenamefont {Ruggenthaler}, \citenamefont
  {Appel},\ and\ \citenamefont {Rubio}}]{Rubio18_992}%
  \BibitemOpen
  \bibfield  {author} {\bibinfo {author} {\bibfnamefont {J.}~\bibnamefont
  {Flick}}, \bibinfo {author} {\bibfnamefont {C.}~\bibnamefont {Schäfer}},
  \bibinfo {author} {\bibfnamefont {M.}~\bibnamefont {Ruggenthaler}}, \bibinfo
  {author} {\bibfnamefont {H.}~\bibnamefont {Appel}}, \ and\ \bibinfo {author}
  {\bibfnamefont {A.}~\bibnamefont {Rubio}},\ }\href {\doibase
  10.1021/acsphotonics.7b01279} {\bibfield  {journal} {\bibinfo  {journal} {ACS
  Photonics}\ }\textbf {\bibinfo {volume} {5}},\ \bibinfo {pages} {992}
  (\bibinfo {year} {2018})}\BibitemShut {NoStop}%
\bibitem [{\citenamefont {Jestädt}\ \emph {et~al.}(2019)\citenamefont
  {Jestädt}, \citenamefont {Ruggenthaler}, \citenamefont {Oliveira},
  \citenamefont {Rubio},\ and\ \citenamefont {Appel}}]{Appel19_225}%
  \BibitemOpen
  \bibfield  {author} {\bibinfo {author} {\bibfnamefont {R.}~\bibnamefont
  {Jestädt}}, \bibinfo {author} {\bibfnamefont {M.}~\bibnamefont
  {Ruggenthaler}}, \bibinfo {author} {\bibfnamefont {M.~J.~T.}\ \bibnamefont
  {Oliveira}}, \bibinfo {author} {\bibfnamefont {A.}~\bibnamefont {Rubio}}, \
  and\ \bibinfo {author} {\bibfnamefont {H.}~\bibnamefont {Appel}},\ }\href
  {\doibase 10.1080/00018732.2019.1695875} {\bibfield  {journal} {\bibinfo
  {journal} {Adv. Phys.}\ }\textbf {\bibinfo {volume} {68}},\ \bibinfo {pages}
  {225} (\bibinfo {year} {2019})}\BibitemShut {NoStop}%
\bibitem [{\citenamefont {Flick}\ and\ \citenamefont
  {Narang}(2020)}]{Narang20_094116}%
  \BibitemOpen
  \bibfield  {author} {\bibinfo {author} {\bibfnamefont {J.}~\bibnamefont
  {Flick}}\ and\ \bibinfo {author} {\bibfnamefont {P.}~\bibnamefont {Narang}},\
  }\href {\doibase 10.1063/5.0021033} {\bibfield  {journal} {\bibinfo
  {journal} {J. Chem. Phys.}\ }\textbf {\bibinfo {volume} {153}},\ \bibinfo
  {pages} {094116} (\bibinfo {year} {2020})}\BibitemShut {NoStop}%
\bibitem [{\citenamefont {Vu}\ \emph {et~al.}(2022)\citenamefont {Vu},
  \citenamefont {McLeod}, \citenamefont {Hanson},\ and\ \citenamefont
  {DePrince}}]{DePrince22_9303}%
  \BibitemOpen
  \bibfield  {author} {\bibinfo {author} {\bibfnamefont {N.}~\bibnamefont
  {Vu}}, \bibinfo {author} {\bibfnamefont {G.~M.}\ \bibnamefont {McLeod}},
  \bibinfo {author} {\bibfnamefont {K.}~\bibnamefont {Hanson}}, \ and\ \bibinfo
  {author} {\bibfnamefont {A.~E.~I.}\ \bibnamefont {DePrince}},\ }\href
  {\doibase 10.1021/acs.jpca.2c07134} {\bibfield  {journal} {\bibinfo
  {journal} {J. Phys. Chem. A}\ }\textbf {\bibinfo {volume} {126}},\ \bibinfo
  {pages} {9303} (\bibinfo {year} {2022})}\BibitemShut {NoStop}%
\bibitem [{\citenamefont {Pavošević}\ and\ \citenamefont
  {Rubio}(2022)}]{Rubio22_094101}%
  \BibitemOpen
  \bibfield  {author} {\bibinfo {author} {\bibfnamefont {F.}~\bibnamefont
  {Pavošević}}\ and\ \bibinfo {author} {\bibfnamefont {A.}~\bibnamefont
  {Rubio}},\ }\href {\doibase 10.1063/5.0095552} {\bibfield  {journal}
  {\bibinfo  {journal} {J. Chem. Phys.}\ }\textbf {\bibinfo {volume} {157}},\
  \bibinfo {pages} {094101} (\bibinfo {year} {2022})}\BibitemShut {NoStop}%
\bibitem [{\citenamefont {Liebenthal}, \citenamefont {Vu},\ and\ \citenamefont
  {{DePrince III}}(2023)}]{DePrince23_5264}%
  \BibitemOpen
  \bibfield  {author} {\bibinfo {author} {\bibfnamefont {M.~D.}\ \bibnamefont
  {Liebenthal}}, \bibinfo {author} {\bibfnamefont {N.}~\bibnamefont {Vu}}, \
  and\ \bibinfo {author} {\bibfnamefont {A.~E.}\ \bibnamefont {{DePrince
  III}}},\ }\href {\doibase 10.1021/acs.jpca.3c01842} {\bibfield  {journal}
  {\bibinfo  {journal} {J. Phys. Chem. A}\ }\textbf {\bibinfo {volume} {127}},\
  \bibinfo {pages} {5264} (\bibinfo {year} {2023})}\BibitemShut {NoStop}%
\bibitem [{\citenamefont {McTague}\ and\ \citenamefont {{Foley
  IV}}(2022)}]{Foley_154103}%
  \BibitemOpen
  \bibfield  {author} {\bibinfo {author} {\bibfnamefont {J.}~\bibnamefont
  {McTague}}\ and\ \bibinfo {author} {\bibfnamefont {J.~J.}\ \bibnamefont
  {{Foley IV}}},\ }\href@noop {} {\bibfield  {journal} {\bibinfo  {journal} {J.
  Chem. Phys.}\ }\textbf {\bibinfo {volume} {156}},\ \bibinfo {pages} {154103}
  (\bibinfo {year} {2022})}\BibitemShut {NoStop}%
\bibitem [{\citenamefont {Haugland}\ \emph {et~al.}(2021)\citenamefont
  {Haugland}, \citenamefont {Schäfer}, \citenamefont {Ronca}, \citenamefont
  {Rubio},\ and\ \citenamefont {Koch}}]{Koch21_094113}%
  \BibitemOpen
  \bibfield  {author} {\bibinfo {author} {\bibfnamefont {T.~S.}\ \bibnamefont
  {Haugland}}, \bibinfo {author} {\bibfnamefont {C.}~\bibnamefont {Schäfer}},
  \bibinfo {author} {\bibfnamefont {E.}~\bibnamefont {Ronca}}, \bibinfo
  {author} {\bibfnamefont {A.}~\bibnamefont {Rubio}}, \ and\ \bibinfo {author}
  {\bibfnamefont {H.}~\bibnamefont {Koch}},\ }\href {\doibase
  10.1063/5.0039256} {\bibfield  {journal} {\bibinfo  {journal} {J. Chem.
  Phys.}\ }\textbf {\bibinfo {volume} {154}},\ \bibinfo {pages} {094113}
  (\bibinfo {year} {2021})}\BibitemShut {NoStop}%
\bibitem [{\citenamefont {Nielsen}\ \emph {et~al.}(2018)\citenamefont
  {Nielsen}, \citenamefont {Schäfer}, \citenamefont {Ruggenthaler},\ and\
  \citenamefont {Rubio}}]{Rubio18_arxiv}%
  \BibitemOpen
  \bibfield  {author} {\bibinfo {author} {\bibfnamefont {S.~E.~B.}\
  \bibnamefont {Nielsen}}, \bibinfo {author} {\bibfnamefont {C.}~\bibnamefont
  {Schäfer}}, \bibinfo {author} {\bibfnamefont {M.}~\bibnamefont
  {Ruggenthaler}}, \ and\ \bibinfo {author} {\bibfnamefont {A.}~\bibnamefont
  {Rubio}},\ }\href {https://arxiv.org/abs/1812.00388} {\bibfield  {journal}
  {\bibinfo  {journal} {arXiv preprint}\ ,\ \bibinfo {pages} {1812.00388}}
  (\bibinfo {year} {2018})}\BibitemShut {NoStop}%
\bibitem [{\citenamefont {Buchholz}\ \emph {et~al.}(2019)\citenamefont
  {Buchholz}, \citenamefont {Theophilou}, \citenamefont {Nielsen},
  \citenamefont {Ruggenthaler},\ and\ \citenamefont {Rubio}}]{Rubio19_2694}%
  \BibitemOpen
  \bibfield  {author} {\bibinfo {author} {\bibfnamefont {F.}~\bibnamefont
  {Buchholz}}, \bibinfo {author} {\bibfnamefont {I.}~\bibnamefont
  {Theophilou}}, \bibinfo {author} {\bibfnamefont {S.~E.~B.}\ \bibnamefont
  {Nielsen}}, \bibinfo {author} {\bibfnamefont {M.}~\bibnamefont
  {Ruggenthaler}}, \ and\ \bibinfo {author} {\bibfnamefont {A.}~\bibnamefont
  {Rubio}},\ }\href {\doibase 10.1021/acsphotonics.9b00648} {\bibfield
  {journal} {\bibinfo  {journal} {ACS Photonics}\ }\textbf {\bibinfo {volume}
  {6}},\ \bibinfo {pages} {2694} (\bibinfo {year} {2019})}\BibitemShut
  {NoStop}%
\bibitem [{\citenamefont {Buchholz}\ \emph {et~al.}(2020)\citenamefont
  {Buchholz}, \citenamefont {Theophilou}, \citenamefont {Giesbertz},
  \citenamefont {Ruggenthaler},\ and\ \citenamefont {Rubio}}]{Rubio20_5601}%
  \BibitemOpen
  \bibfield  {author} {\bibinfo {author} {\bibfnamefont {F.}~\bibnamefont
  {Buchholz}}, \bibinfo {author} {\bibfnamefont {I.}~\bibnamefont
  {Theophilou}}, \bibinfo {author} {\bibfnamefont {K.~J.~H.}\ \bibnamefont
  {Giesbertz}}, \bibinfo {author} {\bibfnamefont {M.}~\bibnamefont
  {Ruggenthaler}}, \ and\ \bibinfo {author} {\bibfnamefont {A.}~\bibnamefont
  {Rubio}},\ }\href {\doibase 10.1021/acs.jctc.0c00469} {\bibfield  {journal}
  {\bibinfo  {journal} {J. Chem. Theory Comput.}\ }\textbf {\bibinfo {volume}
  {16}},\ \bibinfo {pages} {5601} (\bibinfo {year} {2020})}\BibitemShut
  {NoStop}%
\bibitem [{\citenamefont {Cohen}, \citenamefont {Mori-S{\'a}nchez},\ and\
  \citenamefont {Yang}(2008)}]{Yang08_792}%
  \BibitemOpen
  \bibfield  {author} {\bibinfo {author} {\bibfnamefont {A.~J.}\ \bibnamefont
  {Cohen}}, \bibinfo {author} {\bibfnamefont {P.}~\bibnamefont
  {Mori-S{\'a}nchez}}, \ and\ \bibinfo {author} {\bibfnamefont
  {W.}~\bibnamefont {Yang}},\ }\href {\doibase 10.1126/science.1158722}
  {\bibfield  {journal} {\bibinfo  {journal} {Science}\ }\textbf {\bibinfo
  {volume} {321}},\ \bibinfo {pages} {792} (\bibinfo {year}
  {2008})}\BibitemShut {NoStop}%
\bibitem [{\citenamefont {Spohn}(2004)}]{Spohn04_book}%
  \BibitemOpen
  \bibfield  {author} {\bibinfo {author} {\bibfnamefont {H.}~\bibnamefont
  {Spohn}},\ }\href {https://cds.cern.ch/record/803777} {\emph {\bibinfo
  {title} {{Dynamics of charged particles and their radiation field}}}}\
  (\bibinfo  {publisher} {Cambridge Univ. Press},\ \bibinfo {address}
  {Cambridge},\ \bibinfo {year} {2004})\BibitemShut {NoStop}%
\bibitem [{\citenamefont {Ruggenthaler}\ \emph {et~al.}(2018)\citenamefont
  {Ruggenthaler}, \citenamefont {Tancogne-Dejean}, \citenamefont {Flick},
  \citenamefont {Appel},\ and\ \citenamefont {Rubio}}]{Rubio18_0118}%
  \BibitemOpen
  \bibfield  {author} {\bibinfo {author} {\bibfnamefont {M.}~\bibnamefont
  {Ruggenthaler}}, \bibinfo {author} {\bibfnamefont {N.}~\bibnamefont
  {Tancogne-Dejean}}, \bibinfo {author} {\bibfnamefont {J.}~\bibnamefont
  {Flick}}, \bibinfo {author} {\bibfnamefont {H.}~\bibnamefont {Appel}}, \ and\
  \bibinfo {author} {\bibfnamefont {A.}~\bibnamefont {Rubio}},\ }\href
  {\doibase 10.1038/s41570-018-0118} {\bibfield  {journal} {\bibinfo  {journal}
  {Nat. Rev. Chem.}\ }\textbf {\bibinfo {volume} {2}},\ \bibinfo {pages} {0118}
  (\bibinfo {year} {2018})}\BibitemShut {NoStop}%
\bibitem [{\citenamefont {Mandal}, \citenamefont {Vega},\ and\ \citenamefont
  {Huo}(2020)}]{Huo20_9215}%
  \BibitemOpen
  \bibfield  {author} {\bibinfo {author} {\bibfnamefont {A.}~\bibnamefont
  {Mandal}}, \bibinfo {author} {\bibfnamefont {S.~M.}\ \bibnamefont {Vega}}, \
  and\ \bibinfo {author} {\bibfnamefont {P.}~\bibnamefont {Huo}},\ }\href@noop
  {} {\bibfield  {journal} {\bibinfo  {journal} {J. Phys. Chem. Lett.}\
  }\textbf {\bibinfo {volume} {11}},\ \bibinfo {pages} {9215} (\bibinfo {year}
  {2020})}\BibitemShut {NoStop}%
\bibitem [{\citenamefont {Hoffmann}\ \emph {et~al.}(2020)\citenamefont
  {Hoffmann}, \citenamefont {Lacombe}, \citenamefont {Rubio},\ and\
  \citenamefont {Maitra}}]{Hoffman_JCP_2020}%
  \BibitemOpen
  \bibfield  {author} {\bibinfo {author} {\bibfnamefont {N.~M.}\ \bibnamefont
  {Hoffmann}}, \bibinfo {author} {\bibfnamefont {L.}~\bibnamefont {Lacombe}},
  \bibinfo {author} {\bibfnamefont {A.}~\bibnamefont {Rubio}}, \ and\ \bibinfo
  {author} {\bibfnamefont {N.~T.}\ \bibnamefont {Maitra}},\ }\href {\doibase
  10.1063/5.0012723} {\bibfield  {journal} {\bibinfo  {journal} {J. Chem.
  Phys.}\ }\textbf {\bibinfo {volume} {153}},\ \bibinfo {pages} {104103}
  (\bibinfo {year} {2020})},\ \Eprint
  {http://arxiv.org/abs/https://pubs.aip.org/aip/jcp/article-pdf/doi/10.1063/5.0012723/16726857/104103\_1\_online.pdf}
  {https://pubs.aip.org/aip/jcp/article-pdf/doi/10.1063/5.0012723/16726857/104103\_1\_online.pdf}
  \BibitemShut {NoStop}%
\bibitem [{\citenamefont {Vidal}, \citenamefont {Manby},\ and\ \citenamefont
  {Knowles}(2022)}]{Knowles22_204119}%
  \BibitemOpen
  \bibfield  {author} {\bibinfo {author} {\bibfnamefont {M.~L.}\ \bibnamefont
  {Vidal}}, \bibinfo {author} {\bibfnamefont {F.~R.}\ \bibnamefont {Manby}}, \
  and\ \bibinfo {author} {\bibfnamefont {P.~J.}\ \bibnamefont {Knowles}},\
  }\href {\doibase 10.1063/5.0089412} {\bibfield  {journal} {\bibinfo
  {journal} {J. Chem. Phys.}\ }\textbf {\bibinfo {volume} {156}},\ \bibinfo
  {pages} {204119} (\bibinfo {year} {2022})}\BibitemShut {NoStop}%
\bibitem [{\citenamefont {Haugland}\ \emph {et~al.}(2020)\citenamefont
  {Haugland}, \citenamefont {Ronca}, \citenamefont {Kj\o{}nstad}, \citenamefont
  {Rubio},\ and\ \citenamefont {Koch}}]{Koch20_041043}%
  \BibitemOpen
  \bibfield  {author} {\bibinfo {author} {\bibfnamefont {T.~S.}\ \bibnamefont
  {Haugland}}, \bibinfo {author} {\bibfnamefont {E.}~\bibnamefont {Ronca}},
  \bibinfo {author} {\bibfnamefont {E.~F.}\ \bibnamefont {Kj\o{}nstad}},
  \bibinfo {author} {\bibfnamefont {A.}~\bibnamefont {Rubio}}, \ and\ \bibinfo
  {author} {\bibfnamefont {H.}~\bibnamefont {Koch}},\ }\href {\doibase
  10.1103/PhysRevX.10.041043} {\bibfield  {journal} {\bibinfo  {journal}
  {Physical Review X}\ }\textbf {\bibinfo {volume} {10}},\ \bibinfo {pages}
  {041043} (\bibinfo {year} {2020})}\BibitemShut {NoStop}%
\bibitem [{\citenamefont {Riso}\ \emph
  {et~al.}(2022{\natexlab{a}})\citenamefont {Riso}, \citenamefont {Haugland},
  \citenamefont {Ronda},\ and\ \citenamefont {Koch}}]{Koch22_NatComm}%
  \BibitemOpen
  \bibfield  {author} {\bibinfo {author} {\bibfnamefont {R.~R.}\ \bibnamefont
  {Riso}}, \bibinfo {author} {\bibfnamefont {T.~S.}\ \bibnamefont {Haugland}},
  \bibinfo {author} {\bibfnamefont {E.}~\bibnamefont {Ronda}}, \ and\ \bibinfo
  {author} {\bibfnamefont {H.}~\bibnamefont {Koch}},\ }\href {\doibase
  10.1038/s41467-022-29003-2} {\bibfield  {journal} {\bibinfo  {journal} {Nat.
  Commun.}\ }\textbf {\bibinfo {volume} {13}},\ \bibinfo {pages} {1368}
  (\bibinfo {year} {2022}{\natexlab{a}})}\BibitemShut {NoStop}%
\bibitem [{\citenamefont {DePrince}(2021)}]{DePrince21_094112}%
  \BibitemOpen
  \bibfield  {author} {\bibinfo {author} {\bibfnamefont {A.~E.}\ \bibnamefont
  {DePrince}},\ }\href {\doibase 10.1063/5.0038748} {\bibfield  {journal}
  {\bibinfo  {journal} {J. Chem. Phys.}\ }\textbf {\bibinfo {volume} {154}},\
  \bibinfo {pages} {094112} (\bibinfo {year} {2021})}\BibitemShut {NoStop}%
\bibitem [{\citenamefont {Mallory}\ and\ \citenamefont
  {DePrince}(2022)}]{DePrince22_053710}%
  \BibitemOpen
  \bibfield  {author} {\bibinfo {author} {\bibfnamefont {J.~D.}\ \bibnamefont
  {Mallory}}\ and\ \bibinfo {author} {\bibfnamefont {A.~E.}\ \bibnamefont
  {DePrince}},\ }\href {\doibase 10.1103/PhysRevA.106.053710} {\bibfield
  {journal} {\bibinfo  {journal} {Phys. Rev. A}\ }\textbf {\bibinfo {volume}
  {106}},\ \bibinfo {pages} {053710} (\bibinfo {year} {2022})}\BibitemShut
  {NoStop}%
\bibitem [{\citenamefont {Liebenthal}, \citenamefont {Vu},\ and\ \citenamefont
  {DePrince}(2022)}]{DePrince22_054105}%
  \BibitemOpen
  \bibfield  {author} {\bibinfo {author} {\bibfnamefont {M.~D.}\ \bibnamefont
  {Liebenthal}}, \bibinfo {author} {\bibfnamefont {N.}~\bibnamefont {Vu}}, \
  and\ \bibinfo {author} {\bibfnamefont {A.~E.}\ \bibnamefont {DePrince}},\
  }\href {\doibase 10.1063/5.0078795} {\bibfield  {journal} {\bibinfo
  {journal} {J. Chem. Phys.}\ }\textbf {\bibinfo {volume} {156}},\ \bibinfo
  {pages} {054105} (\bibinfo {year} {2022})}\BibitemShut {NoStop}%
\bibitem [{\citenamefont {Flick}\ and\ \citenamefont
  {Narang}(2018)}]{Narang18_113002}%
  \BibitemOpen
  \bibfield  {author} {\bibinfo {author} {\bibfnamefont {J.}~\bibnamefont
  {Flick}}\ and\ \bibinfo {author} {\bibfnamefont {P.}~\bibnamefont {Narang}},\
  }\href@noop {} {\bibfield  {journal} {\bibinfo  {journal} {Phys. Rev. Lett.}\
  }\textbf {\bibinfo {volume} {121}},\ \bibinfo {pages} {113002} (\bibinfo
  {year} {2018})}\BibitemShut {NoStop}%
\bibitem [{\citenamefont {Yang}\ \emph {et~al.}(2021)\citenamefont {Yang},
  \citenamefont {Ou}, \citenamefont {Pei}, \citenamefont {Wang}, \citenamefont
  {Weng}, \citenamefont {Shuai}, \citenamefont {Mullen},\ and\ \citenamefont
  {Shao}}]{Shao21_064107}%
  \BibitemOpen
  \bibfield  {author} {\bibinfo {author} {\bibfnamefont {J.}~\bibnamefont
  {Yang}}, \bibinfo {author} {\bibfnamefont {Q.}~\bibnamefont {Ou}}, \bibinfo
  {author} {\bibfnamefont {Z.}~\bibnamefont {Pei}}, \bibinfo {author}
  {\bibfnamefont {H.}~\bibnamefont {Wang}}, \bibinfo {author} {\bibfnamefont
  {B.}~\bibnamefont {Weng}}, \bibinfo {author} {\bibfnamefont {Z.}~\bibnamefont
  {Shuai}}, \bibinfo {author} {\bibfnamefont {K.}~\bibnamefont {Mullen}}, \
  and\ \bibinfo {author} {\bibfnamefont {Y.}~\bibnamefont {Shao}},\ }\href
  {\doibase 10.1063/5.0057542} {\bibfield  {journal} {\bibinfo  {journal} {J.
  Chem. Phys.}\ }\textbf {\bibinfo {volume} {155}},\ \bibinfo {pages} {064107}
  (\bibinfo {year} {2021})}\BibitemShut {NoStop}%
\bibitem [{\citenamefont {J.~McTague}, \citenamefont {{Foley IV}},\ and\
  \citenamefont {{DePrince
  III}}(2023{\natexlab{a}})}]{Foley23_QED_HF_TUTORIAL}%
  \BibitemOpen
  \bibfield  {author} {\bibinfo {author} {\bibfnamefont {J.}~\bibnamefont
  {J.~McTague}}, \bibinfo {author} {\bibfnamefont {J.~J.}\ \bibnamefont {{Foley
  IV}}}, \ and\ \bibinfo {author} {\bibfnamefont {A.~E.}\ \bibnamefont
  {{DePrince III}}},\ }\href@noop {} {\enquote {\bibinfo {title} {Hartree-fock
  self-consistent field theory for the pauli-fierz hamiltonian (qed-hf)},}\ }
  (\bibinfo {year} {2023}{\natexlab{a}}),\ \bibinfo {note}
  {https://github.com/FoleyLab/psi4polaritonic/blob/cpr/QED-HF\_Tutorial.ipynb}\BibitemShut
  {NoStop}%
\bibitem [{\citenamefont {Kohn}\ and\ \citenamefont
  {Sham}(1965)}]{Sham65_A1133}%
  \BibitemOpen
  \bibfield  {author} {\bibinfo {author} {\bibfnamefont {W.}~\bibnamefont
  {Kohn}}\ and\ \bibinfo {author} {\bibfnamefont {L.~J.}\ \bibnamefont
  {Sham}},\ }\href {\doibase 10.1103/PhysRev.140.A1133} {\bibfield  {journal}
  {\bibinfo  {journal} {Phys. Rev.}\ }\textbf {\bibinfo {volume} {140}},\
  \bibinfo {pages} {A1133} (\bibinfo {year} {1965})}\BibitemShut {NoStop}%
\bibitem [{\citenamefont {Flick}\ \emph {et~al.}(2015)\citenamefont {Flick},
  \citenamefont {Ruggenthaler}, \citenamefont {Appel},\ and\ \citenamefont
  {Rubio}}]{Rubio15_15285}%
  \BibitemOpen
  \bibfield  {author} {\bibinfo {author} {\bibfnamefont {J.}~\bibnamefont
  {Flick}}, \bibinfo {author} {\bibfnamefont {M.}~\bibnamefont {Ruggenthaler}},
  \bibinfo {author} {\bibfnamefont {H.}~\bibnamefont {Appel}}, \ and\ \bibinfo
  {author} {\bibfnamefont {A.}~\bibnamefont {Rubio}},\ }\href {\doibase
  10.1073/pnas.1518224112} {\bibfield  {journal} {\bibinfo  {journal} {Proc.
  Natl. Acad. Sci. U.S.A.}\ }\textbf {\bibinfo {volume} {112}},\ \bibinfo
  {pages} {15285} (\bibinfo {year} {2015})}\BibitemShut {NoStop}%
\bibitem [{\citenamefont {Flick}\ \emph {et~al.}(2019)\citenamefont {Flick},
  \citenamefont {Welakuh}, \citenamefont {Ruggenthaler}, \citenamefont
  {Appel},\ and\ \citenamefont {Rubio}}]{Rubio19_2757}%
  \BibitemOpen
  \bibfield  {author} {\bibinfo {author} {\bibfnamefont {J.}~\bibnamefont
  {Flick}}, \bibinfo {author} {\bibfnamefont {D.~M.}\ \bibnamefont {Welakuh}},
  \bibinfo {author} {\bibfnamefont {M.}~\bibnamefont {Ruggenthaler}}, \bibinfo
  {author} {\bibfnamefont {H.}~\bibnamefont {Appel}}, \ and\ \bibinfo {author}
  {\bibfnamefont {A.}~\bibnamefont {Rubio}},\ }\href {\doibase
  10.1021/acsphotonics.9b00768} {\bibfield  {journal} {\bibinfo  {journal} {ACS
  Photonics}\ }\textbf {\bibinfo {volume} {6}},\ \bibinfo {pages} {2757}
  (\bibinfo {year} {2019})}\BibitemShut {NoStop}%
\bibitem [{\citenamefont {Wang}\ \emph {et~al.}(2021)\citenamefont {Wang},
  \citenamefont {Neuman}, \citenamefont {Flick},\ and\ \citenamefont
  {Narang}}]{Narang21_104109}%
  \BibitemOpen
  \bibfield  {author} {\bibinfo {author} {\bibfnamefont {D.~S.}\ \bibnamefont
  {Wang}}, \bibinfo {author} {\bibfnamefont {T.}~\bibnamefont {Neuman}},
  \bibinfo {author} {\bibfnamefont {J.}~\bibnamefont {Flick}}, \ and\ \bibinfo
  {author} {\bibfnamefont {P.}~\bibnamefont {Narang}},\ }\href@noop {}
  {\bibfield  {journal} {\bibinfo  {journal} {J. Chem. Phys.}\ }\textbf
  {\bibinfo {volume} {154}},\ \bibinfo {pages} {104109} (\bibinfo {year}
  {2021})}\BibitemShut {NoStop}%
\bibitem [{\citenamefont {Dimitrov}\ \emph {et~al.}(2017)\citenamefont
  {Dimitrov}, \citenamefont {Flick}, \citenamefont {Ruggenthaler},\ and\
  \citenamefont {Rubio}}]{Rubio17_113036}%
  \BibitemOpen
  \bibfield  {author} {\bibinfo {author} {\bibfnamefont {T.}~\bibnamefont
  {Dimitrov}}, \bibinfo {author} {\bibfnamefont {J.}~\bibnamefont {Flick}},
  \bibinfo {author} {\bibfnamefont {M.}~\bibnamefont {Ruggenthaler}}, \ and\
  \bibinfo {author} {\bibfnamefont {A.}~\bibnamefont {Rubio}},\ }\href@noop {}
  {\bibfield  {journal} {\bibinfo  {journal} {New J. Phys.}\ }\textbf {\bibinfo
  {volume} {19}},\ \bibinfo {pages} {113036} (\bibinfo {year}
  {2017})}\BibitemShut {NoStop}%
\bibitem [{\citenamefont {Schäfer}\ \emph {et~al.}(2021)\citenamefont
  {Schäfer}, \citenamefont {Buchholz}, \citenamefont {Penz}, \citenamefont
  {Ruggenthaler},\ and\ \citenamefont {Rubio}}]{Rubio21_e2110464118}%
  \BibitemOpen
  \bibfield  {author} {\bibinfo {author} {\bibfnamefont {C.}~\bibnamefont
  {Schäfer}}, \bibinfo {author} {\bibfnamefont {F.}~\bibnamefont {Buchholz}},
  \bibinfo {author} {\bibfnamefont {M.}~\bibnamefont {Penz}}, \bibinfo {author}
  {\bibfnamefont {M.}~\bibnamefont {Ruggenthaler}}, \ and\ \bibinfo {author}
  {\bibfnamefont {A.}~\bibnamefont {Rubio}},\ }\href {\doibase
  10.1073/pnas.2110464118} {\bibfield  {journal} {\bibinfo  {journal} {Proc.
  Natl. Acad. Sci. U.S.A.}\ }\textbf {\bibinfo {volume} {118}},\ \bibinfo
  {pages} {e2110464118} (\bibinfo {year} {2021})}\BibitemShut {NoStop}%
\bibitem [{\citenamefont {Flick}(2022)}]{Flick22_143201}%
  \BibitemOpen
  \bibfield  {author} {\bibinfo {author} {\bibfnamefont {J.}~\bibnamefont
  {Flick}},\ }\href {\doibase 10.1103/PhysRevLett.129.143201} {\bibfield
  {journal} {\bibinfo  {journal} {Phys. Rev. Lett.}\ }\textbf {\bibinfo
  {volume} {129}},\ \bibinfo {pages} {143201} (\bibinfo {year}
  {2022})}\BibitemShut {NoStop}%
\bibitem [{\citenamefont {Flick}\ \emph {et~al.}(2017)\citenamefont {Flick},
  \citenamefont {Ruggenthaler}, \citenamefont {Appel},\ and\ \citenamefont
  {Rubio}}]{Rubio17_3026}%
  \BibitemOpen
  \bibfield  {author} {\bibinfo {author} {\bibfnamefont {J.}~\bibnamefont
  {Flick}}, \bibinfo {author} {\bibfnamefont {M.}~\bibnamefont {Ruggenthaler}},
  \bibinfo {author} {\bibfnamefont {H.}~\bibnamefont {Appel}}, \ and\ \bibinfo
  {author} {\bibfnamefont {A.}~\bibnamefont {Rubio}},\ }\href@noop {}
  {\bibfield  {journal} {\bibinfo  {journal} {Proc. Natl. Acad. Sci. USA}\
  }\textbf {\bibinfo {volume} {114}},\ \bibinfo {pages} {3026} (\bibinfo {year}
  {2017})}\BibitemShut {NoStop}%
\bibitem [{\citenamefont {Tokatly}(2018)}]{Tokatly18_235123}%
  \BibitemOpen
  \bibfield  {author} {\bibinfo {author} {\bibfnamefont {I.~V.}\ \bibnamefont
  {Tokatly}},\ }\href {\doibase 10.1103/PhysRevB.98.235123} {\bibfield
  {journal} {\bibinfo  {journal} {Phys. Rev. B}\ }\textbf {\bibinfo {volume}
  {98}},\ \bibinfo {pages} {235123} (\bibinfo {year} {2018})}\BibitemShut
  {NoStop}%
\bibitem [{\citenamefont {Malave}\ \emph
  {et~al.}(2022{\natexlab{a}})\citenamefont {Malave}, \citenamefont {Ahrens},
  \citenamefont {Pitagora}, \citenamefont {Covington},\ and\ \citenamefont
  {Varga}}]{Varga22_194106}%
  \BibitemOpen
  \bibfield  {author} {\bibinfo {author} {\bibfnamefont {J.}~\bibnamefont
  {Malave}}, \bibinfo {author} {\bibfnamefont {A.}~\bibnamefont {Ahrens}},
  \bibinfo {author} {\bibfnamefont {D.}~\bibnamefont {Pitagora}}, \bibinfo
  {author} {\bibfnamefont {C.}~\bibnamefont {Covington}}, \ and\ \bibinfo
  {author} {\bibfnamefont {K.}~\bibnamefont {Varga}},\ }\href {\doibase
  10.1063/5.0123909} {\bibfield  {journal} {\bibinfo  {journal} {J. Chem.
  Phys.}\ }\textbf {\bibinfo {volume} {157}},\ \bibinfo {pages} {194106}
  (\bibinfo {year} {2022}{\natexlab{a}})}\BibitemShut {NoStop}%
\bibitem [{\citenamefont {Sidler}\ \emph {et~al.}(2020)\citenamefont {Sidler},
  \citenamefont {Schäfer}, \citenamefont {Ruggenthaler},\ and\ \citenamefont
  {Rubio}}]{Rubio20_508}%
  \BibitemOpen
  \bibfield  {author} {\bibinfo {author} {\bibfnamefont {D.}~\bibnamefont
  {Sidler}}, \bibinfo {author} {\bibfnamefont {C.}~\bibnamefont {Schäfer}},
  \bibinfo {author} {\bibfnamefont {M.}~\bibnamefont {Ruggenthaler}}, \ and\
  \bibinfo {author} {\bibfnamefont {A.}~\bibnamefont {Rubio}},\ }\href@noop {}
  {\bibfield  {journal} {\bibinfo  {journal} {J. Phys. Chem. Lett.}\ }\textbf
  {\bibinfo {volume} {12}},\ \bibinfo {pages} {508} (\bibinfo {year}
  {2020})}\BibitemShut {NoStop}%
\bibitem [{\citenamefont {Yang}\ \emph {et~al.}(2022)\citenamefont {Yang},
  \citenamefont {Pei}, \citenamefont {Leon}, \citenamefont {Wickizer},
  \citenamefont {Weng}, \citenamefont {Mao}, \citenamefont {Ou},\ and\
  \citenamefont {Shao}}]{Shao22_124104}%
  \BibitemOpen
  \bibfield  {author} {\bibinfo {author} {\bibfnamefont {J.}~\bibnamefont
  {Yang}}, \bibinfo {author} {\bibfnamefont {Z.}~\bibnamefont {Pei}}, \bibinfo
  {author} {\bibfnamefont {E.~C.}\ \bibnamefont {Leon}}, \bibinfo {author}
  {\bibfnamefont {C.}~\bibnamefont {Wickizer}}, \bibinfo {author}
  {\bibfnamefont {B.}~\bibnamefont {Weng}}, \bibinfo {author} {\bibfnamefont
  {Y.}~\bibnamefont {Mao}}, \bibinfo {author} {\bibfnamefont {Q.}~\bibnamefont
  {Ou}}, \ and\ \bibinfo {author} {\bibfnamefont {Y.}~\bibnamefont {Shao}},\
  }\href@noop {} {\bibfield  {journal} {\bibinfo  {journal} {J. Chem. Phys.}\
  }\textbf {\bibinfo {volume} {156}},\ \bibinfo {pages} {124104} (\bibinfo
  {year} {2022})}\BibitemShut {NoStop}%
\bibitem [{\citenamefont {J.~McTague}, \citenamefont {{Foley IV}},\ and\
  \citenamefont {{DePrince
  III}}(2023{\natexlab{b}})}]{Foley23_QED_CIS_TUTORIAL}%
  \BibitemOpen
  \bibfield  {author} {\bibinfo {author} {\bibfnamefont {J.}~\bibnamefont
  {J.~McTague}}, \bibinfo {author} {\bibfnamefont {J.~J.}\ \bibnamefont {{Foley
  IV}}}, \ and\ \bibinfo {author} {\bibfnamefont {A.~E.}\ \bibnamefont
  {{DePrince III}}},\ }\href@noop {} {\enquote {\bibinfo {title} {Configuration
  interaction with single electronic and photonic excitations applied to the
  pauli-fierz hamiltonian},}\ } (\bibinfo {year} {2023}{\natexlab{b}}),\
  \bibinfo {note}
  {https://github.com/FoleyLab/psi4polaritonic/blob/cpr/QED-CIS-1.ipynb}\BibitemShut
  {NoStop}%
\bibitem [{\citenamefont {{\v{C}}{\'\i}{\v{z}}ek}(1966)}]{Cizek66_4256}%
  \BibitemOpen
  \bibfield  {author} {\bibinfo {author} {\bibfnamefont {J.}~\bibnamefont
  {{\v{C}}{\'\i}{\v{z}}ek}},\ }\href {\doibase 10.1063/1.1727484} {\bibfield
  {journal} {\bibinfo  {journal} {J. Chem. Phys.}\ }\textbf {\bibinfo {volume}
  {45}},\ \bibinfo {pages} {4256} (\bibinfo {year} {1966})}\BibitemShut
  {NoStop}%
\bibitem [{\citenamefont {{\v{C}}{\'\i}{\v{z}}ek}\ and\ \citenamefont
  {Paldus}(1971)}]{Paldus71_359}%
  \BibitemOpen
  \bibfield  {author} {\bibinfo {author} {\bibfnamefont {J.}~\bibnamefont
  {{\v{C}}{\'\i}{\v{z}}ek}}\ and\ \bibinfo {author} {\bibfnamefont
  {J.}~\bibnamefont {Paldus}},\ }\href@noop {} {\bibfield  {journal} {\bibinfo
  {journal} {Int. J. Quantum Chem.}\ }\textbf {\bibinfo {volume} {5}},\
  \bibinfo {pages} {359} (\bibinfo {year} {1971})}\BibitemShut {NoStop}%
\bibitem [{\citenamefont {Shavitt}\ and\ \citenamefont
  {Bartlett}(2009)}]{Bartlett09_book}%
  \BibitemOpen
  \bibfield  {author} {\bibinfo {author} {\bibfnamefont {I.}~\bibnamefont
  {Shavitt}}\ and\ \bibinfo {author} {\bibfnamefont {R.}~\bibnamefont
  {Bartlett}},\ }\href {https://books.google.com/books?id=SWw6ac1NHZYC} {\emph
  {\bibinfo {title} {Many-Body Methods in Chemistry and Physics: MBPT and
  Coupled-Cluster Theory}}},\ Cambridge Molecular Science\ (\bibinfo
  {publisher} {Cambridge University Press},\ \bibinfo {year}
  {2009})\BibitemShut {NoStop}%
\bibitem [{\citenamefont {Bartlett}\ and\ \citenamefont
  {Musia\l{}}(2007)}]{Musial07_291}%
  \BibitemOpen
  \bibfield  {author} {\bibinfo {author} {\bibfnamefont {R.~J.}\ \bibnamefont
  {Bartlett}}\ and\ \bibinfo {author} {\bibfnamefont {M.}~\bibnamefont
  {Musia\l{}}},\ }\href {\doibase 10.1103/RevModPhys.79.291} {\bibfield
  {journal} {\bibinfo  {journal} {Rev. Mod. Phys.}\ }\textbf {\bibinfo {volume}
  {79}},\ \bibinfo {pages} {291} (\bibinfo {year} {2007})}\BibitemShut
  {NoStop}%
\bibitem [{\citenamefont {Stanton}\ and\ \citenamefont
  {Bartlett}(1993)}]{Bartlett93_7029}%
  \BibitemOpen
  \bibfield  {author} {\bibinfo {author} {\bibfnamefont {J.~F.}\ \bibnamefont
  {Stanton}}\ and\ \bibinfo {author} {\bibfnamefont {R.~J.}\ \bibnamefont
  {Bartlett}},\ }\href {\doibase 10.1063/1.464746} {\bibfield  {journal}
  {\bibinfo  {journal} {J. Chem. Phys.}\ }\textbf {\bibinfo {volume} {98}},\
  \bibinfo {pages} {7029} (\bibinfo {year} {1993})}\BibitemShut {NoStop}%
\bibitem [{\citenamefont {Bartlett}(2012)}]{Bartlett12_126}%
  \BibitemOpen
  \bibfield  {author} {\bibinfo {author} {\bibfnamefont {R.~J.}\ \bibnamefont
  {Bartlett}},\ }\href {\doibase 10.1002/wcms.76} {\bibfield  {journal}
  {\bibinfo  {journal} {WIREs Computational Molecular Science}\ }\textbf
  {\bibinfo {volume} {2}},\ \bibinfo {pages} {126} (\bibinfo {year} {2012})},\
  \Eprint
  {http://arxiv.org/abs/https://onlinelibrary.wiley.com/doi/pdf/10.1002/wcms.76}
  {https://onlinelibrary.wiley.com/doi/pdf/10.1002/wcms.76} \BibitemShut
  {NoStop}%
\bibitem [{\citenamefont {Krylov}(2008)}]{Krylov08_433}%
  \BibitemOpen
  \bibfield  {author} {\bibinfo {author} {\bibfnamefont {A.~I.}\ \bibnamefont
  {Krylov}},\ }\href {\doibase 10.1146/annurev.physchem.59.032607.093602}
  {\bibfield  {journal} {\bibinfo  {journal} {Annual Review of Physical
  Chemistry}\ }\textbf {\bibinfo {volume} {59}},\ \bibinfo {pages} {433}
  (\bibinfo {year} {2008})}\BibitemShut {NoStop}%
\bibitem [{\citenamefont {Monkhorst}(1977)}]{Monkhorst77_421}%
  \BibitemOpen
  \bibfield  {author} {\bibinfo {author} {\bibfnamefont {H.~J.}\ \bibnamefont
  {Monkhorst}},\ }\href@noop {} {\bibfield  {journal} {\bibinfo  {journal}
  {Int. J. Quantum Chem.}\ }\textbf {\bibinfo {volume} {12}},\ \bibinfo {pages}
  {421} (\bibinfo {year} {1977})}\BibitemShut {NoStop}%
\bibitem [{\citenamefont {Mukherjee}\ and\ \citenamefont
  {Mukherjee}(1979)}]{Mukherjee79_325}%
  \BibitemOpen
  \bibfield  {author} {\bibinfo {author} {\bibfnamefont {D.}~\bibnamefont
  {Mukherjee}}\ and\ \bibinfo {author} {\bibfnamefont {P.}~\bibnamefont
  {Mukherjee}},\ }\href@noop {} {\bibfield  {journal} {\bibinfo  {journal}
  {Chem. Phys.}\ }\textbf {\bibinfo {volume} {39}},\ \bibinfo {pages} {325}
  (\bibinfo {year} {1979})}\BibitemShut {NoStop}%
\bibitem [{\citenamefont {Dalgaard}\ and\ \citenamefont
  {Monkhorst}(1983)}]{Monkhorst83_1217}%
  \BibitemOpen
  \bibfield  {author} {\bibinfo {author} {\bibfnamefont {E.}~\bibnamefont
  {Dalgaard}}\ and\ \bibinfo {author} {\bibfnamefont {H.~J.}\ \bibnamefont
  {Monkhorst}},\ }\href {\doibase 10.1103/PhysRevA.28.1217} {\bibfield
  {journal} {\bibinfo  {journal} {Phys. Rev. A}\ }\textbf {\bibinfo {volume}
  {28}},\ \bibinfo {pages} {1217} (\bibinfo {year} {1983})}\BibitemShut
  {NoStop}%
\bibitem [{\citenamefont {Koch}\ and\ \citenamefont
  {J{\o}rgensen}(1990)}]{Jorgensen90_3333}%
  \BibitemOpen
  \bibfield  {author} {\bibinfo {author} {\bibfnamefont {H.}~\bibnamefont
  {Koch}}\ and\ \bibinfo {author} {\bibfnamefont {P.}~\bibnamefont
  {J{\o}rgensen}},\ }\href@noop {} {\bibfield  {journal} {\bibinfo  {journal}
  {J. Chem. Phys.}\ }\textbf {\bibinfo {volume} {93}},\ \bibinfo {pages} {3333}
  (\bibinfo {year} {1990})}\BibitemShut {NoStop}%
\bibitem [{\citenamefont {Koch}\ \emph {et~al.}(1990)\citenamefont {Koch},
  \citenamefont {Jensen}, \citenamefont {J{\o}rgensen},\ and\ \citenamefont
  {Helgaker}}]{Helgaker90_3345}%
  \BibitemOpen
  \bibfield  {author} {\bibinfo {author} {\bibfnamefont {H.}~\bibnamefont
  {Koch}}, \bibinfo {author} {\bibfnamefont {H.~J.~A.}\ \bibnamefont {Jensen}},
  \bibinfo {author} {\bibfnamefont {P.}~\bibnamefont {J{\o}rgensen}}, \ and\
  \bibinfo {author} {\bibfnamefont {T.}~\bibnamefont {Helgaker}},\ }\href
  {\doibase 10.1063/1.458815} {\bibfield  {journal} {\bibinfo  {journal} {J.
  Chem. Phys.}\ }\textbf {\bibinfo {volume} {93}},\ \bibinfo {pages} {3345}
  (\bibinfo {year} {1990})}\BibitemShut {NoStop}%
\bibitem [{\citenamefont {Pedersen}\ and\ \citenamefont
  {Koch}(1997)}]{Koch97_8059}%
  \BibitemOpen
  \bibfield  {author} {\bibinfo {author} {\bibfnamefont {T.~B.}\ \bibnamefont
  {Pedersen}}\ and\ \bibinfo {author} {\bibfnamefont {H.}~\bibnamefont
  {Koch}},\ }\href@noop {} {\bibfield  {journal} {\bibinfo  {journal} {J. Chem.
  Phys.}\ }\textbf {\bibinfo {volume} {106}},\ \bibinfo {pages} {8059}
  (\bibinfo {year} {1997})}\BibitemShut {NoStop}%
\bibitem [{\citenamefont {Christiansen}, \citenamefont {Koch},\ and\
  \citenamefont {J{\o}rgensen}(1995)}]{Jorgensen95_7429}%
  \BibitemOpen
  \bibfield  {author} {\bibinfo {author} {\bibfnamefont {O.}~\bibnamefont
  {Christiansen}}, \bibinfo {author} {\bibfnamefont {H.}~\bibnamefont {Koch}},
  \ and\ \bibinfo {author} {\bibfnamefont {P.}~\bibnamefont {J{\o}rgensen}},\
  }\href {\doibase 10.1063/1.470315} {\bibfield  {journal} {\bibinfo  {journal}
  {J. Chem. Phys.}\ }\textbf {\bibinfo {volume} {103}},\ \bibinfo {pages}
  {7429} (\bibinfo {year} {1995})}\BibitemShut {NoStop}%
\bibitem [{\citenamefont {Paw{\l}owski}, \citenamefont {Olsen},\ and\
  \citenamefont {J{\o}rgensen}(2019)}]{Jorgensen19_134109}%
  \BibitemOpen
  \bibfield  {author} {\bibinfo {author} {\bibfnamefont {F.}~\bibnamefont
  {Paw{\l}owski}}, \bibinfo {author} {\bibfnamefont {J.}~\bibnamefont {Olsen}},
  \ and\ \bibinfo {author} {\bibfnamefont {P.}~\bibnamefont {J{\o}rgensen}},\
  }\href {\doibase 10.1063/1.5053167} {\bibfield  {journal} {\bibinfo
  {journal} {J. Chem. Phys.}\ }\textbf {\bibinfo {volume} {150}},\ \bibinfo
  {pages} {134109} (\bibinfo {year} {2019})}\BibitemShut {NoStop}%
\bibitem [{\citenamefont {Mordovina}\ \emph {et~al.}(2020)\citenamefont
  {Mordovina}, \citenamefont {Bungey}, \citenamefont {Appel}, \citenamefont
  {Knowles}, \citenamefont {Rubio},\ and\ \citenamefont
  {Manby}}]{Manby20_023262}%
  \BibitemOpen
  \bibfield  {author} {\bibinfo {author} {\bibfnamefont {U.}~\bibnamefont
  {Mordovina}}, \bibinfo {author} {\bibfnamefont {C.}~\bibnamefont {Bungey}},
  \bibinfo {author} {\bibfnamefont {H.}~\bibnamefont {Appel}}, \bibinfo
  {author} {\bibfnamefont {P.~J.}\ \bibnamefont {Knowles}}, \bibinfo {author}
  {\bibfnamefont {A.}~\bibnamefont {Rubio}}, \ and\ \bibinfo {author}
  {\bibfnamefont {F.~R.}\ \bibnamefont {Manby}},\ }\href {\doibase
  10.1103/PhysRevResearch.2.023262} {\bibfield  {journal} {\bibinfo  {journal}
  {Phys. Rev. Res.}\ }\textbf {\bibinfo {volume} {2}},\ \bibinfo {pages}
  {023262} (\bibinfo {year} {2020})}\BibitemShut {NoStop}%
\bibitem [{\citenamefont {Purvis}\ and\ \citenamefont
  {Bartlett}(1982)}]{Bartlett82_1910}%
  \BibitemOpen
  \bibfield  {author} {\bibinfo {author} {\bibfnamefont {G.~D.}\ \bibnamefont
  {Purvis}}\ and\ \bibinfo {author} {\bibfnamefont {R.~J.}\ \bibnamefont
  {Bartlett}},\ }\href@noop {} {\bibfield  {journal} {\bibinfo  {journal} {J.
  Chem. Phys.}\ }\textbf {\bibinfo {volume} {76}},\ \bibinfo {pages} {1910}
  (\bibinfo {year} {1982})}\BibitemShut {NoStop}%
\bibitem [{\citenamefont {Pavošević}\ and\ \citenamefont
  {Flick}(2021)}]{Flick21_9100}%
  \BibitemOpen
  \bibfield  {author} {\bibinfo {author} {\bibfnamefont {F.}~\bibnamefont
  {Pavošević}}\ and\ \bibinfo {author} {\bibfnamefont {J.}~\bibnamefont
  {Flick}},\ }\href {\doibase 10.1021/acs.jpclett.1c02659} {\bibfield
  {journal} {\bibinfo  {journal} {J. Phys. Chem. Lett.}\ }\textbf {\bibinfo
  {volume} {12}},\ \bibinfo {pages} {9100} (\bibinfo {year}
  {2021})}\BibitemShut {NoStop}%
\bibitem [{\citenamefont {Peruzzo}\ \emph {et~al.}(2014)\citenamefont
  {Peruzzo}, \citenamefont {McClean}, \citenamefont {Shadbolt}, \citenamefont
  {Yung}, \citenamefont {Zhou}, \citenamefont {Love}, \citenamefont
  {Aspuru-Guzik},\ and\ \citenamefont {O'brien}}]{Obrien14_4213}%
  \BibitemOpen
  \bibfield  {author} {\bibinfo {author} {\bibfnamefont {A.}~\bibnamefont
  {Peruzzo}}, \bibinfo {author} {\bibfnamefont {J.}~\bibnamefont {McClean}},
  \bibinfo {author} {\bibfnamefont {P.}~\bibnamefont {Shadbolt}}, \bibinfo
  {author} {\bibfnamefont {M.-H.}\ \bibnamefont {Yung}}, \bibinfo {author}
  {\bibfnamefont {X.-Q.}\ \bibnamefont {Zhou}}, \bibinfo {author}
  {\bibfnamefont {P.~J.}\ \bibnamefont {Love}}, \bibinfo {author}
  {\bibfnamefont {A.}~\bibnamefont {Aspuru-Guzik}}, \ and\ \bibinfo {author}
  {\bibfnamefont {J.~L.}\ \bibnamefont {O'brien}},\ }\href@noop {} {\bibfield
  {journal} {\bibinfo  {journal} {Nature Communications}\ }\textbf {\bibinfo
  {volume} {5}},\ \bibinfo {pages} {4213} (\bibinfo {year} {2014})}\BibitemShut
  {NoStop}%
\bibitem [{\citenamefont {Yung}\ \emph {et~al.}(2014)\citenamefont {Yung},
  \citenamefont {Casanova}, \citenamefont {Mezzacapo}, \citenamefont {McClean},
  \citenamefont {Lamata}, \citenamefont {Aspuru-Guzik},\ and\ \citenamefont
  {Solano}}]{Solano14_3589}%
  \BibitemOpen
  \bibfield  {author} {\bibinfo {author} {\bibfnamefont {M.-H.}\ \bibnamefont
  {Yung}}, \bibinfo {author} {\bibfnamefont {J.}~\bibnamefont {Casanova}},
  \bibinfo {author} {\bibfnamefont {A.}~\bibnamefont {Mezzacapo}}, \bibinfo
  {author} {\bibfnamefont {J.}~\bibnamefont {McClean}}, \bibinfo {author}
  {\bibfnamefont {L.}~\bibnamefont {Lamata}}, \bibinfo {author} {\bibfnamefont
  {A.}~\bibnamefont {Aspuru-Guzik}}, \ and\ \bibinfo {author} {\bibfnamefont
  {E.}~\bibnamefont {Solano}},\ }\href {\doibase 10.1038/srep03589} {\bibfield
  {journal} {\bibinfo  {journal} {Scientific Reports}\ }\textbf {\bibinfo
  {volume} {4}},\ \bibinfo {pages} {3589} (\bibinfo {year} {2014})}\BibitemShut
  {NoStop}%
\bibitem [{\citenamefont {McClean}\ \emph {et~al.}(2016)\citenamefont
  {McClean}, \citenamefont {Romero}, \citenamefont {Babbush},\ and\
  \citenamefont {Aspuru-Guzik}}]{Aspuru-Guzik16_023023}%
  \BibitemOpen
  \bibfield  {author} {\bibinfo {author} {\bibfnamefont {J.~R.}\ \bibnamefont
  {McClean}}, \bibinfo {author} {\bibfnamefont {J.}~\bibnamefont {Romero}},
  \bibinfo {author} {\bibfnamefont {R.}~\bibnamefont {Babbush}}, \ and\
  \bibinfo {author} {\bibfnamefont {A.}~\bibnamefont {Aspuru-Guzik}},\ }\href
  {\doibase 10.1088/1367-2630/18/2/023023} {\bibfield  {journal} {\bibinfo
  {journal} {New Journal of Physics}\ }\textbf {\bibinfo {volume} {18}},\
  \bibinfo {pages} {023023} (\bibinfo {year} {2016})}\BibitemShut {NoStop}%
\bibitem [{\citenamefont {Riso}\ \emph
  {et~al.}(2022{\natexlab{b}})\citenamefont {Riso}, \citenamefont {Haugland},
  \citenamefont {Ronca},\ and\ \citenamefont {Koch}}]{Koch22_234103}%
  \BibitemOpen
  \bibfield  {author} {\bibinfo {author} {\bibfnamefont {R.~R.}\ \bibnamefont
  {Riso}}, \bibinfo {author} {\bibfnamefont {T.~S.}\ \bibnamefont {Haugland}},
  \bibinfo {author} {\bibfnamefont {E.}~\bibnamefont {Ronca}}, \ and\ \bibinfo
  {author} {\bibfnamefont {H.}~\bibnamefont {Koch}},\ }\href {\doibase
  10.1063/5.0091119} {\bibfield  {journal} {\bibinfo  {journal} {J. Chem.
  Phys.}\ }\textbf {\bibinfo {volume} {156}},\ \bibinfo {pages} {234103}
  (\bibinfo {year} {2022}{\natexlab{b}})}\BibitemShut {NoStop}%
\bibitem [{\citenamefont {Pavošević}\ \emph {et~al.}(2022)\citenamefont
  {Pavošević}, \citenamefont {Hammes-Schiffer}, \citenamefont {Rubio},\ and\
  \citenamefont {Flick}}]{Flick22_4995}%
  \BibitemOpen
  \bibfield  {author} {\bibinfo {author} {\bibfnamefont {F.}~\bibnamefont
  {Pavošević}}, \bibinfo {author} {\bibfnamefont {S.}~\bibnamefont
  {Hammes-Schiffer}}, \bibinfo {author} {\bibfnamefont {A.}~\bibnamefont
  {Rubio}}, \ and\ \bibinfo {author} {\bibfnamefont {J.}~\bibnamefont
  {Flick}},\ }\href {\doibase 10.1021/jacs.1c13201} {\bibfield  {journal}
  {\bibinfo  {journal} {J. Amer. Chem. Soc.}\ }\textbf {\bibinfo {volume}
  {144}},\ \bibinfo {pages} {4995} (\bibinfo {year} {2022})}\BibitemShut
  {NoStop}%
\bibitem [{\citenamefont {White}\ \emph {et~al.}(2020)\citenamefont {White},
  \citenamefont {Gao}, \citenamefont {Minnich},\ and\ \citenamefont
  {Chan}}]{Chan20_224112}%
  \BibitemOpen
  \bibfield  {author} {\bibinfo {author} {\bibfnamefont {A.~F.}\ \bibnamefont
  {White}}, \bibinfo {author} {\bibfnamefont {Y.}~\bibnamefont {Gao}}, \bibinfo
  {author} {\bibfnamefont {A.~J.}\ \bibnamefont {Minnich}}, \ and\ \bibinfo
  {author} {\bibfnamefont {G.~K.-L.}\ \bibnamefont {Chan}},\ }\href {\doibase
  10.1063/5.0033132} {\bibfield  {journal} {\bibinfo  {journal} {J. Chem.
  Phys.}\ }\textbf {\bibinfo {volume} {153}},\ \bibinfo {pages} {224112}
  (\bibinfo {year} {2020})}\BibitemShut {NoStop}%
\bibitem [{\citenamefont {Philbin}\ \emph {et~al.}(2022)\citenamefont
  {Philbin}, \citenamefont {Haugland}, \citenamefont {Ghosh}, \citenamefont
  {Ronca}, \citenamefont {Chen}, \citenamefont {Narang},\ and\ \citenamefont
  {Koch}}]{Koch22_arXiv:2209.07956}%
  \BibitemOpen
  \bibfield  {author} {\bibinfo {author} {\bibfnamefont {J.~P.}\ \bibnamefont
  {Philbin}}, \bibinfo {author} {\bibfnamefont {T.~S.}\ \bibnamefont
  {Haugland}}, \bibinfo {author} {\bibfnamefont {T.~K.}\ \bibnamefont {Ghosh}},
  \bibinfo {author} {\bibfnamefont {E.}~\bibnamefont {Ronca}}, \bibinfo
  {author} {\bibfnamefont {M.}~\bibnamefont {Chen}}, \bibinfo {author}
  {\bibfnamefont {P.}~\bibnamefont {Narang}}, \ and\ \bibinfo {author}
  {\bibfnamefont {H.}~\bibnamefont {Koch}},\ }\href@noop {} {\bibfield
  {journal} {\bibinfo  {journal} {arXiv preprint}\ ,\ \bibinfo {pages}
  {arXiv:2209.07956}} (\bibinfo {year} {2022})}\BibitemShut {NoStop}%
\bibitem [{\citenamefont {Riso}\ \emph
  {et~al.}(2022{\natexlab{c}})\citenamefont {Riso}, \citenamefont {Grazioli},
  \citenamefont {Ronca}, \citenamefont {Giovannini},\ and\ \citenamefont
  {Koch}}]{Koch22_chiral}%
  \BibitemOpen
  \bibfield  {author} {\bibinfo {author} {\bibfnamefont {R.~R.}\ \bibnamefont
  {Riso}}, \bibinfo {author} {\bibfnamefont {L.}~\bibnamefont {Grazioli}},
  \bibinfo {author} {\bibfnamefont {E.}~\bibnamefont {Ronca}}, \bibinfo
  {author} {\bibfnamefont {T.}~\bibnamefont {Giovannini}}, \ and\ \bibinfo
  {author} {\bibfnamefont {H.}~\bibnamefont {Koch}},\ }\href {\doibase
  10.48550/ARXIV.2209.01987} {\bibfield  {journal} {\bibinfo  {journal} {arXiv
  preprint}\ ,\ \bibinfo {pages} {arXiv:2209.01987}} (\bibinfo {year}
  {2022}{\natexlab{c}})}\BibitemShut {NoStop}%
\bibitem [{\citenamefont {Nooijen}\ and\ \citenamefont
  {Bartlett}(1995)}]{Bartlett95_3629}%
  \BibitemOpen
  \bibfield  {author} {\bibinfo {author} {\bibfnamefont {M.}~\bibnamefont
  {Nooijen}}\ and\ \bibinfo {author} {\bibfnamefont {R.~J.}\ \bibnamefont
  {Bartlett}},\ }\href {\doibase 10.1063/1.468592} {\bibfield  {journal}
  {\bibinfo  {journal} {J. Chem. Phys.}\ }\textbf {\bibinfo {volume} {102}},\
  \bibinfo {pages} {3629} (\bibinfo {year} {1995})}\BibitemShut {NoStop}%
\bibitem [{\citenamefont {Fregoni}\ \emph {et~al.}(2021)\citenamefont
  {Fregoni}, \citenamefont {Haugland}, \citenamefont {Pipolo}, \citenamefont
  {Giovannini}, \citenamefont {Koch},\ and\ \citenamefont
  {Corni}}]{Corni21_6664}%
  \BibitemOpen
  \bibfield  {author} {\bibinfo {author} {\bibfnamefont {J.}~\bibnamefont
  {Fregoni}}, \bibinfo {author} {\bibfnamefont {T.~S.}\ \bibnamefont
  {Haugland}}, \bibinfo {author} {\bibfnamefont {S.}~\bibnamefont {Pipolo}},
  \bibinfo {author} {\bibfnamefont {T.}~\bibnamefont {Giovannini}}, \bibinfo
  {author} {\bibfnamefont {H.}~\bibnamefont {Koch}}, \ and\ \bibinfo {author}
  {\bibfnamefont {S.}~\bibnamefont {Corni}},\ }\href {\doibase
  10.1021/acs.nanolett.1c02162} {\bibfield  {journal} {\bibinfo  {journal}
  {Nano Letters}\ }\textbf {\bibinfo {volume} {21}},\ \bibinfo {pages} {6664}
  (\bibinfo {year} {2021})}\BibitemShut {NoStop}%
\bibitem [{\citenamefont {Mennucci}\ and\ \citenamefont
  {Corni}(2019)}]{Corni19_315}%
  \BibitemOpen
  \bibfield  {author} {\bibinfo {author} {\bibfnamefont {B.}~\bibnamefont
  {Mennucci}}\ and\ \bibinfo {author} {\bibfnamefont {S.}~\bibnamefont
  {Corni}},\ }\href {\doibase 10.1038/s41570-019-0092-4} {\bibfield  {journal}
  {\bibinfo  {journal} {Nat. Rev. Chem.}\ }\textbf {\bibinfo {volume} {3}},\
  \bibinfo {pages} {315} (\bibinfo {year} {2019})}\BibitemShut {NoStop}%
\bibitem [{\citenamefont {Galego}\ \emph {et~al.}(2019)\citenamefont {Galego},
  \citenamefont {Climent}, \citenamefont {Garcia-Vidal},\ and\ \citenamefont
  {Feist}}]{Feist19_021057}%
  \BibitemOpen
  \bibfield  {author} {\bibinfo {author} {\bibfnamefont {J.}~\bibnamefont
  {Galego}}, \bibinfo {author} {\bibfnamefont {C.}~\bibnamefont {Climent}},
  \bibinfo {author} {\bibfnamefont {F.~J.}\ \bibnamefont {Garcia-Vidal}}, \
  and\ \bibinfo {author} {\bibfnamefont {J.}~\bibnamefont {Feist}},\ }\href
  {\doibase 10.1103/PhysRevX.9.021057} {\bibfield  {journal} {\bibinfo
  {journal} {Physical Review X}\ }\textbf {\bibinfo {volume} {9}},\ \bibinfo
  {pages} {021057} (\bibinfo {year} {2019})}\BibitemShut {NoStop}%
\bibitem [{\citenamefont {Feist}, \citenamefont {Fernández-Domínguez},\ and\
  \citenamefont {García-Vidal}(2021)}]{Fesit21_477}%
  \BibitemOpen
  \bibfield  {author} {\bibinfo {author} {\bibfnamefont {J.}~\bibnamefont
  {Feist}}, \bibinfo {author} {\bibfnamefont {A.~I.}\ \bibnamefont
  {Fernández-Domínguez}}, \ and\ \bibinfo {author} {\bibfnamefont {F.~J.}\
  \bibnamefont {García-Vidal}},\ }\href@noop {} {\bibfield  {journal}
  {\bibinfo  {journal} {Nanophotonics}\ }\textbf {\bibinfo {volume} {10}},\
  \bibinfo {pages} {477} (\bibinfo {year} {2021})}\BibitemShut {NoStop}%
\bibitem [{\citenamefont {Romanelli}\ \emph {et~al.}(2023)\citenamefont
  {Romanelli}, \citenamefont {Riso}, \citenamefont {Haugland}, \citenamefont
  {Ronca}, \citenamefont {Corni},\ and\ \citenamefont
  {Koch}}]{Koch23_2302.05381}%
  \BibitemOpen
  \bibfield  {author} {\bibinfo {author} {\bibfnamefont {M.}~\bibnamefont
  {Romanelli}}, \bibinfo {author} {\bibfnamefont {R.~R.}\ \bibnamefont {Riso}},
  \bibinfo {author} {\bibfnamefont {T.~S.}\ \bibnamefont {Haugland}}, \bibinfo
  {author} {\bibfnamefont {E.}~\bibnamefont {Ronca}}, \bibinfo {author}
  {\bibfnamefont {S.}~\bibnamefont {Corni}}, \ and\ \bibinfo {author}
  {\bibfnamefont {H.}~\bibnamefont {Koch}},\ }\href@noop {} {\bibfield
  {journal} {\bibinfo  {journal} {arXiv preprint}\ ,\ \bibinfo {pages}
  {arXiv:2302.05381}} (\bibinfo {year} {2023})}\BibitemShut {NoStop}%
\bibitem [{\citenamefont {Nascimento}\ and\ \citenamefont {{DePrince
  III}}(2015)}]{DePrince15_214104}%
  \BibitemOpen
  \bibfield  {author} {\bibinfo {author} {\bibfnamefont {D.~R.}\ \bibnamefont
  {Nascimento}}\ and\ \bibinfo {author} {\bibfnamefont {A.~E.}\ \bibnamefont
  {{DePrince III}}},\ }\href {\doibase 10.1063/1.4936348} {\bibfield  {journal}
  {\bibinfo  {journal} {J. Chem. Phys.}\ }\textbf {\bibinfo {volume} {143}},\
  \bibinfo {pages} {214104} (\bibinfo {year} {2015})}\BibitemShut {NoStop}%
\bibitem [{\citenamefont {Bauer}\ and\ \citenamefont
  {Dreuw}(2023)}]{Dreuw23_124128}%
  \BibitemOpen
  \bibfield  {author} {\bibinfo {author} {\bibfnamefont {M.}~\bibnamefont
  {Bauer}}\ and\ \bibinfo {author} {\bibfnamefont {A.}~\bibnamefont {Dreuw}},\
  }\href {\doibase 10.1063/5.0142403} {\bibfield  {journal} {\bibinfo
  {journal} {J. Chem. Phys.}\ }\textbf {\bibinfo {volume} {158}},\ \bibinfo
  {pages} {124128} (\bibinfo {year} {2023})}\BibitemShut {NoStop}%
\bibitem [{\citenamefont {Haugland}\ \emph {et~al.}(2023)\citenamefont
  {Haugland}, \citenamefont {Philbin}, \citenamefont {Ghosh}, \citenamefont
  {Chen}, \citenamefont {Koch},\ and\ \citenamefont
  {Narang}}]{Narang23_arXiv:2307.14822}%
  \BibitemOpen
  \bibfield  {author} {\bibinfo {author} {\bibfnamefont {T.~S.}\ \bibnamefont
  {Haugland}}, \bibinfo {author} {\bibfnamefont {J.~P.}\ \bibnamefont
  {Philbin}}, \bibinfo {author} {\bibfnamefont {T.~K.}\ \bibnamefont {Ghosh}},
  \bibinfo {author} {\bibfnamefont {M.}~\bibnamefont {Chen}}, \bibinfo {author}
  {\bibfnamefont {H.}~\bibnamefont {Koch}}, \ and\ \bibinfo {author}
  {\bibfnamefont {P.}~\bibnamefont {Narang}},\ }\href@noop {} {\bibfield
  {journal} {\bibinfo  {journal} {arXiv preprint}\ ,\ \bibinfo {pages}
  {arXiv:2307.14822}} (\bibinfo {year} {2023})}\BibitemShut {NoStop}%
\bibitem [{\citenamefont {Li}, \citenamefont {Subotnik},\ and\ \citenamefont
  {Nitzan}(2020)}]{Li_PNAS_2022}%
  \BibitemOpen
  \bibfield  {author} {\bibinfo {author} {\bibfnamefont {T.~E.}\ \bibnamefont
  {Li}}, \bibinfo {author} {\bibfnamefont {J.~E.}\ \bibnamefont {Subotnik}}, \
  and\ \bibinfo {author} {\bibfnamefont {A.}~\bibnamefont {Nitzan}},\ }\href
  {\doibase 10.1073/pnas.2009272117} {\bibfield  {journal} {\bibinfo  {journal}
  {Proc. Nat. Acad. Sci. USA}\ }\textbf {\bibinfo {volume} {117}},\ \bibinfo
  {pages} {18324} (\bibinfo {year} {2020})}\BibitemShut {NoStop}%
\bibitem [{\citenamefont {Pscherer}\ \emph {et~al.}(2021)\citenamefont
  {Pscherer}, \citenamefont {Meierhofer}, \citenamefont {Wang}, \citenamefont
  {Kelkar}, \citenamefont {Mart\'{\i}n-Cano}, \citenamefont {Utikal},
  \citenamefont {G\"otzinger},\ and\ \citenamefont
  {Sandoghdar}}]{Sandoghdar_PRL_2021}%
  \BibitemOpen
  \bibfield  {author} {\bibinfo {author} {\bibfnamefont {A.}~\bibnamefont
  {Pscherer}}, \bibinfo {author} {\bibfnamefont {M.}~\bibnamefont
  {Meierhofer}}, \bibinfo {author} {\bibfnamefont {D.}~\bibnamefont {Wang}},
  \bibinfo {author} {\bibfnamefont {H.}~\bibnamefont {Kelkar}}, \bibinfo
  {author} {\bibfnamefont {D.}~\bibnamefont {Mart\'{\i}n-Cano}}, \bibinfo
  {author} {\bibfnamefont {T.}~\bibnamefont {Utikal}}, \bibinfo {author}
  {\bibfnamefont {S.}~\bibnamefont {G\"otzinger}}, \ and\ \bibinfo {author}
  {\bibfnamefont {V.}~\bibnamefont {Sandoghdar}},\ }\href {\doibase
  10.1103/PhysRevLett.127.133603} {\bibfield  {journal} {\bibinfo  {journal}
  {Phys. Rev. Lett.}\ }\textbf {\bibinfo {volume} {127}},\ \bibinfo {pages}
  {133603} (\bibinfo {year} {2021})}\BibitemShut {NoStop}%
\bibitem [{\citenamefont {Malave}\ \emph
  {et~al.}(2022{\natexlab{b}})\citenamefont {Malave}, \citenamefont {Ahrens},
  \citenamefont {Pitagora}, \citenamefont {Covington},\ and\ \citenamefont
  {Varga}}]{Varga_JCP_2022}%
  \BibitemOpen
  \bibfield  {author} {\bibinfo {author} {\bibfnamefont {J.}~\bibnamefont
  {Malave}}, \bibinfo {author} {\bibfnamefont {A.}~\bibnamefont {Ahrens}},
  \bibinfo {author} {\bibfnamefont {D.}~\bibnamefont {Pitagora}}, \bibinfo
  {author} {\bibfnamefont {C.}~\bibnamefont {Covington}}, \ and\ \bibinfo
  {author} {\bibfnamefont {K.}~\bibnamefont {Varga}},\ }\href {\doibase
  10.1063/5.0123909} {\bibfield  {journal} {\bibinfo  {journal} {J. Chem.
  Phys.}\ }\textbf {\bibinfo {volume} {157}},\ \bibinfo {pages} {194106}
  (\bibinfo {year} {2022}{\natexlab{b}})}\BibitemShut {NoStop}%
\end{thebibliography}%

\end{document}